\documentclass[usenatbib,iop,numberedappendix]{aeb_emulateapj_2010}
\usepackage{epsfig}
\usepackage{xspace}

  \SetSymbolFont{symbols}{bold}{OMS}{cmsy}{b}{n}
  \DeclareSymbolFont{bmisymbols}{OML}{cmm}{b}{it}
  \DeclareMathSymbol{\balpha}{0}{bmisymbols}{"0B}
  \DeclareMathSymbol{\bbeta}{0}{bmisymbols}{"0C}
  \DeclareMathSymbol{\bgamma}{0}{bmisymbols}{"0D}
  \DeclareMathSymbol{\bdelta}{0}{bmisymbols}{"0E}
  \DeclareMathSymbol{\bepsilon}{0}{bmisymbols}{"0F}
  \DeclareMathSymbol{\bzeta}{0}{bmisymbols}{"10}
  \DeclareMathSymbol{\boldeta}{0}{bmisymbols}{"11}
  \DeclareMathSymbol{\btheta}{0}{bmisymbols}{"12}
  \DeclareMathSymbol{\biota}{0}{bmisymbols}{"13}
  \DeclareMathSymbol{\bkappa}{0}{bmisymbols}{"14}
  \DeclareMathSymbol{\blambda}{0}{bmisymbols}{"15}
  \DeclareMathSymbol{\bmu}{0}{bmisymbols}{"16}
  \DeclareMathSymbol{\bnu}{0}{bmisymbols}{"17}
  \DeclareMathSymbol{\bxi}{0}{bmisymbols}{"18}
  \DeclareMathSymbol{\bpi}{0}{bmisymbols}{"19}
  \DeclareMathSymbol{\brho}{0}{bmisymbols}{"1A}
  \DeclareMathSymbol{\bsigma}{0}{bmisymbols}{"1B}
  \DeclareMathSymbol{\btau}{0}{bmisymbols}{"1C}
  \DeclareMathSymbol{\bupsilon}{0}{bmisymbols}{"1D}
  \DeclareMathSymbol{\bphi}{0}{bmisymbols}{"1E}
  \DeclareMathSymbol{\bchi}{0}{bmisymbols}{"1F}
  \DeclareMathSymbol{\bpsi}{0}{bmisymbols}{"20}
  \DeclareMathSymbol{\bomega}{0}{bmisymbols}{"21}
  \DeclareMathSymbol{\bvarepsilon}{0}{bmisymbols}{"22}
  \DeclareMathSymbol{\bvartheta}{0}{bmisymbols}{"23}
  \DeclareMathSymbol{\bvarpi}{0}{bmisymbols}{"24}
  \DeclareMathSymbol{\bvarrho}{0}{bmisymbols}{"25}
  \DeclareMathSymbol{\bvarsigma}{0}{bmisymbols}{"26}
  \DeclareMathSymbol{\bvarphi}{0}{bmisymbols}{"27}

\def\YM{$Y$-$M$\xspace}

\newcommand{\mathbfit}[1]{\textbf{\textit{#1}}}

\newcommand{\elct}{\mathrm{e}}
\newcommand{\rmn}{\mathrm}
\newcommand{\dd}{\mathrm{d}}
\newcommand{\vecbf}{\mathbfit}

\newcommand{\bvel}{\bupsilon}
\newcommand{\hsev}{h_{70}}

\newcommand{\e}{{\rm e}}
\newcommand{\p}{{\rm p}}
\newcommand{\KU}{K/U}
\newcommand{\ca}{c/a}
\newcommand{\ba}{b/a}

\def\avrg#1{{\langle #1 \rangle}}

\def\Trace{{\rm Tr}}

\def\spose#1{\hbox to 0pt{#1\hss}}
\def\lta{\mathrel{\spose{\lower 3pt\hbox{$\mathchar"218$}}
     \raise 2.0pt\hbox{$\mathchar"13C$}}}
\def\gta{\mathrel{\spose{\lower 3pt\hbox{$\mathchar"218$}}
     \raise 2.0pt\hbox{$\mathchar"13E$}}}

\slugcomment{Submitted to ApJ}

\shorttitle{On the Cluster Physics of Sunyaev Zel'dovich Surveys I}
\shortauthors{{Battaglia, Bond, Pfrommer, Sievers}}

\begin{document}

\title{On the Cluster Physics of Sunyaev-Zel'dovich Surveys I:\\
The Influence of Feedback, Non-thermal Pressure and Cluster Shapes on \YM Scaling Relations}

\author{N. Battaglia\altaffilmark{1,2,3}, J. R. Bond\altaffilmark{2}, C. Pfrommer\altaffilmark{4,2},  J. L. Sievers\altaffilmark{2,5}}

\altaffiltext{1}{Department of Astronomy and Astrophysics, University of Toronto, 50 St George, Toronto ON, Canada, M5S 3H4}
\altaffiltext{2}{Canadian Institute for Theoretical Astrophysics, 60 St George, Toronto ON, Canada, M5S 3H8}
\altaffiltext{3}{McWilliams Center for Cosmology, Carnegie Mellon University,
Department of Physics, 5000 Forbes Ave., Pittsburgh PA, USA, 15213}
\altaffiltext{4}{Heidelberg Institute for Theoretical Studies, Schloss-Wolfsbrunnenweg 35, D-69118 Heidelberg, Germany}
\altaffiltext{5}{Joseph Henry Laboratories of Physics, Jadwin Hall, Princeton University, Princeton NJ, USA, 08544}

\begin{abstract}

  The utility of large Sunyaev Zel'dovich (SZ) surveys for determining
  cosmological parameters from cluster abundances is limited by the theoretical
  uncertainties in the integrated SZ-flux-to-mass relation, \YM. We explore how
  non-thermal pressure and the anisotropic shape of the gas distribution of the
  intracluster medium (ICM) impacts \YM scaling using a suite of hydrodynamical
  TreePM-SPH simulations of the cosmic web in large periodic boxes. We contrast
  results for models with different treatments of entropy injection and
  transport, varying radiative cooling, star formation and accompanying
  supernova feedback, cosmic rays, and energetic feedback from active galactic
  nuclei (AGN) and/or starbursts. We find that the gas kinetic-to-thermal
  pressure ratio from internal bulk motions depends on the cluster mass, and
  increases in the outer-cluster due to enhanced substructure, as does the
  asphericity of the ICM gas (which is substantially more pronounced for the
  dark matter). The asphericity is less dependent on the mass and on variations
  in the simulated physics. With only a $\sim 5-10$\% correction to projected
  (observable) ellipticities, we can infer the 3D ellipticities.  We find radii
  around $R_{500}$ -- within which the mean density is 500 times the critical
  density -- are the most robust for studying virial properties of clusters,
  being far enough out to avoid the complex ``short-distance'' physics of the
  cluster core, having a relatively low non-thermal to thermal pressure ratio
  ($\sim20$\%), and having the smallest variance of gas ellipticity as cluster
  mass and redshift vary.  Our simulated \YM-slope roughly follows the
  self-similar $Y\sim M^{5/3}$ prediction, except for a steepening due to a
  deficit of gas in lower mass clusters at low redshift in our AGN-feedback
  simulations.  The overall \YM amplitudes with AGN feedback and radiative
  cooling are lower than for the shock-heating-only case, by $\sim 30$\%. AGN
  feedback enhances slightly the overall \YM-scatter, from $\sim 11$\% to $\sim
  13$\%, a reflection of accretion history variations due to cluster merging.
  The scatter falls back to $\sim 11$\% if we select clusters with lower kinetic
  pressure. If we split the cluster system into lower, middle and upper bands of
  $P_\rmn{kin}/P_\rmn{th}$, we find a $\sim 10$\% effect on \YM.  A 3-split on
  asymmetry as measured by the long-to-short axis ratio has a $<10$\% effect on
  \YM, but using 3D-sphericalized estimates instead of projected (cylindrical)
  has a a $\sim 30$\% effect. Identifying observable second parameters related
  to internal bulk flows and anisotropy for cluster-selection to minimize
  \YM-scatter in a (fuzzy) ``fundamental plane'' would allow tighter
  cosmological parameter constraints.

\end{abstract}

\keywords{Cosmic Microwave Background --- Cosmology: Theory ---
  Galaxies: Clusters: General --- Large-Scale Structure of Universe
   --- Methods: Numerical}

\section{Introduction}

Clusters are the largest gravitationally-collapsed objects in the universe,
forming at sites of constructive interference of long waves in the primordial
density fluctuations, the coherent peak-patches \citep{1986ApJ...304...15B,1996ApJS..103....1B}. The interiors
are separated from the Hubble-flow, but maintain contact with the nearby cosmic
web through ongoing accretion and mergers as they evolve.  Although there is a
strong internal baryon-to-dark-matter density bias, a consequence of
collisional-to-collisionless physics, when cluster-scale-averaged
the smoothed densities are nearly in the universal Hubble-volume-smoothed
proportion. Clusters have proven to be useful cosmological probes as the rarest
collapsed-event tracers of the growth of structure in the universe, with a
well-defined number count that steeply falls as mass and redshift increase. The
number density tail is very sensitive to changes in cosmological parameters and
primordial non-Gaussianity.  In clusters most of the baryons are in the form of
a hot diffuse plasma, the intracluster medium (ICM). The remaining baryons are
housed in the cluster's numerous stars and galaxies. Observations of gas in the
cluster system not only reveal the detailed astrophysical processes at work in
the ICM, but the counts derived as a function of the global-cluster-observables
such as thermal energy content can allow for a high precision probe of the
cosmological parameters defining the count density shape and amplitude.

This scheme of using the cluster system for cosmology must rely on simulations capturing the physics at work: the clusters have been revealed to be too complex for simple sphericalized analytical modelling as the observations have been progressively refined and resolutions improved. We hope that the basic global observables will be sufficiently robust to the high resolution complexities that the cluster/group system can be cosmologically useful, but that must be demonstrated by detailed theoretical work with a necessarily heavy computational component.  Direct observation of mass or gravitational energy or overall binding energy of clusters would be ideal, but we are stuck with what can be observed, in the optical, X-ray and microwave/radio/sub-mm. Each derived observable from these windows into clusters is fraught with complication that requires a computational understanding. A thermal Sunyaev-Zel'dovich (SZ)   \citep{1970Ap&SS...7....3S} probe directly observes the integrated Compton-$y$ parameter which is a measure of the cluster's global gas heat-energy content, a volume-average of the thermal gas pressure, and this is related to gravitational energy through the virial relation, so it might be expected to provide a robust probe. In this paper, and the following sequence, BBPS2,3,4 \citep{battinprep2,battinprep3,battinprep4}, a follow-on to \citep{2010ApJ...725...91B}, we focus on the SZ effect, the Compton
up-scattering of cosmic microwave background (CMB) photons by hot
electrons with its unique signature of a spatially-varying distortion  of the CMB spectrum, a
decrement in thermodynamic
temperature at frequencies below $\sim 220$~GHz, and an excess above. The SZ signal is
proportional to the integrated electron pressure, so the hot gas of the ICM
dominates the effect.  The SZ surface brightness is independent of
the redshift a specific cluster is at. Hence SZ surveys have a different selection function in redshift and
in mass than X-ray and optical cluster surveys do, being generically more sensitive to higher redshift clusters. The combination of the three probes can provide  
more robust, tighter constraints on cosmological parameters than any one can alone. 

In a large cluster survey there is a wealth of information contained on
cosmology and structure formation. The abundance of clusters, their distribution
in redshift, and their spatial clustering should be determined purely by the
geometry of the universe, the power spectrum of initial density fluctuations,
and cosmological parameters such as 
the {\it rms} amplitude of the (linear) density power spectrum on cluster-mass
scales, $\sigma_8$, the mass-energy density in baryons, dark matter, and dark energy, and the equation of state of the latter. In SZ surveys, the number counts as a function of the total SZ flux (integrated Compton-$y$ parameter $Y$) and redshift
and the angular power spectrum are two complementary probes of cosmology
\citep[e.g.,][]{1999PhR...310...97B, 2002ARA&A..40..643C}.  Identifying clusters
through blind SZ surveys and measuring their integrated power spectrum have been
long term goals in CMB research, and are reaching fruition through,
\textit{e.g.}, the South Pole Telescope, SPT
\citep[e.g.,][]{2010ApJ...719.1045L,2010arXiv1012.4788S,2011arXiv1105.3182K,2010ApJ...722.1180V},
the Atacama Cosmology Telescope, ACT
\citep[e.g.,][]{2010arXiv1001.2934T,2010arXiv1009.0866D,2010arXiv1010.1065M},
and the {\em Planck} satellite
\citep[e.g.,][]{2011arXiv1101.2024P,2011arXiv1101.2043P,2011arXiv1101.2026P}.
To determine cosmological parameters from number
counts requires understanding the relationship of SZ observables such as total SZ flux to 
fundamental cluster properties such as mass $M$. And to determine them from the SZ power spectrum 
requires knowing the sum of the squares of pressure profiles 
of unresolved groups and clusters, as well as of resolved ones, weighted by the counts. Both depend sensitively on $\sigma_8$,  hence can provide an independent measure of it. Such extraction from the SZ probes is inevitably entangled with the 
uncertainties in the astrophysical properties of the ICM. This paper deconstructs the influence of various physical processes on the
\YM scaling relation. BBPS2 does the same for the SZ power spectrum.

Previous work attempted to calibrate the \YM scaling relation through
observations
\citep[e.g.,][]{2004ApJ...617..829B,2008ApJ...675..106B,2009ApJ...701L.114M,2010arXiv1006.3068A,2011ApJ...728...39S,2011arXiv1107.5115M},
self-calibration techniques
\citep[e.g.,][]{2003ApJ...585..603M,2004ApJ...613...41M,2004PhRvD..70d3504L,2011ApJ...728L..41C,2011arXiv1105.2826N},
simulations
\citep[e.g.,][]{2004MNRAS.348.1401D,2005ApJ...623L..63M,2006MNRAS.370.1309S,2006MNRAS.370.1713S,2007MNRAS.378.1248B}
and analytical approaches \citep{2007ApJ...663..139B,2008ApJ...686..206S,2011ApJ...728L..35M}.  Combining
 \YM scaling relations so determined with the survey selection function and marginalizing
over associated statistical and systematic uncertainties can enable
 accurate determination of cosmological parameters.  Using a small sample of SZ-clusters, SPT
\citep{2010ApJ...722.1180V} and ACT \citep{2011ApJ...732...44S}  determined some
cosmological constraints, e.g.,  on $\sigma_8$. However, the errors on $\sigma_8$ are dominated by systematic
uncertainties in the underlying cluster physics, making this approach not 
competitive with  other cosmological probes. Hence in order to improve
upon the determination of cosmological parameters,  a better
understanding of the mass proxies and their scatter is needed 
\citep{2006ApJ...650..538N,2008ApJ...686..206S,2010ApJ...715.1508S,2010ApJ...725.1124Y,2011arXiv1107.5740K}.

\subsection{ICM processes}

Clusters have been increasingly revealed to be complex systems as the data has progressively improved, necessitating a revision of the simplified pictures popular  in the eighties for interpreting the data. For example, pioneering work by \citet{1986MNRAS.222..323K} assumed that clusters were
self-similar systems with the mass determining their ICM thermodynamic
properties. As shown by subsequent X-ray observations, this self-similar description is
broken, especially on group scales; low-mass systems are less luminous in
comparison to the self-similar expectation \citep[see][for a review]{Voit2005}.
Studying  how non-thermal processes such as
magnetic fields, cosmic rays, active galactic nuclei (AGN), star formation,
radiative cooling and bulk motions contribute to the energy balance and
thermodynamic stability within clusters  is a very active research field.  It remains unclear how these
processes vary with cluster  radius or dynamical state. State-of-the-art simulations are about the only tool available for building a consistent
picture of clusters.  Here we contribute to these non-thermal studies by using our simulations to explore the three effects
 that influence the \YM scaling relation. These are the feedback
processes that appear to be necessary to explain the thermodynamic
characteristics of the ICM and avoid a cooling catastrophe leading to too much star formation, 
non-thermal pressure support from bulk motions internal to the clusters that are a natural consequence of a dynamically evolving structure formation 
hierarchy, and deviations from spherical symmetry. 

\subsection{Energetic feedback}

In many clusters the ICM cooling times are much shorter than a Hubble
time \citep{1994ARA&A..32..277F, Cavagnolo+2009}, which should cause extremely
high star formation rates that are well beyond what is observed. However,
current simulations with {\em only} radiative cooling and star formation
excessively over-cool cluster centers
\citep[e.g.,][]{1998ApJ...507...16S,2000ApJ...536..623L,2000MNRAS.317.1029P},
even with the addition of supernova feedback. This leads to too many stars in
the cluster cores, an unphysical rearrangement of the thermal and hydrodynamic
structure, and creates problems when comparing simulations to observations, in
particular for the entropy and pressure profiles.  Self-regulated, inhomogeneous
energy feedback mechanisms by, e.g., AGN are very successful in globally
stabilizing the group and cluster "atmospheres", and, in particular, preventing the
cooling catastrophe \citep{Churazov2001}.  Observations of cool core galaxy
clusters show evidence for AGN-moderation of the cooling and AGN feedback can potentially 
heat the surrounding ICM from kpc sized bubbles to hundreds of kpc sized outbursts
\citep{2005Natur.433...45M}. In hydrodynamical simulations, it has been shown
that incorporating a sub-grid for AGN feedback can resolve the over-cooling
problem
\citep[e.g.,][]{2007MNRAS.380..877S,Sijacki+2008,2010ApJ...725...91B,2011MNRAS.412.1965M}. The
effects of AGN feedback on the ICM will mainly alter the cluster and group cores, where
the actual physics is poorly resolved and understood.  These effects can seen to be
dramatic in X-ray observations \citep[e.g.,][]{2003MNRAS.344L..43F}, since the emission is 
proportional to gas density squared.  Since the SZ signal is proportional to the
gas pressure, these effects are smaller. Hence, AGN feedback should only perturb
the integrated thermal SZ signal, with an amplitude that is not yet known.

\subsection{Non-thermal pressure support}

Studying non-thermal pressure support from bulk motion in clusters has a long
history and was first noticed in simulations by \citet{1990ApJ...363..349E}, who
showed that estimates for the binding mass of a cluster using a hydrostatic
isothermal $\beta$-model in comparison to a fit to the surface brightness
profile differed by 15\%.  They found that inclusion of velocity dispersion in
the hydrostatic isothermal $\beta$-model reconciled this difference between
binding masses.  Including the support from residual gas motions in the
hydrostatic cluster mass estimator improved the match with the true cluster mass
\citep{2004MNRAS.351..237R}, with increasing kinetic pressure at larger cluster
radii \citep{2009ApJ...705.1129L}. The amount of energy in these bulk motions
are of the order of 20\% to 30\% at radii of interest for cosmology
\citep{2010ApJ...725...91B,2010ApJ...721.1105B}. However, kinetic pressure
support has only recently been included in analytical and semi-analytical
templates for the thermal SZ power spectrum
\citep{2010ApJ...725.1452S,2011ApJ...727...94T}. Of course cosmological
hydrodynamical simulations fully include this contribution and thus do not
require additional modeling of kinetic pressure effects. While of importance for
correctly interpreting SZ measurements, the X-ray observations of clusters have
been calibrated to partly take this effect into account when determining mass
from the X-ray inferred total thermal energy \citep[e.g., using the $Y_{X}$-$M$
relation,][]{2006ApJ...650..128K}. 

In this paper, our focus is on the effects of bulk motions within clusters.
These dominate the total kinetic pressure budget since there is generally a
smaller fraction of energy in a hydrodynamical turbulent cascade compared to the
energy on the injection scale which is well resolved for the relevant large
scale motions.  Quantifying turbulence in clusters is becoming feasible with
simulations that include the modeling of sub-grid turbulence
\citep[e.g.,][]{2008MNRAS.388.1089I} as well as simulations with
magnetohydrodynamics and anisotropic thermal conduction
\citep{2011arXiv1109.1285P}.

The method of smoothed particle hydrodynamics (SPH) that we are using for
solving the inviscid Euler equations in this work is perfectly suited for
studying large-scale bulk motions which dominate the kinetic pressure support
due to its Lagrangian and conservative nature. However, it is known that SPH in
its standard implementation poorly resolves hydrodynamical instabilities, such
as of Kelvin-Helmholtz or Rayleigh-Taylor type \citep{2007MNRAS.380..963A}. In
non-radiative simulations of cluster formation, adaptively-refined mesh codes
generate a larger core entropy level in comparison to SPH simulations which is
presumably due to the difference in the amount of mixing in SPH and mesh codes
and possibly related to a different treatment of vorticity in the simulations
\citep[e.g.,][]{Frenk+1999,2009MNRAS.395..180M, 2011arXiv1106.2159V}. A recent
comparison of a galaxy formation simulation with the SPH technique and the
recently developed moving mesh code AREPO \citep{2010MNRAS.401..791S} enabled --
for the first time -- to test the nature of the hydrodynamic solver with
otherwise identical implementations of the gravity solver, the sub-resolution
physics, and the detailed form of the initial conditions.  The moving mesh
calculations resulted in more disk-like galaxy morphologies in comparison to
SPH. This difference originated from an artificially high heating rate with SPH
in the outer parts of haloes, caused by viscous dissipation of inherent sonic
velocity noise of neighboring SPH particles, an efficient damping of subsonic
turbulence injected in the halo infall region potentially by the artificial
viscosity employed by SPH, and because of a higher efficiency of gas stripping
in AREPO \citep{2011arXiv1109.1281V}. Based on these results it may be
questionable to analyse small-scale velocity power spectra and 2-point structure
functions in SPH simulations; however, the dominating large-scale bulk
motions which is our primary interest should be followed accurately. This
theoretical expectation is confirmed by analyzing otherwise identical cluster
simulations run with AREPO and GADGET/SPH that show an equal kinetic pressure
contribution outside the core region ($r>0.05 R_{200}$) for the two numerical
techniques (Puchwein \& Springel, priv. comm., 2011).

\subsection{ICM shapes}

DM halo shapes have been studied extensively. For SZ observations, the shape of
the gas distribution of the ICM is important, especially in the far field of the
intracluster medium which contributes substantially to the total integrated SZ
flux \citep{2010ApJ...725...91B}. The assumption of spherical symmetry is often
made when calculating cluster properties from observations and in analytical
prescriptions so we would like to assess its
validity. Semi-analytic models that employ the full three-dimensional
information of a dissipationless dark-matter-only simulation use the shape of the resulting
gravitational cluster potentials, so it is important to study how such shapes compare with those that include the dissipational gas component. 
Recent numerical work has shown the impact of cooling and star-formation
on the properties of ICM shape for a sample size of 16 clusters
\citep{2011ApJ...734...93L}, however, such a study has not been extended to a
larger sample. Furthermore, the question of how energetic feedback in the
cluster cores affects ICM shapes has not been addressed.

\subsection{Overview}

In this work we explore a large statistical sample of simulated clusters with
identical initial conditions but employing different models for sub-grid
physics. We quantify the importance of non-thermal pressure support and ICM
shapes on the SZ \YM scaling relation.  In Section~\ref{sec:sim}, we briefly
describe the simulations and sub-grid physics used. We present our results for
non-thermal pressure support from bulk motions and ICM shapes in Sections
\ref{sec:Pkin} and \ref{sec:Elpt}, respectively. The impact of these processes
and changes in the simulated physics on the \YM scaling relation is presented in
Section \ref{sec:SZ}. In Section~\ref{sec:conclusion} we summarize our results
and conclude.

\section{Cosmological simulations and cluster data set}
\label{sec:sim}

We described the basic suite of hydrodynamical models used in
\citet{2010ApJ...725...91B}. We simulate tens of large-scale boxes of
the cosmic web in order to improve our statistics of the number of
objects while simultaneously aiming for a sufficiently high mass
resolution to map out the core regions of those clusters and groups
which are the target of current SZ cluster surveys and which dominate
the SZ power spectrum signal on scales larger than 1'. Here we
characterize the average behaviour of the properties of the ICM over a
large mass and redshift range using a modified version of the GADGET-2
\citep{2005MNRAS.364.1105S} code which employs SPH and treePM for the
gravity solver.  For each modeled physics, we simulate a sequence of
10 boxes of side length $165\,h^{-1}\,\rmn{Mpc} = 200\, \rmn{Mpc} $
with periodic boundary conditions, encompassing
$N_\rmn{DM}=N_{\mathrm{gas}} = 256^3$ DM and gas particles.  This
gives an initial gas particle mass of $m_{\mathrm{gas}}= 3.2\times
10^9\, h^{-1}\,\mathrm{M}_{\sun}$ and a DM particle mass of
$m_{\mathrm{DM}}= 1.54\times 10^{10}\, h^{-1}\,\mathrm{M}_{\sun}$. We
adopt a minimum comoving (Plummer) gravitational smoothing length of
$\varepsilon_\rmn{s}=20\, h^{-1}\,$kpc.  Our SPH densities are
computed with 32 neighbours. For our standard calculations, we adopt a
tilted $\Lambda$CDM cosmology, with total matter density (in units of
the critical)
$\Omega_{\mathrm{m}}=\Omega_{\mathrm{DM}}+\Omega_{\mathrm{b}} = 0.25$,
baryon density $\Omega_{\mathrm{b}}$ = 0.043, cosmological constant
$\Omega_{\Lambda}$ = 0.75, a present day Hubble constant of $H_0 = 100
h \mbox{ km s}^{-1} \mbox{ Mpc}^{-1}$ with $h=0.7$, a spectral index
of the primordial power-spectrum $n_{\rmn{s}}$ = 0.96 and $\sigma_8$ =
0.8.

We compare results for three variants of simulated physics: (1) the
classic non-radiative `adiabatic' case with only gravitational
formation {\it shock heating}; (2) an extended {\it radiative cooling}
case with star formation, supernova (SN) energy feedback and cosmic
rays (CRs) from structure formation shocks \citep[for more information
on CRs,
see][]{2006MNRAS.367..113P,2007MNRAS.378..385P,2007A&A...473...41E,2008A&A...481...33J};
(3) {\it AGN feedback} in addition to radiative cooling, star
formation, and SN feedback. Radiative cooling and heating were
computed assuming an optically thin gas of primordial composition in a
time-dependent, spatially uniform ultraviolet background. Star
formation and supernovae feedback were modelled using the hybrid
multiphase model for the interstellar medium of
\citet{2003MNRAS.339..289S}. The CR population is modelled as a
relativistic population of protons described by an isotropic power-law
distribution function in momentum space with a spectral index of
$\alpha=2.3$, following \citet{2007A&A...473...41E}. With those
parameters, the CR pressure causes a small reduction in the integrated
Compton-$y$ parameter \citep{2007MNRAS.378..385P}, but can result in
interesting modifications of the local intracluster $y$-map.

The AGN feedback prescription we adopt for our standard simulations \citep[for more
details see][]{2010ApJ...725...91B} allows for lower resolution and hence can be
applied to large-scale structure simulations.  It couples the black hole
accretion rate to the global star formation rate (SFR) of the cluster, as
suggested by \citet{2005ApJ...630..167T}. If the SFR is larger than an
observationally motivated threshold, $\dot{M}_*>5\rmn{M}_\sun\,\rmn{yr}^{-1}$,
the thermal energy is injected into the ICM at a rate which is proportional to
the SFR within a given spherical region.  The AGN feedback in these box
simulations injects approximately one third of total injected energy in the
cluster formation phases at $z >2$ (analogous to high-z QSO like feedback),
another third in the redshift range $1<z<2$, and the final third below $z=1$
(analogous to jet/bubble like feedback). These fractions depend moderately on
the the numerical resolution; increasing the resolution enables to resolve the
growth of smaller halos at earlier times and causes a higher fraction of energy
injection at higher redshifts \citep[see][for a discussion]{{2010ApJ...725...91B}}.

\begin{figure*}[thbp]
  \resizebox{0.5\hsize}{!}{\includegraphics{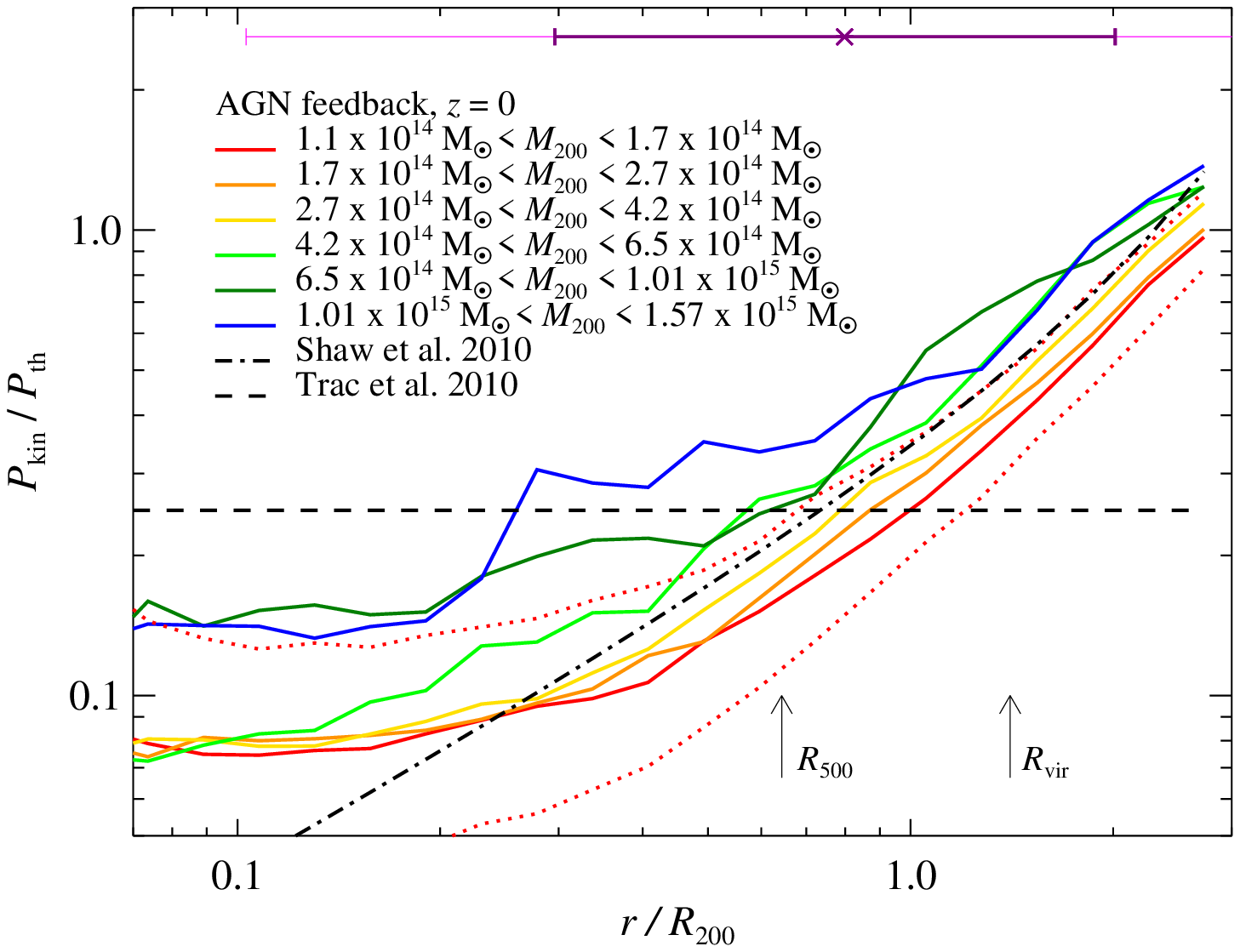}}%
  \resizebox{0.5\hsize}{!}{\includegraphics{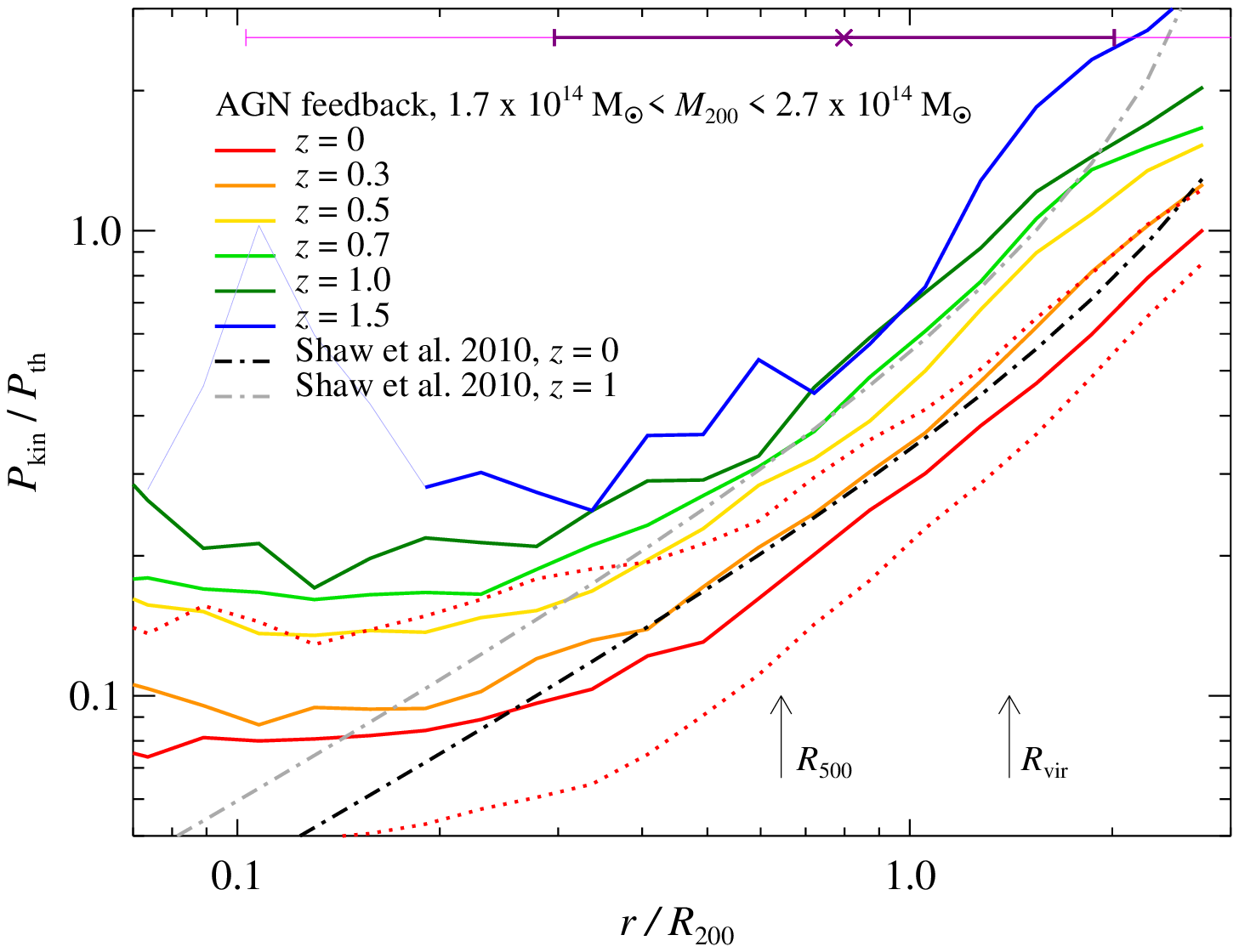}}\\
  \caption{The ratio of kinetic and thermal pressure support,
    $P_{\mathrm{kin}}/P_{\mathrm{th}}$, depends on mass and redshift. We show
    the median of $P_{\mathrm{kin}}/P_{\mathrm{th}}$ as a function of radius for
    the AGN feedback simulations for various mass bins at $z=0$ (left) and as a
    function of redshift for a fixed mass bin (right). We additionally show the
    25$^{\rmn{th}}$ and 75$^{\rmn{th}}$ percentile values for the lowest mass
    bin at $z=0$ (dotted).  In both panels we illustrate the 1 and 2 $\sigma$
    contributions to $Y_{\Delta}$ centered on the median for the feedback
    simulation by horizontal purple and pink error bars which extends out to
    $4R_{200}$ \citep{2010ApJ...725...91B}. Two analytical models for the
    $P_{\mathrm{kin}}$ by \citet{2010ApJ...725.1452S} and
    \citet{2011ApJ...727...94T} are shown with the dash dot and dashed lines,
    respectively. The \citet{2010ApJ...725.1452S} model matches our result in
    the mass bin $2.7\times10^{14} \,\mathrm{M}_{\sun} \le M_{200} \le
    4.2\times10^{14} \,\mathrm{M}_{\sun}$ at intermediate cluster radii (this
    mass bin best represents the mean mass of their sample at redshift zero),
    but also illustrates the need for a mass dependence in future analytical
    models. The dependence of $P_{\mathrm{kin}}/P_{\mathrm{th}}$ on cluster mass
    is driven by the variation of $P_{\rmn{kin}}$ with mass (see
    Fig. \ref{fig:pfit_mz} below).}
\label{fig:pratio_mass}
\end{figure*}

We define the virial radius of a cluster, $R_{\Delta}$, as the radius at
which the mean interior density equals $\Delta$ times the {\em critical
  density}, $\rho_\rmn{cr}(z)$ (e.g., for $\Delta =200$ or 500). For comparison,
we will use an alternative definition of the virial radius,
$R_{\Delta,\rmn{m}}$, where the mean interior density is compared to the {\em
  mean matter density}, $\bar{\rho}_\rmn{m}(z)$. For clarity the critical
density and the mean matter density are,
\begin{eqnarray}
\rho_\rmn{cr}(z) &=& \frac{3H_0^2} {8\pi G}
\left[\Omega_{\rmn{m}}(1 + z)^3 + \Omega_{\Lambda} \right],\\
\bar{\rho}_\rmn{m}(z) &=& \frac{3H_0^2} {8\pi G}\,
\Omega_{\rmn{m}}(1 + z)^3.
\end{eqnarray}
Here we have assumed a flat universe ($\Omega_\rmn{m}+\Omega_\Lambda = 1$ ) and are
only interested in times after the matter-radiation equality, i.e., the
radiation term with $\Omega_{\rmn{r}}$ is negligible. We chose to define the
virial radius with respect to the critical density in continuity with recent
cluster measurements. The merits and utilities of both these definitions
are discussed later in Appendix \ref{sec:rad}.

We apply the following two-step algorithm to compute the virial mass of a
cluster in our simulations. First, we find all clusters in a given snapshot
using a friends-of-friends (FOF) algorithm \citep{1982ApJ...257..423H}. Then,
using a spherical overdensity method with the FOF values as starting estimates,
we recursively calculate the center of mass, the virial radius, $R_{\Delta}$,
and mass, $M_{\Delta}$, contained within $R_{\Delta}$, and compute the radially
averaged profiles of a given quantity with radii scaled by $R_{\Delta}$. We then
form a weighted average of these profiles for the entire sample of clusters at a
given redshift unless stated otherwise. We use the integrated Compton
$y$-parameter as our weighting function,
\begin{equation} 
Y_{\Delta} = \frac{\sigma_{\rmn{T}}}{m_\elct
c^2}\int^{R_{\Delta}}_0 P_{\elct}(r) 4\pi r^2\, \dd r \,  \propto E_{\rmn{th}} (< R_{\Delta})\, , 
\label{eq:Ydelta}
\end{equation}

\noindent where $\sigma_{\rmn{T}}$ is the Thompson cross-section, $m_\elct$ is
the electron mass and $P_{\elct}$ is electron pressure. For a fully ionized
medium of primordial abundance, the thermal pressure $P = P_{\elct}
({5X_{\rmn{H}} + 3}) / 2(X_{\rmn{H}} + 1) = 1.932\, P_{\elct}$, where $X_{\rmn{H}}
= 0.76$ is the primordial hydrogen mass fraction.

\section{Non-thermal cluster profiles}
\label{sec:Pkin}

\begin{figure*}[thbp]
  \resizebox{0.5\hsize}{!}{\includegraphics{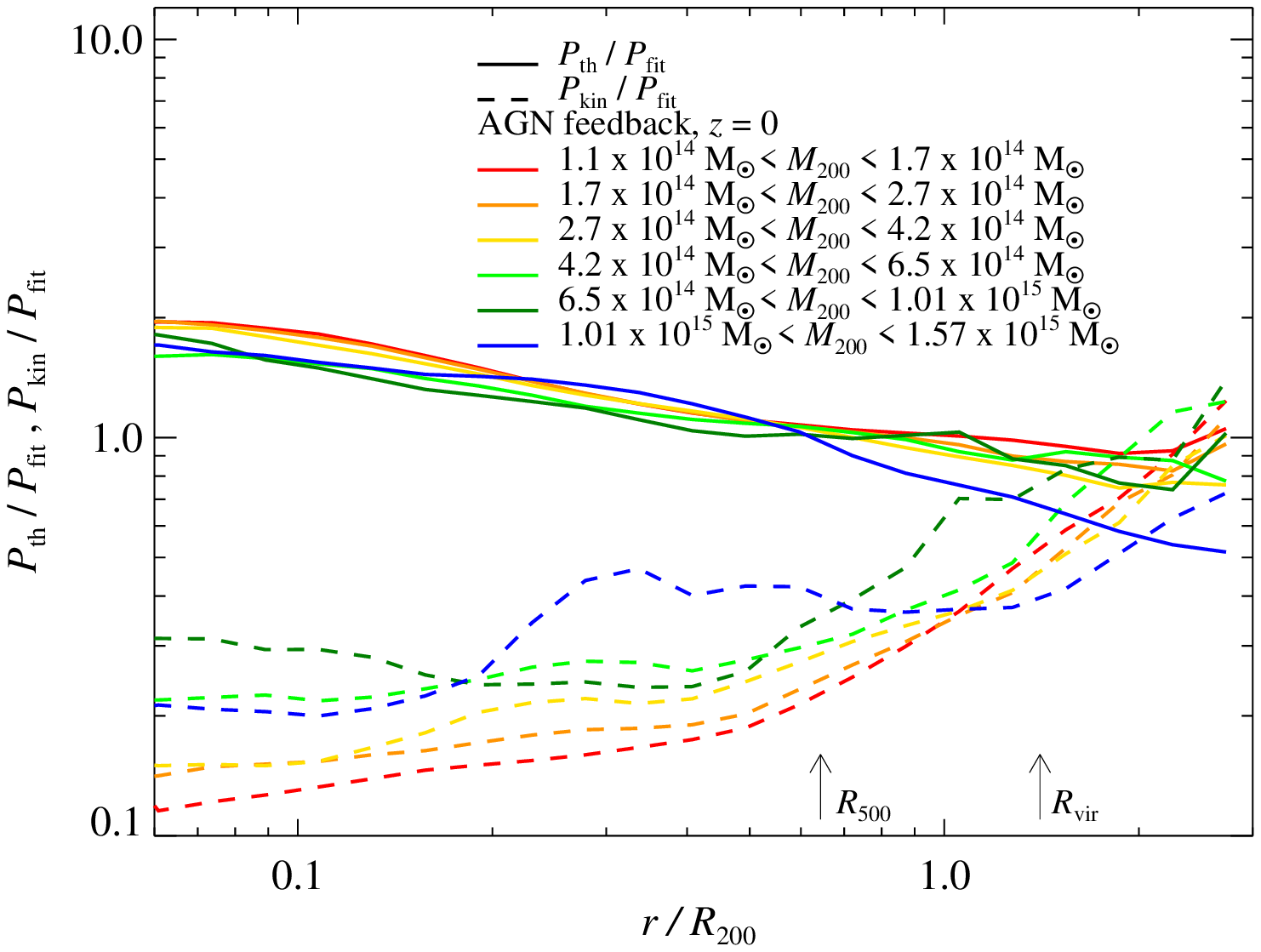}}%
  \resizebox{0.5\hsize}{!}{\includegraphics{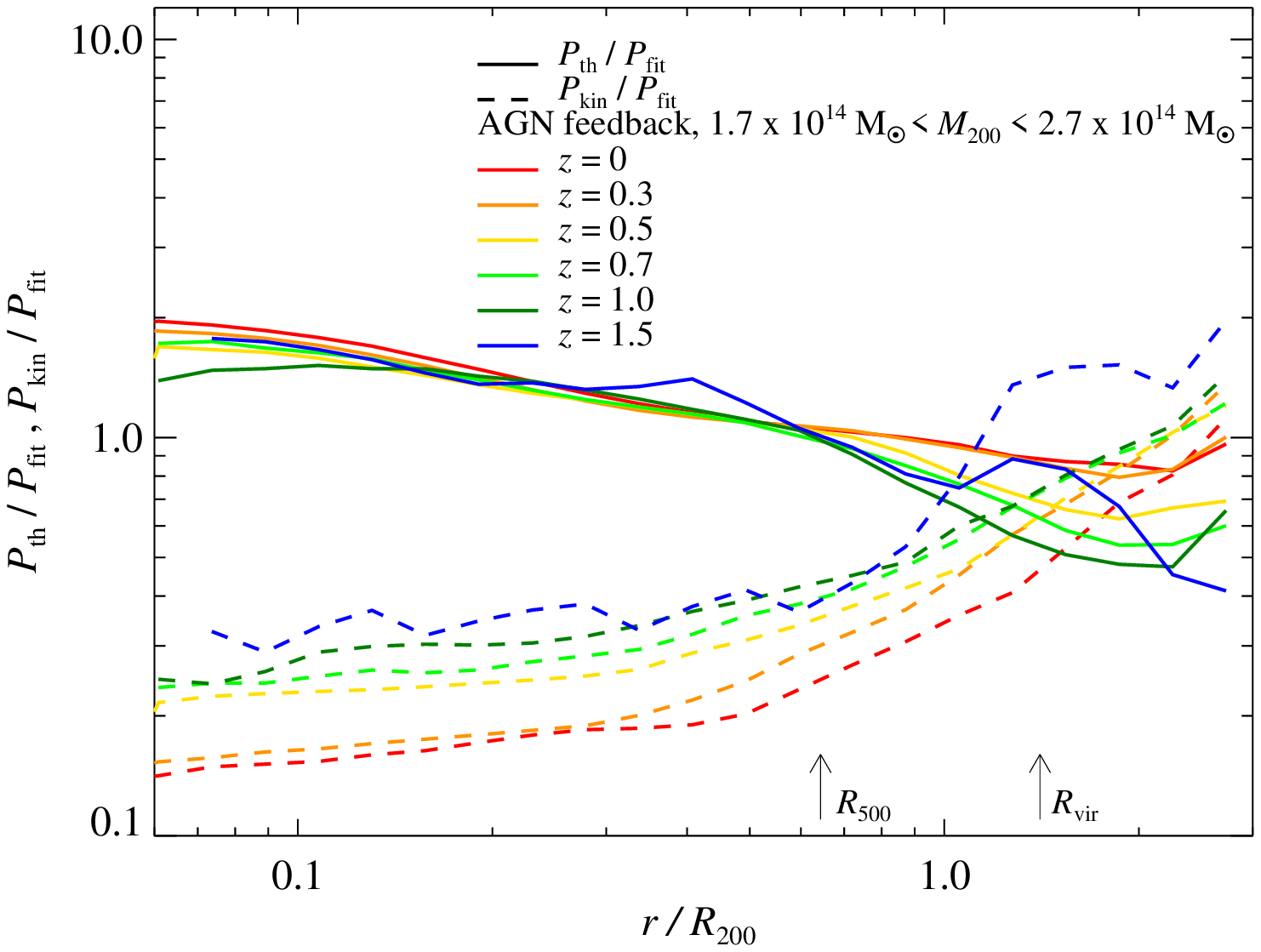}}\\
  \caption{Mass and redshift dependence of $P_{\mathrm{th}}$ and
    $P_{\mathrm{kin}}$, normalized to an empirical fit from BBPS2, $P_\rmn{fit}$, to the
    scaled thermal pressure, $P_\rmn{th}/P_\Delta$. We show the mean scaled
    thermal and kinetic pressure profiles at $z=0$ as a function of radius for
    the AGN feedback simulations in various mass bins (left), and for various
    redshifts at fixed mass bin (right). }
\label{fig:pfit_mz}
\end{figure*}

Several simulations
\citep{1990ApJ...363..349E,2004MNRAS.351..237R,2009ApJ...705.1129L} showed that
the kinetic pressure from bulk motions contributes a small but still significant
amount of energy within $R_{500}$ and this importance increases for larger
cluster radii
\citep{2009ApJ...705.1129L,2010ApJ...725...91B,2010ApJ...721.1105B}. Thus, it is
important to accurately quantify the kinetic pressure contribution as it biases
the hydrostatic cluster masses and is significant to the total energy budget within clusters. There are two kinetic pressure contributions,
namely large-scale, unvirialized bulk motions and subsonic turbulence. For a
Kolmogorov power spectrum of turbulence, the energy is dominated by the largest
scales which we resolved and characterize in our simulations. Hence we believe
that our approach captures the majority of the kinetic pressure contribution.

\subsection{Kinetic pressure support}
\label{sec:pres_support}

The internal bulk motions in the medium can be quantified by the mass-averaged
velocity fluctuation tensor, 
$\avrg{\delta V_i \delta V_j }$, which is associate with the kinetic pressure
(stress) tensor, 
\begin{eqnarray}
P_{\mathrm{kin},ij} &=& \rho \, \avrg{\delta V_i \delta V_j } \, , \nonumber\\
P_{\mathrm{kin}} &\equiv& \Trace\, {\bf P}_{\mathrm{kin}}/3 = \rho\, \avrg{ \delta
{\mathbfit V} \cdot  \delta {\mathbfit V}}  / 3, \nonumber\\
\delta {\mathbfit V} &=& a\,\left(\bvel - \bar{\bvel}\right) +
a\,H(z)\left(\vecbf{x}-\bar{\vecbf{x}}\right) \, .
\label{eq:pkin}
\end{eqnarray}
In this paper we focus on the trace,  $P_{\mathrm{kin}}$, which we refer to as the
kinetic pressure. Issues associated with the anisotropic stress tensor will be made
explicit where they appear, and are also explored in more detail in BBPS3. 
The code uses comoving peculiar velocities, which are translated into the internal
cluster velocities relative to the overall mean cluster velocity in the Hubble flow
by the relation given, where $H(z)$ is the
Hubble function, $a$ is the scale factor,  $\bvel$ ($=\dd \vecbf{x}\,/\,\dd t$) is
the peculiar velocity  and $\vecbf{x}$ is the comoving
position of each particle.  The gas-particle-averaged cluster bulk flow within
$R_{200}$ is $\bar{\bvel}$ and the center of
mass within $R_{200}$ is $\bar{\vecbf{x}}$. 

The radial profiles of the kinetic-to-thermal pressure,
$P_{\mathrm{kin}}/P_{\mathrm{th}}$, shown in Fig. \ref{fig:pratio_mass} for
various mass bins demonstrate an overall mass dependence at all cluster radii,
predominately driven by the variation of $P_{\rmn{kin}}$ and not $P_{\rmn{th}}$
with mass. We show this explicitly in Fig. \ref{fig:pfit_mz} where we scale
$P_{\rmn{kin}}$ with the virial analogue of the thermal pressure,
\begin{equation}
P_{\Delta} \equiv G M_{\Delta} \Delta\, \rho_\rmn{cr}(z)
f_{\mathrm{b}}/(2 R_{\Delta}), \ f_{\mathrm{b}} = \Omega_{\mathrm{b}} /
\Omega_{\mathrm{m}}.
\end{equation}
This behavior reflects the average formation history of galaxy groups and
clusters which, according to the hierarchical picture of structure formation,
sit atop the mass hierarchy, with the most massive clusters forming and
virializing near to the present time. In contrast, the median galaxy group
($M_{200}=10^{13} \rmn{M}_\sun$) has stopped forming today as can be seen by the
dramatically decreasing mass accretion rates implying that the associated
virializing shocks have dissipated the energy associated with the growth of
these objects and hence decreasing the kinetic pressure support
\citep{Wechsler+2002,Zhao+2009,PCB}.  The semi-analytic model for non-thermal
pressure support by \citet{2010ApJ...725.1452S} falls in the middle of the mass
bins chosen since this model results from a sample of 16 high resolutions
adaptive mesh refinement (AMR) simulations of individual galaxy clusters
\citep{2009ApJ...705.1129L} which have a similar mass range. We provide a simple
fit for the mass dependence of $P_{\mathrm{kin}}/P_{\mathrm{th}}$ in
Appendix~\ref{sec:fit_Pkin_overPtot}.
We find that the radius at which $P_{\mathrm{kin}}=P_{\mathrm{th}}$ is just
beyond the spherical collapse definition for $R_\rmn{vir}$
from \citet{1998ApJ...495...80B},
\begin{equation}
R_\rmn{vir} = \left(\frac{3\,M_\rmn{vir}}{4\pi\,\Delta_{\rmn{cr}}(z)\,\rho_{\rmn{cr}}(z)}\right)^{1/3},
\end{equation}
\noindent where $\Delta_{\rmn{cr}}(z) = 18\pi^2 + 82[\Omega (z)- 1] -
39[\Omega (z)- 1]^2$ and $\Omega (z) =
\Omega_{\rmn{m}}(1+z)^3\,\left[\Omega_{\rmn{m}}(1+z)^3 +
\Omega_{\Lambda}\right]^{-1}$.  Hence, this radius represents a
possible physical definition for the virialized boundary of clusters.

The redshift evolution of $P_{\mathrm{kin}}/P_{\mathrm{th}}$ is dramatic. At
higher redshift, $P_{\mathrm{kin}}$ is increasing faster than $P_{\mathrm{th}}$
over all radii (cf. Fig. \ref{fig:pfit_mz}), such that at $z=1$,
$P_{\mathrm{kin}}/P_{\mathrm{th}}$ is approximately twice that at $z=0$.  In the
picture of hierarchical structure formation, at any given redshift the most
massive objects are currently assembled and hence show the largest kinetic
pressure contribution in comparison to smaller objects that formed on average
earlier. Or equivalently, at fixed cluster mass, the relative contribution from
kinetic pressure and the relative amount of substructure increases with
redshift.  In particular, the relative mass accretion rates increase from $z=0$
to $z=2$ by a factor 3 for clusters ($M_{200}=10^{15} \rmn{M}_\sun$) and 10 for
groups ($M_{200}=10^{13} \rmn{M}_\sun$) \citep[see][]{PCB,Gottloeber+2001}. This strong evolution in $P_{\mathrm{kin}}/P_{\mathrm{th}}$ is
lessened by a different choice of scaling radius, i.e., if we normalize by
$R_{200,\rmn{m}}$ instead of $R_{200}$ (cf. Appendix \ref{sec:rad}).  Although this ratio
cannot be observed, we will use it as an indicator for the dynamical state of
clusters in our simulations. Results from \citet{2009ApJ...705.1129L}
find a similar correlation between $P_{\mathrm{kin}}$ and the X-ray definition
of dynamical state, from a smaller sample of 16 clusters. At $z=1$, the
\citet{2010ApJ...725.1452S} semi-analytic model for non-thermal pressure support
does not match our simulations as well as it does at redshift zero.\footnote{Our
  kinetic pressure contribution is larger at the center compared to that in the
  model by \citet{2010ApJ...725.1452S}. This discrepancy is probably a
  manifestation of the well-known core entropy problem in numerical
  simulations. In (adaptive) grid codes there is a larger level of core entropy
  generated in comparison to SPH codes implying that the enhanced entropy (which
  results from dissipating gas motions) is accompanied by a smaller amount of
  kinetic pressure. This is presumably due to the difference in the amount of
  mixing in SPH and mesh codes and possibly related to a different treatment of
  vorticity in the simulations \citep[e.g.,][]{Frenk+1999,2009MNRAS.395..180M,
    2011arXiv1106.2159V}.}

\begin{figure}
\epsscale{1.20}
\plotone{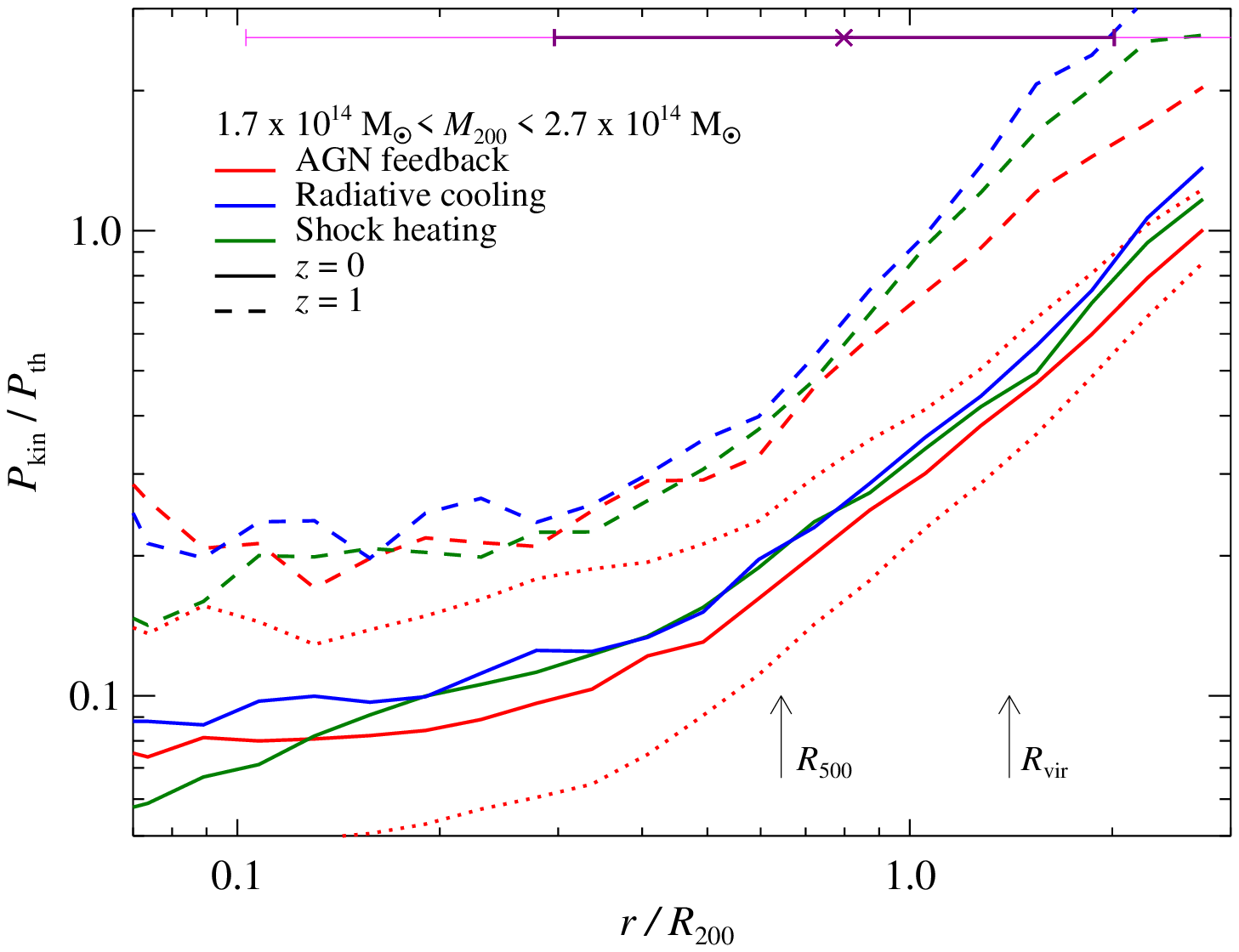}
\caption{Shown is the median of $P_{\mathrm{kin}}/P_{\mathrm{th}}$ as a function
  of radius for different physics models at $z=0$ (solid) and $z=1$ (dashed)
  with the 25$^{\rmn{th}}$ and 75$^{\rmn{th}}$ percentile values shown for the
  AGN feedback simulations at $z=0$ (dotted). Results are shown for the mass bin
  $1.7\times10^{14} \,\mathrm{M}_{\sun} \le M_{200} \le 2.7\times10^{14}
  \,\mathrm{M}_{\sun}$ to take out the dependence on mass of
  $P_{\mathrm{kin}}/P_{\mathrm{th}}$.  The kinetic pressure contribution is
  similar for our differently simulated physics, suggesting that gravitational
  processes dictate that contribution (while AGN feedback slightly decreases the
  kinetic pressure contribution, especially for higher redshifts).  The
  horizontal purple and pink error bars have the same meaning as in
  Fig. \ref{fig:pratio_mass}.}
\label{fig:pratio_phys}
\end{figure}

\begin{figure}
\epsscale{1.20}
\plotone{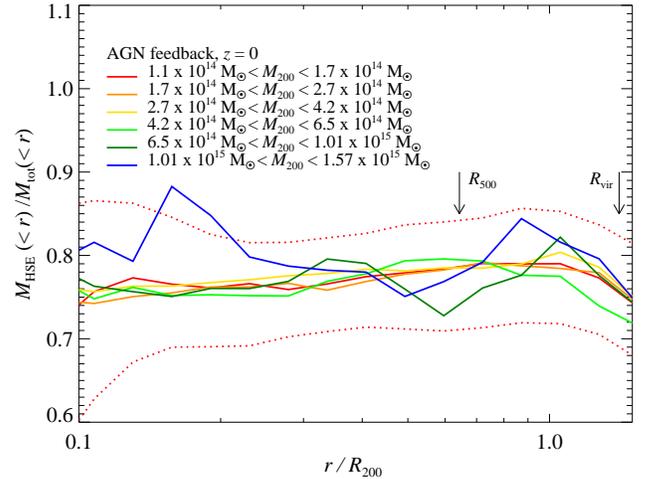}
\caption{Assessing the bias in hydrostatic masses, $M_{\mathrm{HSE}}$, due to
  the kinetic pressure.  The median of $M_{\mathrm{HSE}}/M_{\mathrm{tot}}$ as a
  function of radius for AGN feedback simulations for various mass bins, with
  the 25$^{\rmn{th}}$ and 75$^{\rmn{th}}$ percentile values shown for the
  smallest mass bin (dotted). Assuming hydrostatic equilibrium for all clusters
  of a given mass will bias the mass values low by 20 to 25\%. The scatter about
  the median -- represented by the 25$^{\rmn{th}}$ and 75$^{\rmn{th}}$
  percentiles -- amounts to approximately 5\%. This bias is not representative
  for a relaxed cluster sample which will likely have a smaller bias since the
  calibration of such a sample against numerical cluster simulations shows
  \citep{2006ApJ...650..128K}.}
\label{fig:HSEmass}
\end{figure}

The formation of clusters and the associated accretion of substructure
are driven by the depth of the cluster gravitational potential. Therefore, it is
not surprising that we find kinetic pressure support to be ubiquitous in the
three differently simulated physics cases
(cf. Fig. \ref{fig:pratio_phys}). Looking at the median of this non-thermal
pressure support we find similar radial profiles within the $25^{\rmn{th}}$ and
$75^{\rmn{th}}$ percentiles of the complete distribution of clusters.  In the
AGN feedback simulations we find marginally lower values for
$P_{\mathrm{kin}}/P_{\mathrm{th}}$. These differences are well within the
$25^{\rmn{th}}$ and $75^{\rmn{th}}$ percentiles implying consistency across
differently modeled physics. Thus, our model of AGN feedback does not
significantly alter the kinetic pressure support at low redshift although there
seems to be a hint that this may be the case at larger radii at redshifts
$z\sim1$ which approach the peak of the AGN luminosity density.

We have so far focused on the trace of the kinetic pressure
tensor. We treat in detail the velocity anisotropy in BBPS3. The main
results are that the core regions near the center, the velocity
distribution starts to become isotropic for the gas in groups and (to
a lesser extent) for the DM and gas in larger clusters. The positive
values of velocity anisotropy around the virial radius indicate
(radial) infall, whereas the strong decrease at even larger radii
(very noticeably in the DM) is caused by the turn-around of earlier
collapsed shells, which minimizes the radial velocity component such
that the tangential components dominate the velocity.

\subsection{Hydrostatic Masses}

Even if clusters are in hydrostatic equilibrium, balancing the gravitational
force to the pressure gradient yields
\begin{equation}
\nabla P = \rho {\mathbfit g} \rightarrow  -\rho GM(<r){\hat{\mathbfit{r}}}/ r^2 \ \mbox{for spherical symmetry}, 
\label{eq:HSE}
\end{equation}
and using this relation to estimate cluster masses will give the wrong result 
if one does not include non-thermal pressure, in particular the kinetic pressure. 

\noindent 
As others have shown
\citep{1990ApJ...363..349E,2004MNRAS.351..237R,2009ApJ...705.1129L}, assuming
that all the pressure in Eq.~(\ref{eq:HSE}) is thermal ($P = P_{\mathrm{th}}$)
is incorrect; for clarity we define $M_{\rmn{HSE}}$ to be the mass derived using
$P = P_{\mathrm{th}}$.  Comparing $M_{\rmn{HSE}}$ to the true mass inside a
given radius, $M_{\rmn{tot}}$, we find that $M_{\rmn{HSE}}$ on average
underestimates $M_{\rmn{tot}}$ by $20$--$25$\% depending on the radius (cf. Fig
\ref{fig:HSEmass}). This bias is almost independent of cluster mass out to
$R_{500}$, the current maximum radius typically observed by the X-ray telescopes
{\em Chandra} and {\em XMM-Newton}.  We can understand this weak mass dependence
by rewriting the total pressure $P = P (P_\rmn{th}/P + P_\rmn{kin}/P)$. Since
$P_\rmn{kin}/P\propto M_{200}^{1/5}$ (see Fig.~\ref{fig:pkinfit}), the
hydrostatic mass estimates inherit a similarly weak dependence on mass. Thus, an
overall correction to the hydrostatic mass is reasonable for these measurements.

Individual clusters can stray from this generalization, since each cluster has a
unique dynamical state and formation history. These deviations are suggested by
the scatter of $\sim5$\% between the $25^{\rmn{th}}$ and $75^{\rmn{th}}$
percentiles of the complete distribution. For cluster samples that are selected
against major mergers (for which the assumption of spherical symmetry will also
be questionable), the correction factor will necessarily be smaller, e.g., for
quality X-ray data of a Chandra sample, the hydrostatic mass correction was
found to be of the order $M_{\rmn{HSE}}\sim 10$--$15$\%
\citep{2006ApJ...650..128K}.

\section{Cluster Shapes}
\label{sec:Elpt}

\begin{figure*}
\begin{center}
  \hfill
  \resizebox{0.5\hsize}{!}{\includegraphics{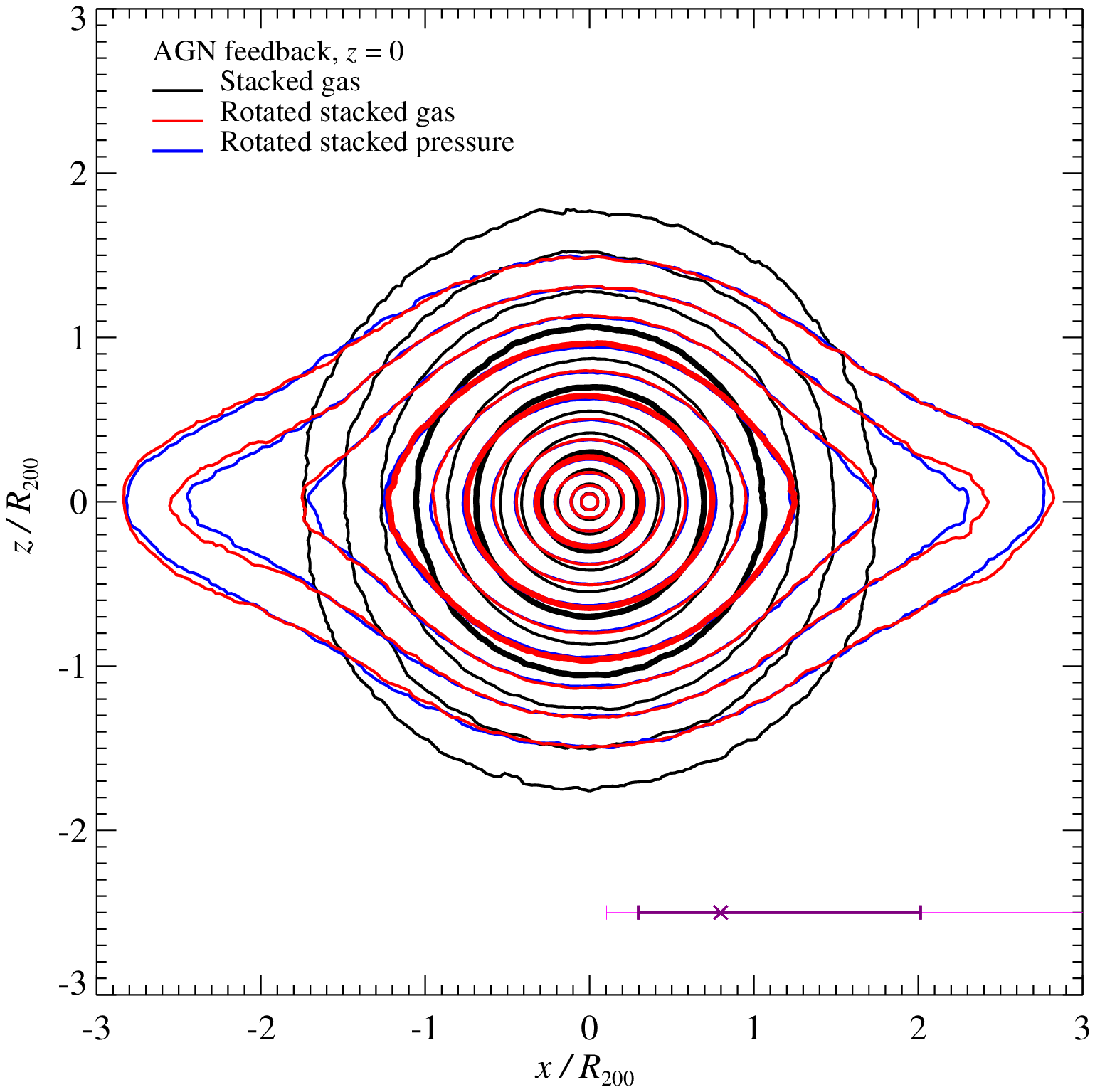}}%
  \resizebox{0.5\hsize}{!}{\includegraphics{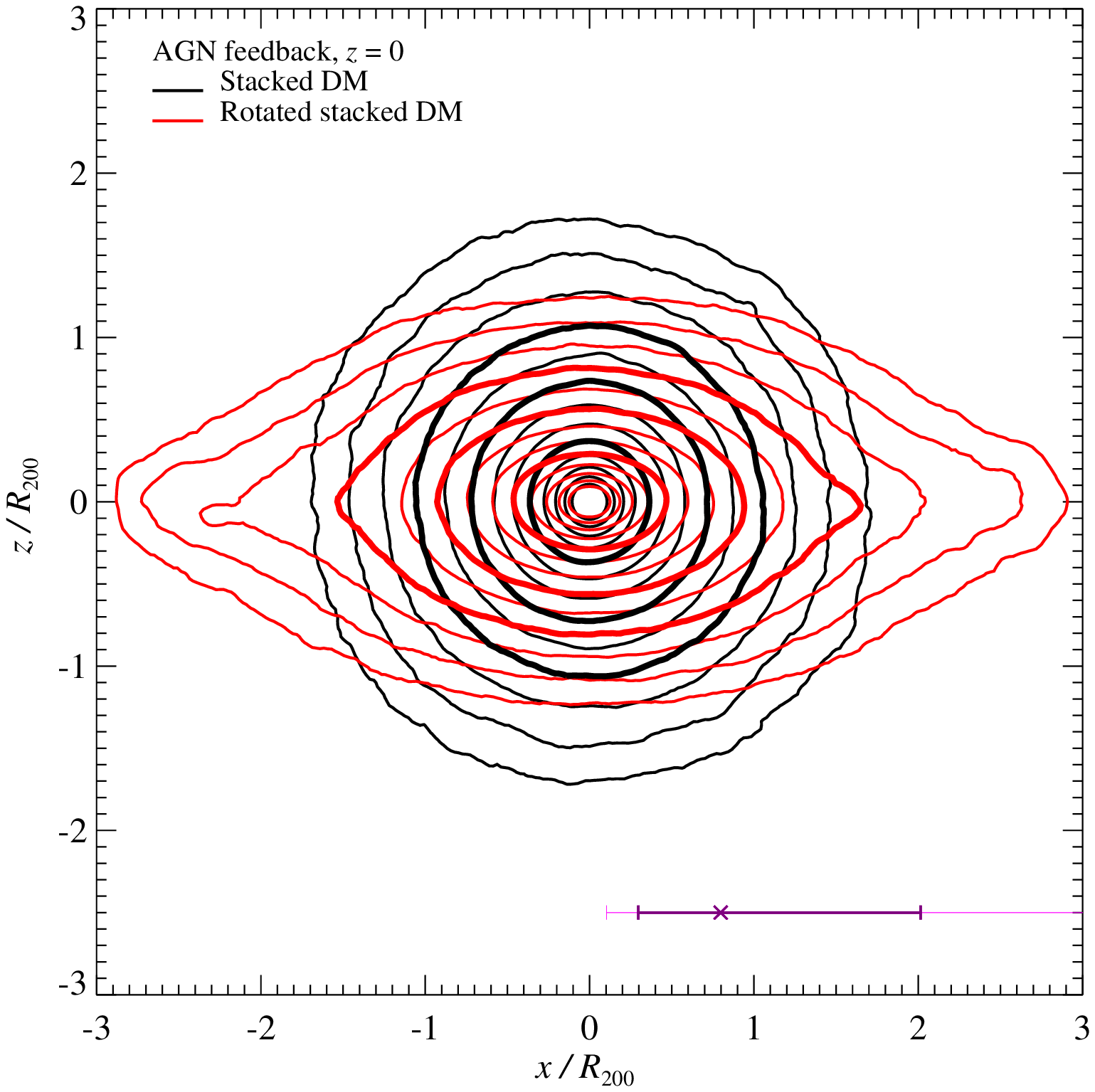}}\\
\end{center}
\caption{Stacked density and pressure distributions with and without rotations
  into the principle axis frame of the correspondingly weighted
  moment-of-inertia tensors at $z=0$. Left: We compare rotated distributions of
  the gas density (red) and pressure (blue) to the non-rotated stacked gas
  density (black) at $z=0$. Right: Shown is the same as on the left for DM.  The
  non-rotated clusters average out to form spherical iso-density contours, while
  the rotated clusters clearly show elongations along the major axis (defined
  here as the $x$-axis).  The thicker lines approximately show the radii
  $R_{2500}$, $R_{500}$ and $R_{200}$ from the inside out. These contours have
  been smoothed to a pixel size of 0.09$R_{200}$. The horizontal purple and pink
  error bars have the same meaning as in Fig. \ref{fig:pratio_mass}.}
\label{fig:circstack}
\end{figure*}

\begin{figure*}
\begin{center}
  \hfill
  \resizebox{0.5\hsize}{!}{\includegraphics{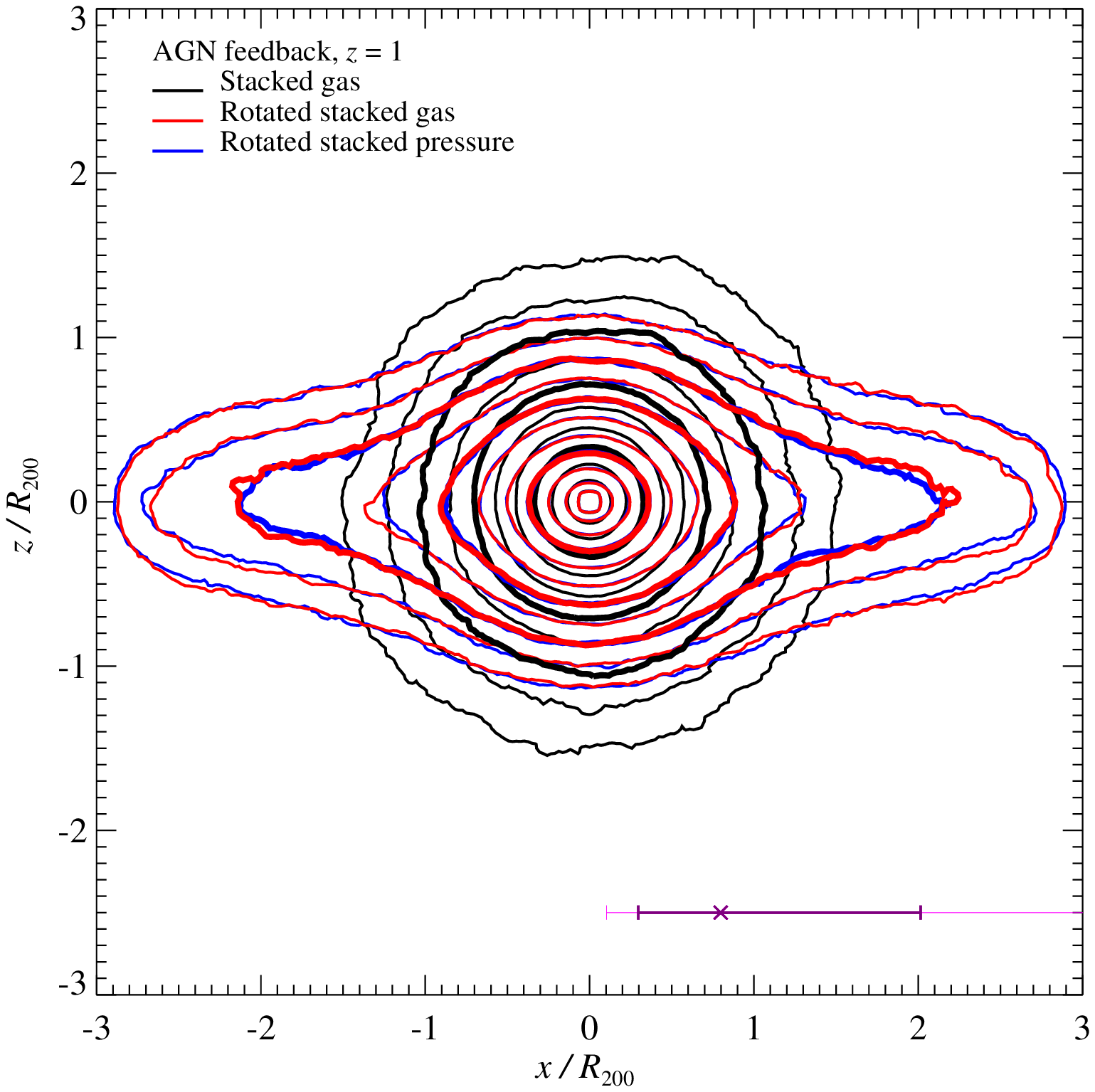}}%
  \resizebox{0.5\hsize}{!}{\includegraphics{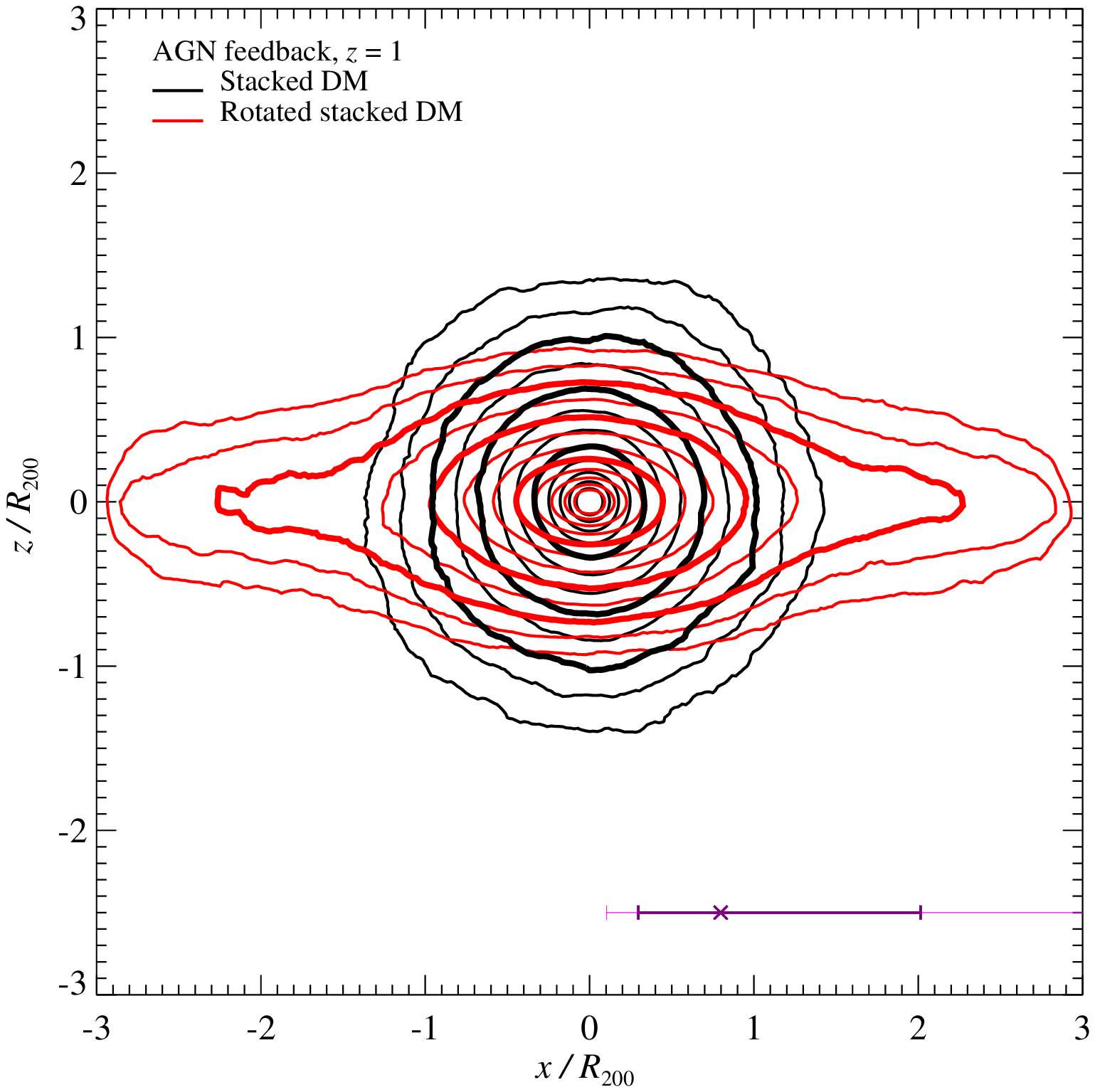}}\\
\end{center}
\caption{Same as in Fig. \ref{fig:circstack}, but at $z=1$.}
\label{fig:circstack2}
\end{figure*}

Generally, we expect clusters to be triaxial since they grow by accretion
and through merging along filamentary structures that impose tidal gravitational
forces upon the forming clusters. Following \citet{1991ApJ...378..496D}, we
estimate this non-sphericity of cluster gas and dark matter (DM) by computing
the normalized moment-of-inertia tensor,
\begin{equation}
\avrg{\delta x \delta x}_{ij} (r<R\vert w) = \frac{\sum_{\alpha} w_{\alpha}(x_{i,\alpha} - \bar{x}_i)(x_{j,\alpha} - \bar{x}_j)}{\sum_{\alpha} w_{\alpha}} ,
\label{eq:Itens}
\end{equation}
for several weightings $w_\alpha$ of the contribution of particle $\alpha$ that
lies within a given radius $R$.  The tensor measures the variance in the spatial
fluctuations within $R$, with $\delta x_i = x_i- \bar{x}_i$ the deviation of the
particle position $x_i$ from the centre-of-mass $\bar{x}_i (<R)$ of the region.
Using mass weighting, $w_{\alpha} = m_{\alpha}$, for the dark matter or the SPH
gas particles gives the moment-of-inertia in its usual form. It has an effective
$\rho (x) x^5 d\ln x$ reach in its probe of the unit vector combination
$\hat{\delta x}_i \hat{\delta x}_j$, hence preferentially feels the outskirts,
near $R$.  In the Appendix \ref{sec:wr2}, we explore how our results are
modified with a weight $w_{\alpha} = m_{\alpha}/x_{\alpha}^2$ less sensitive to
the outskirts: this weight just mass-averages the unit vector product, hence
emphasizes the more isotropic interior.  Since the tSZ signal is of primary
interest to us, we also consider thermal-energy weighting, with $w_{\alpha} =
m_{\alpha}T_{\alpha}$ the product of the mass and temperature. 

We quantify the asphericity of cluster gas and DM by two parametrizations: the
axis ratios (in particular the ratio of the largest-to-smallest main axis,
$\ca$); and the three-dimensional asymmetry parameters for symmetric tensors
introduced by BBKS \citep{1986ApJ...304...15B}. Both use the eigenvalues
$\lambda_i$ of the moment-of-inertia tensor at a prescribed radius, ordered by
$\lambda_1 < \lambda_2 < \lambda_3$.  The ellipsoid associated with the tensor
has axis lengths $a = \sqrt{\lambda_1}$, $b = \sqrt{\lambda_2}$, and $c =
\sqrt{\lambda_3}$.  (We note that \citet{2011ApJ...734...93L} express their
results in terms of eigenvalue ratios, defining $a' = \lambda_1$, $b' =
\lambda_2$, and $c' = \lambda_3$).  The eigenvectors $\vecbf{E}_i$ associated
with $\lambda_i$ are also used to rotate the clusters to their principal axes
and to explore alignment variations with radius.

The BBKS-style asymmetry parameters are defined by 
\begin{eqnarray}
&& e = \frac{\lambda_1 -\lambda_3}{2\bar{\lambda}}, \cr
&& p = \frac{\lambda_1 - 2\lambda_2 +\lambda_3}{2\bar{\lambda}}, \cr
&& \bar{\lambda} \equiv \lambda_1 +\lambda_2 +\lambda_3\, .
\end{eqnarray}
Following BBKS \citep{1986ApJ...304...15B}, we refer to $e$ as the ellipticity and $p$ as the prolaticity. 
When $p$ is positive the
clusters are prolate, and when $p$ is negative the clusters are oblate. Thus we also define the oblateness $o \equiv -p \theta (-p)$, equal to $\vert p \vert$ for negative $p$ and zero for positive $p$. Although $e$ is the most striking indicator of cluster elongation, the degree of prolateness or oblateness are also necessary to specify the general triaxiality of the  morphological configuration on a given smoothing scale. 

\subsection{Overall shapes and their profiles}

\begin{figure*}[]
\begin{center}
  \hfill
  \resizebox{0.5\hsize}{!}{\includegraphics{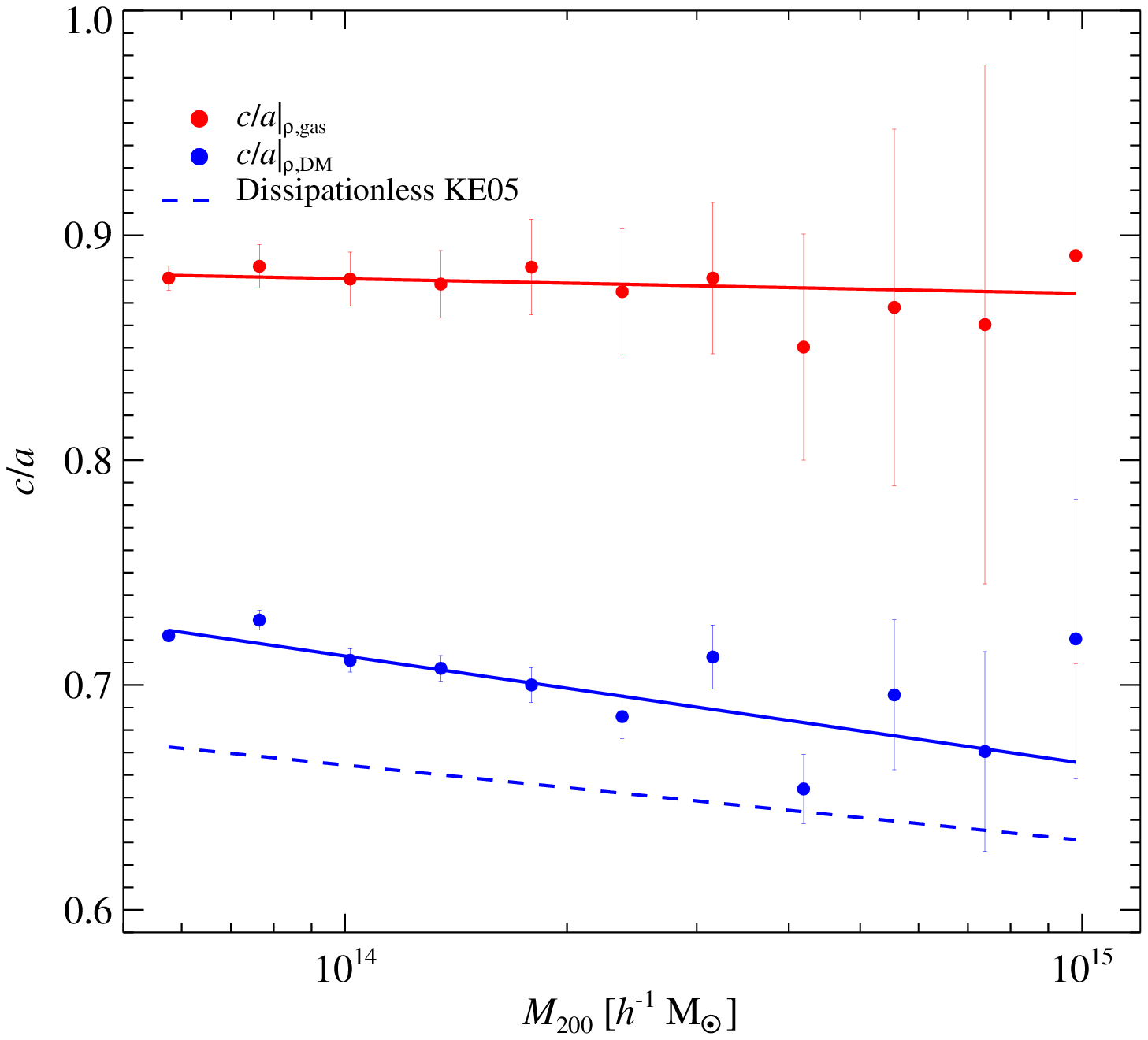}}%
  \resizebox{0.5\hsize}{!}{\includegraphics{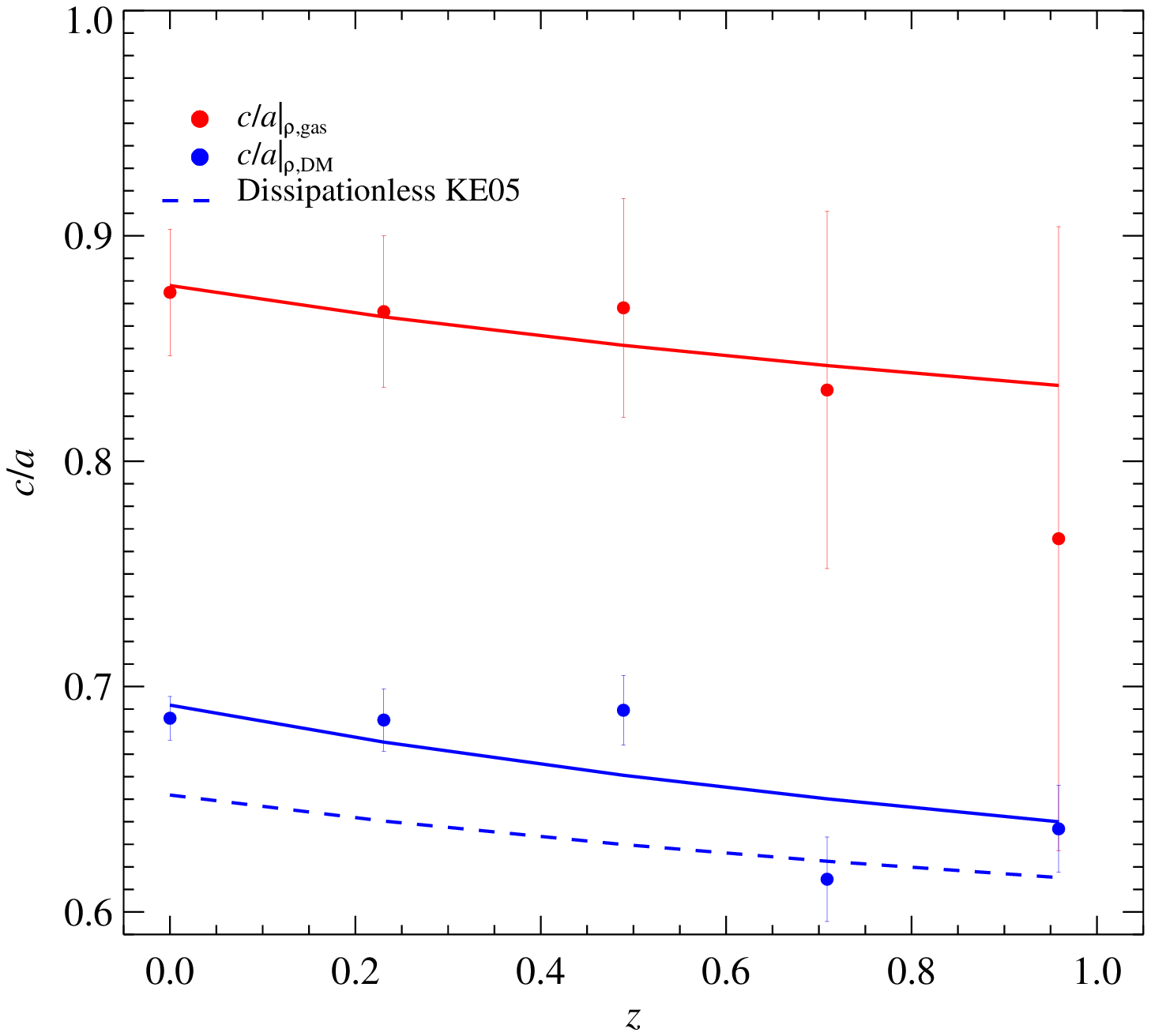}}\\
\end{center}
\caption{We show axis ratios of clusters that are obtained by
  computing the moment-of-inertia tensor of the gas (red) and DM mass
  distributions (blue) within $R_{200}$ and stacking those in bins of
  cluster mass and redshift.  The resulting mean and standard
  deviation of the axis ratio $\ca$ is shown as a function of
  $M_{200}$ at $z=0$ (left panel) and at fixed average mass bin of
  $M_{200} = 2.4\times10^{14} h^{-1}\,M_{\sun}$ as a function $z$
  (right panel). See Table \ref{tab:C_Afits} for fit values; here, we
  have chosen to quote $h^{-1}\,M_{\sun}$ to compare directly with the
  dissipationless simulations by \citet[][
  KE05]{2005ApJ...629..781K}. Shocks dissipate the kinetic energy of
  the gas which causes larger axis ratios/smaller ellipticities, this couples
  through gravity to the DM distribution and sphericalize their axis
  ratios, resulting in smaller ellipticities in comparison to
  dissipationless simulations alone, e.g., by KE05 (that do not follow
  the hydrodynamics of the gas).}
\label{fig:CA_M_z}
\end{figure*}

\begin{figure*}
\begin{center}
  \hfill
  \resizebox{0.5\hsize}{!}{\includegraphics{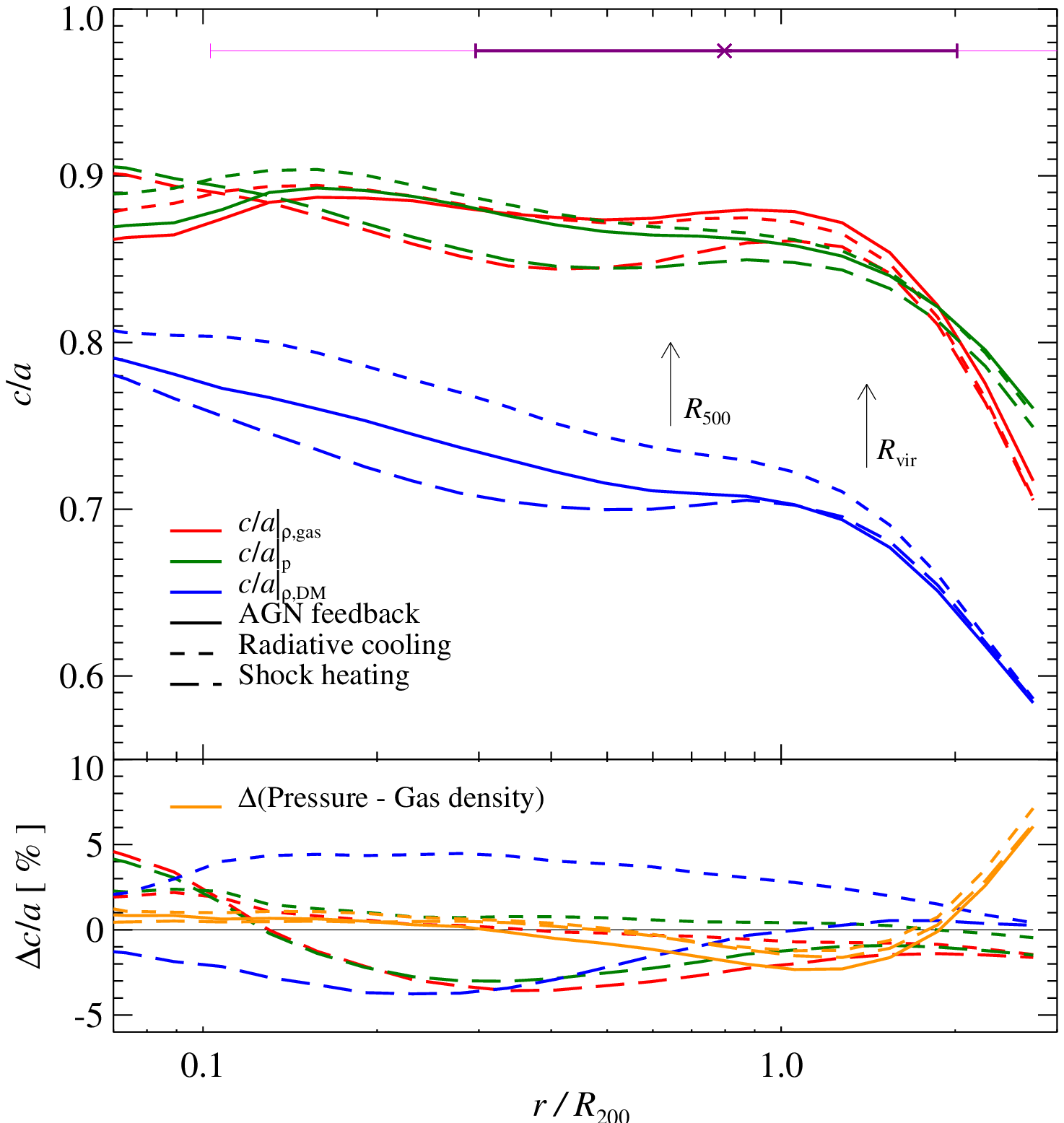}}%
  \resizebox{0.5\hsize}{!}{\includegraphics{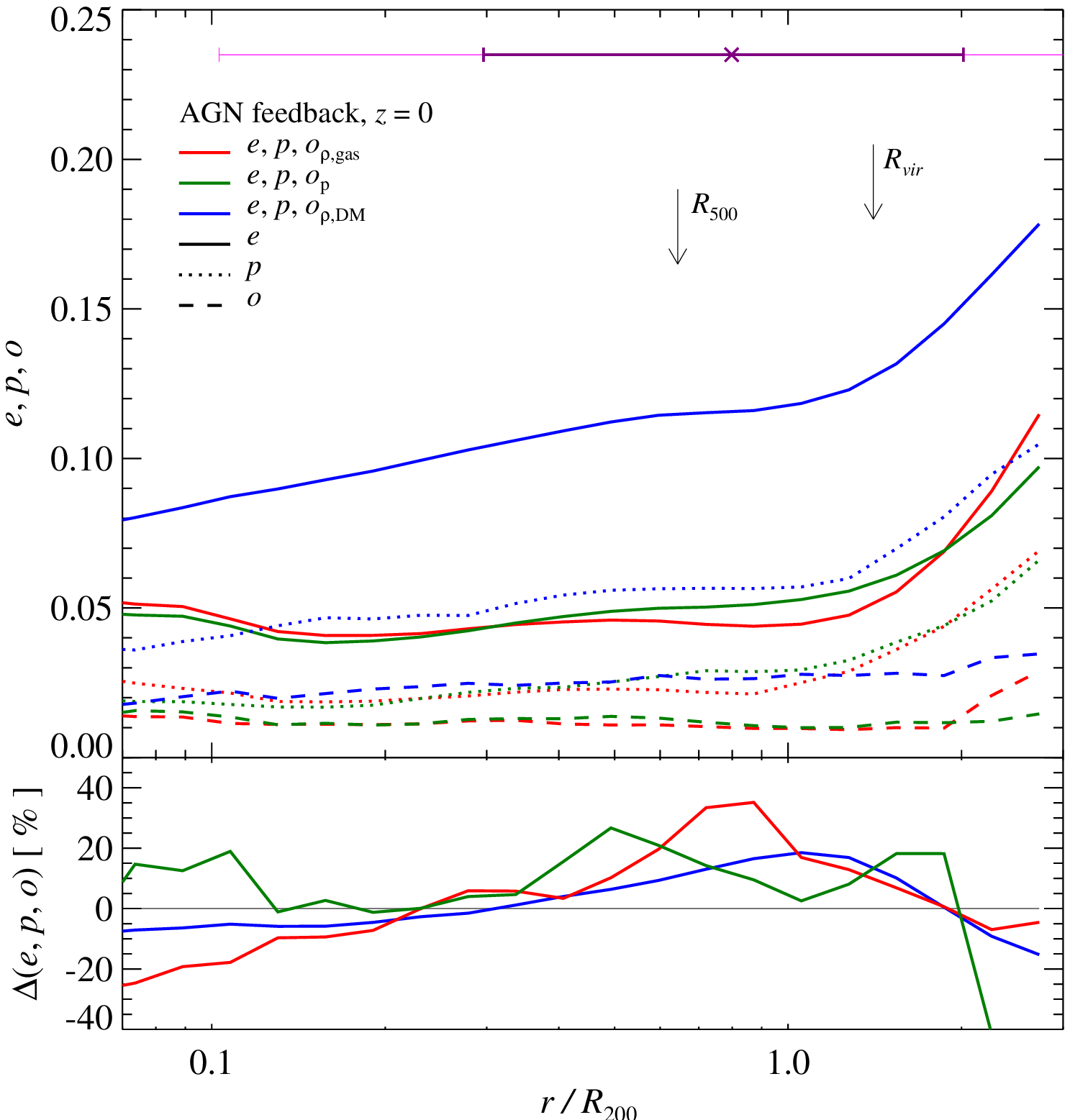}}\\
\end{center}
\caption{Average cluster axis ratios and ellipticities for the DM mass (blue),
  gas mass (red) and pressure (green) distributions within a scaled radius
  $r/R_{200}$. Left: Shown is the axis ratio $\ca$ as a function of scaled
  radius for all simulated physics models. In the bottom panel we show the
  relative differences of
  the shock heating (long-dashed) and the radiative cooling simulations (short-dashed) with respect to AGN feedback simulations.  Right: We show the ellipticity, prolaticity
  and oblaticity as a function of scaled radius. The bottom panel shows the
  relative differences between the pressure and gas density weightings of the
  moment-of-inertia tensor. The axis ratio $\ca$ and ellipticity show the same
  trends. We find clusters to be more prolate then oblate. In the regions beyond
  $R_{\rmn{vir}}$ the sudden decrease in the axis ratios can be attributed to
  other nearby groups and collapsed objects (that is also seen as an enhanced
  density clumping at these radii, \citealt{2011ApJ...731L..10N, battinprep4}). The pressure-weighted
  shapes tightly track the density-weighted shapes with deviations of less than
  5\%.  The horizontal purple and pink error bars have the same meaning as in
  Fig. \ref{fig:pratio_mass}. }
\label{fig:CAelipt}
\end{figure*}

We rotate all clusters into the moment-of-inertia tensor frame using the
eigenvector matrix $\vecbf{E}$, so $x'=\vecbf{E}x$.  The output ordering is
arbitrary; we choose the convention that the major axis is aligned with the
x-axis and the minor axis is aligned with the z-axis.  In
Figs.~\ref{fig:circstack} and \ref{fig:circstack2} we show the results for DM
and gas which have been obtained by computing the weighted moment-of-inertia
tensor within $3R_{200}$, rotating into the moment-of-inertia tensor frame, and
stacking the respective distributions, i.e., gas density and pressure as well as
DM density. The rotated contours show obvious elongations along the major axis;
with the ellipticity being larger at $z=1$ in comparison to $z=0$. The
elongation is larger for the DM distribution in comparison to the gas density
and pressure which show very similar behaviour. Even in the rotated stacked
distributions, the innermost contour lines become more spherical because they
are intrinsically less elliptical (see below) and because the main axes of the
inner distributions are twisted relative to those at $3R_{200}$ so that their
ellipticity partially averages out to become more spherical (see
Sect.~\ref{sec:twist}).

In order to quantify these results, we show the mass dependence and redshift
evolution of the ellipticity within $R_{200}$ in Fig.~\ref{fig:CA_M_z}. Due to
the dissipationless nature of DM, its ellipticity is larger (smaller ratio of
$c/a$) in comparison to that of the gas. This is because the kinetic energy of
the accreted gas is dissipated at cluster accretion shocks -- a process that
erases part of the memory of the geometry of the surrounding large scale
structures and their tidal force field. Those accretion shocks are typically
forming at radii $> R_{200}$ as suggested by numerical simulations
\citep{Miniati+2000,Ryu+2003,Pfrommer+2006,Skillman+2008,Vazza+2009} or
indirectly by the action of shock waves on radio plasma bubbles, which
represents a novel method of finding formation shocks
\citep[e.g.,][]{2001ApJ...549L..39E,2011ApJ...730...22P}.  Following these
qualitative considerations, it is not surprising that the ellipticity of the gas
distribution does not show any mass dependence while the DM distribution of more
massive clusters shows a larger ratio of $\ca$ in comparison to smaller
systems. However, the ellipticity of the gas and DM distribution are increasing
as a function of redshift, at about the same rate. This can be understood by the
fact that 1) a given mass range of clusters shows a larger degree of
morphological disturbances/merging at higher redshifts which probe on average
dynamically younger objects and 2) the redshift evolution of the velocity
anisotropy (cf. BBPS3) which shows that the average location of accretion shocks
moves to smaller radii (if scaled by $R_{200}$). Hence at larger redshifts, also
the gas distribution probes the infall/pre-accretion shock region that is shaped
by the tides exerted by the far-field of clusters.

We compare the results from our simulations directly with those of
\citet{2005ApJ...629..781K} in Figure \ref{fig:CA_M_z} and Table
\ref{tab:C_Afits}. Other work
\citep{2006MNRAS.367.1781A,2007ApJ...664..117G,2008MNRAS.391.1940M,2011ApJ...734...93L}
on DM and gas shapes have used different mass definitions, axis definitions and
cosmologies than we do, making quantitative comparisons difficult, but we note
that these various results were shown to be consistent with
\citet{2005ApJ...629..781K}.  For the mass and redshift functional fits,
\citet{2005ApJ...629..781K} define $\ca (M) =
B_{\rmn{M}}(1+A_{\rmn{M}}\,\ln[M/10^{15}h^{-1}\,M_{\sun}])$ and $\ca (z) =
B_{\rmn{z}} (1 + z)^{A_{\rmn{z}}}$, respectively. The axis ratios that we find
for the DM mass dependence and redshift evolution have slopes consistent with
\citet{2005ApJ...629..781K}, but as an overall trend, our axis ratios are more
spherical than theirs. Some of this may be traced to differing cosmological
parameters, but some may be because the less aspherical baryons may have an
impact on the DM ellipticity. The effect of baryons on DM has been explored in,
e.g., \citet{2008ApJ...672...19R}.

\begin{figure}
\epsscale{1.20}
\plotone{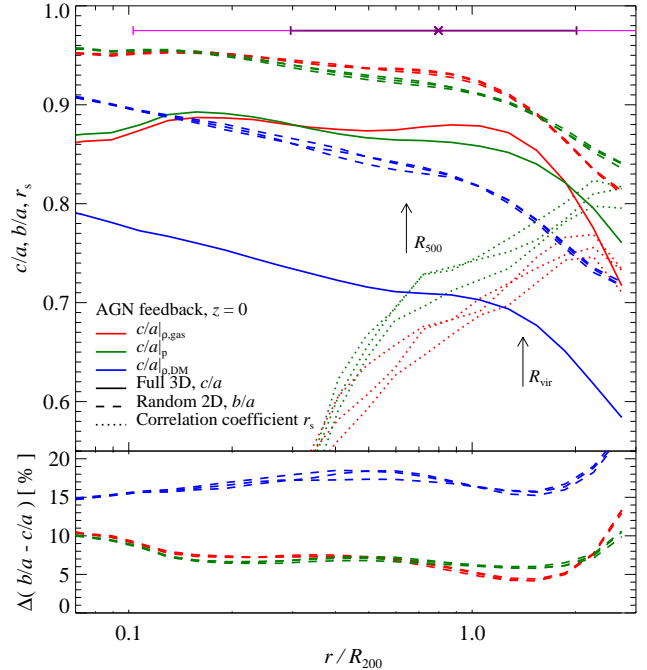}
\caption{Top: We compare the average of the 2D axis ratios (dashed) of three
  random orthogonal projections to the 3D axis ratios (solid) for
  the DM mass (blue), gas mass (red) and pressure (green) from the clusters in the
  AGN feedback simulations. Additionally, we show the linear correlation
  coefficient, $r_{\rmn{s}}$ (dotted line), between the projected 2D and the
  3D axis ratios.  The horizontal purple and pink error bars have the same
  meaning as in Fig. \ref{fig:pratio_mass}. Bottom: Shown is the relative
  differences between the projected 2D and 3D axis ratios. While the
  relative difference between the projected 2D and 3D axis ratios
  vary between 15--20\% for the DM mass distribution (with the 2D axis ratios
  being more spherical), the relative differences are smaller for the mass and
  pressure distribution of the gas, with values between 5--10\%. As expected,
  the projected 2D and the 3D axis ratios are correlated with an increasing
  correlation coefficient at larger radii which suggests that the substructure
  distribution that drives the asphericity also causes this correlation.}
\label{fig:2Dcompelipt}
\end{figure}

\begin{table}
\begin{center}
\caption{Axis ratio fits for cluster as a function of mass and redshift.}
\label{tab:C_Afits}
 \begin{tabular}{l|c|c|c|c}
  \hline
  \hline
    &  $B_{\rmn{M}}$ & $A_{\rmn{M}}$ & $B_{\rmn{z}}$ & $A_{\rmn{z}}$ \\
  \hline
  DM & $0.665 \pm 0.009$  & $-0.031 \pm 0.006$ & $0.692 \pm 0.009$ & $-0.12 \pm 0.04$ \\
  KE05 & $0.631 \pm 0.001$  & $-0.023 \pm 0.002$ & $0.652^a \pm 0.001$ & $-0.086 \pm 0.004$ \\
  Gas & $0.87 \pm 0.02$  & $-0.003 \pm 0.01$ & $0.88 \pm 0.03$ & $-0.08 \pm 0.12$ \\
  \hline
\end{tabular}
\begin{quote} 
$^a$ We use a re-normalized value from $M_{200} = 1 \times
  10^{15}h^{-1}\,M_{\sun}$ to $M_{200} = 2.4 \times
  10^{14}h^{-1}\,M_{\sun}$.
\end{quote}
\end{center}
\end{table}

In the following, we will show radial profiles of the axis ratios and asymmetric
parameters that are obtained by computing the moment-of-inertia tensor at 30 different
radii for each cluster. In Figure~\ref{fig:CAelipt}, we report on the overall radial distribution of
$\ca$ and ellipticity in the gas and DM distributions.  Within $R_{200}$, the
ellipticities of gas density and pressure are rather flat at a level of $\ca
\simeq 0.85-0.9$. As laid out above, this is because dissipation effects at
the accretion shocks cause an effective sphericalization and erase the memory of
large-scale tidal fields. In contrast, ellipticities are increasing for the DM
as a function of radius due to the dissipationless nature of DM, i.e. $\ca$
decreases from values around 0.8 in the center to 0.7 at $R_{200}$. The radial
behaviour may be due to increased tidal effects on DM substructures at small
radii which causes a dramatic drop of their central mass density
\citep{2008MNRAS.391.1685S, 2008Natur.456...73S,
  2011arXiv1105.3240P}. Effectively this causes a redistribution of a clumped
(elliptical) to a smooth distribution that is able to couple more efficiently to
the (more spherical) gas distribution. Studying the asymmetric parameters, we find
that if a cluster is prolate, it is on average more elliptical than an oblate
one that is always close to spherically symmetric.

We find that the average axis ratios and ellipticities have a pronounced break
in their slopes at $r\sim 1.5 R_{200}$. The break in the ellipticity arises from
substructure in the cluster outskirts.  Recent X-ray observations of the Perseus
cluster find a strong signature of clumping in gas density
\citep{2011Sci...331.1576S}; qualitatively consistent with the findings in
simulations but not quantitatively \citep{2011ApJ...731L..10N}. This gas density
clumping is a direct tracer of substructure and becomes important at roughly the
same radius where we find the break in the ellipticity. Interestingly, this
effect is not only seen in DM and gas but also in pressure, which suggests that
the pressure is clumped in a similar fashion as the gas density.  In order to
accurately model the outskirts of clusters, semi-analytic models will need to
properly deal with the substructure.  In the Appendix \ref{sec:wr2}, we show
that one can attempt to counteract or lessen the impact of substructure on the
gas, pressure and DM shapes by including an $r^{-2}$-weighting when calculating
the moment-of-inertia tensor (cf. Eq.~(\ref{eq:Itens})). In future work, we will
further explore the issue of substructure.

\subsection{How shape profiles depend on modeled physics}

We also address the influence that changes in the simulated physics has on
cluster shapes in Fig.~\ref{fig:CAelipt}. While the ellipticity of the gas is
slightly larger in non-radiative models, it is very similar for the gas
distribution in our radiative models (radiative cooling and star formation with
and without AGN feedback). Dissipating accretion shocks seem to explain the
overall behavior rather well and the different physical models only marginally
change the cluster shapes in the gas.  In the DM, however, there is still a
pronounced difference among our two radiative physics models with the
ellipticities of the AGN feedback model being larger that in our pure radiative
model.  This small ellipticity is a remnant of overcooling that our pure
radiative model suffers with an associated star formation rate that is
unphysically high.  Most of these stars form out of the cold, dense gas in the
core region which causes a decreasing central pressure support so that gas at
larger radii moves in adiabatically and causes a deeper potential which in turn
causes the DM to adiabatically contract. Enhanced dissipation processes in the
gas sphericalize the potential which is then communicated to the DM during this
central settling.  We find that including AGN feedback counteracts the
overcooling issue and modifies the DM shapes on the level of 5\% in comparison
to our pure radiative simulations (cf. Fig \ref{fig:CAelipt}).

Our general trends are similar to those reported by \citet{2011ApJ...734...93L}
who also find that the DM distribution is more spherical for radiative
simulations in comparison to non-radiative models.  However, the differences
between radiative and non-radiative simulations are not as extreme as those
found in \citet{2011ApJ...734...93L}, since our radiative simulations do not
have the level of (catastrophic) cooling in the central regions, because their
simulations have higher resolution and include cooling from metals which we do
not.  The AGN feedback stabilizes the cooling and, thereby, lessens this
sphericalizing effect on the DM ellipticity.

\begin{figure*}
\begin{center}
  \hfill
  \resizebox{0.5\hsize}{!}{\includegraphics{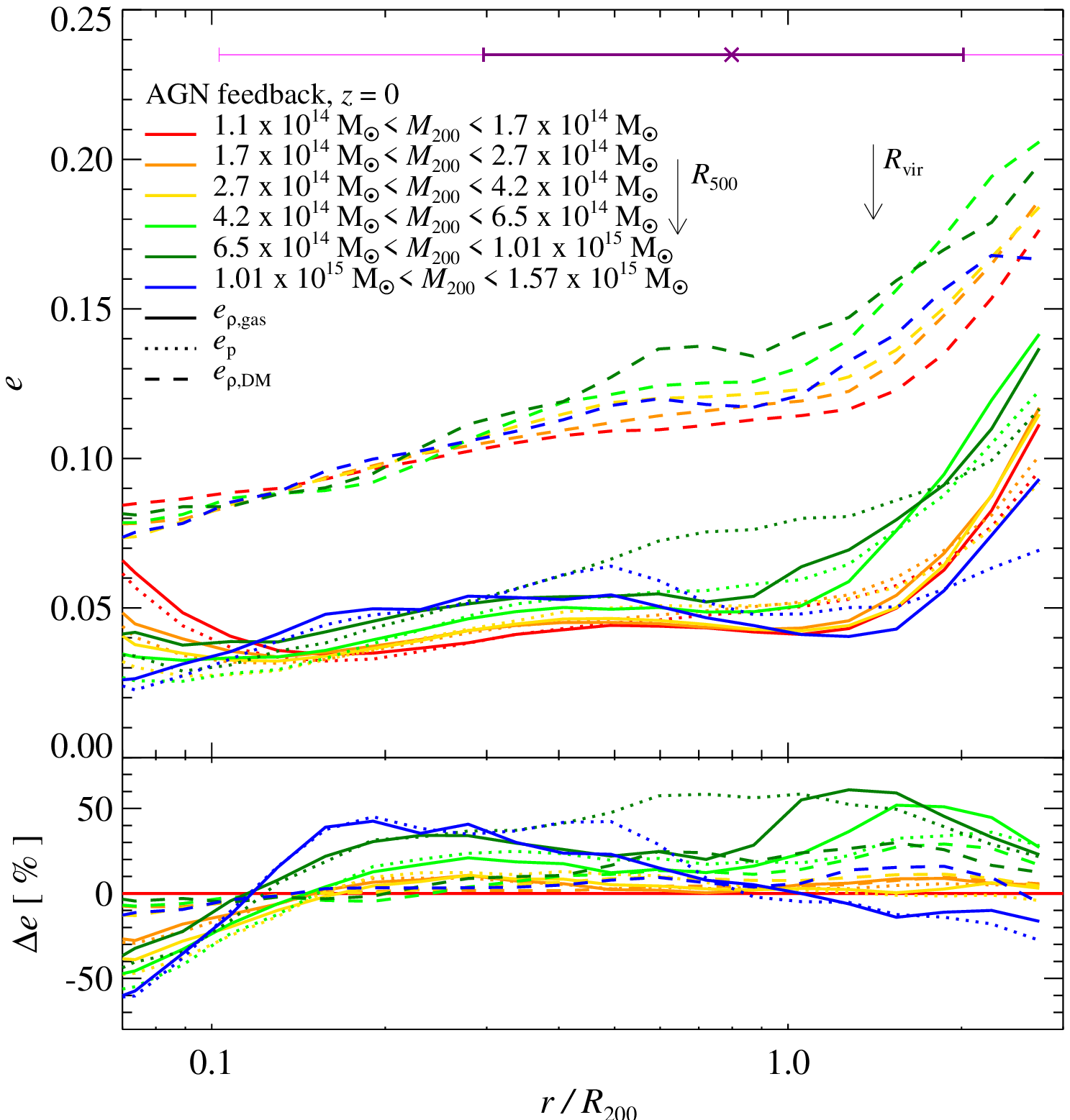}}%
  \resizebox{0.5\hsize}{!}{\includegraphics{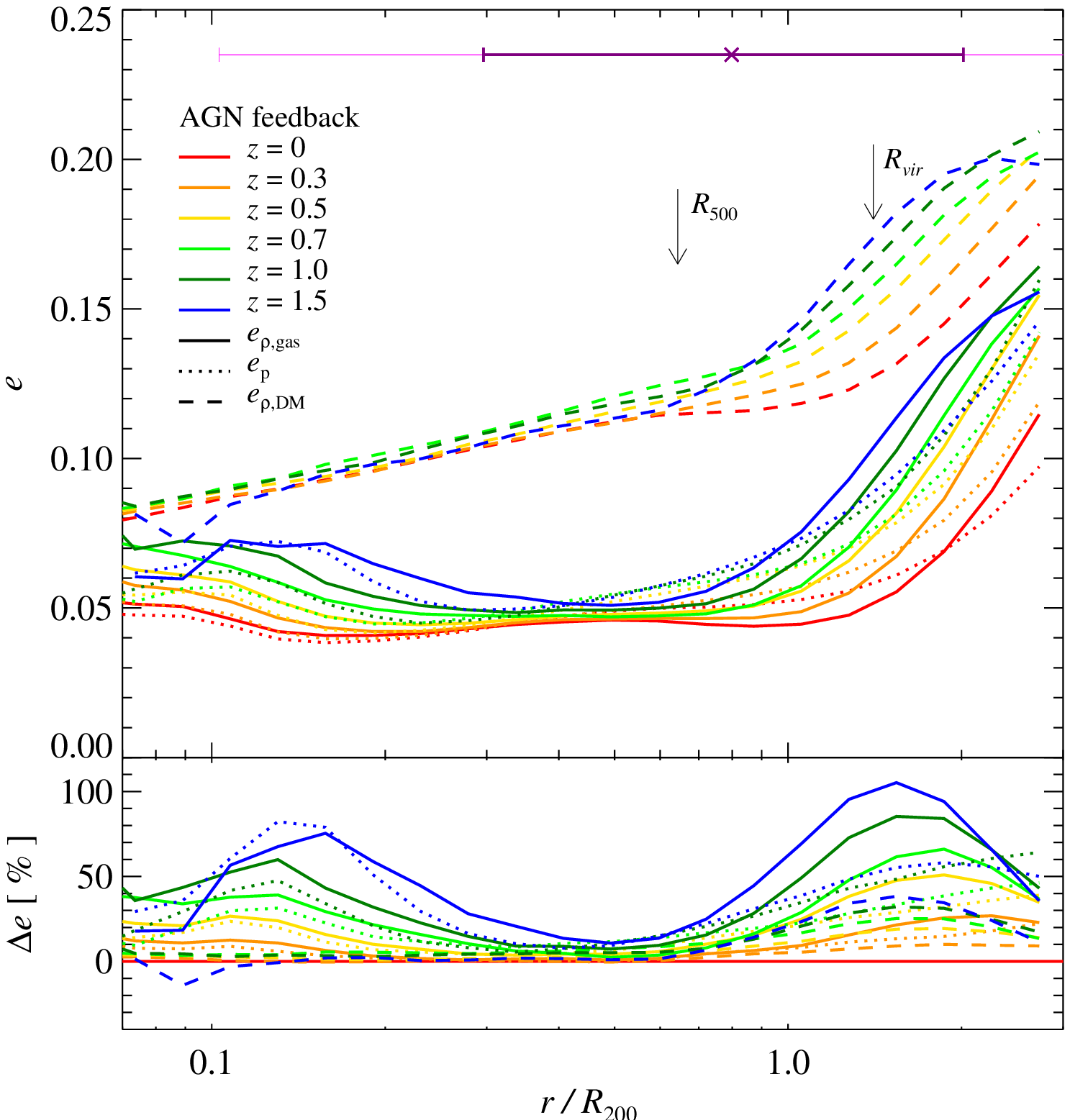}}\\
\end{center}
\caption{The dependence on mass and the redshift evolution of the cumulative ellipticity
  profile as a function of $r/R_{200}$. Left: Shown is the ellipticity profile
  at $z=0$ for various mass bins.  Bottom left: Shown are the relative
  differences in ellipticity to the lowest mass bin ($1.1\, \times 10^{14}\,
  \mathrm{M}_{\sun} \,<\, M_{200}\, <\, 1.7\, \times 10^{14}\,
  \mathrm{M}_{\sun}$ ). Over this mass range, the cluster ellipticities show a
  noticeable but not substantial mass dependence within $R_{500}$, in contrast
  to the stronger dependence on $P_{\mathrm{kin}}/P_{\mathrm{th}}$. Right: Shown
  is the ellipticity profile for various redshift bins. The horizontal purple
  and pink error bars have the same meaning as in
  Fig. \ref{fig:pratio_mass}. Bottom right: Shown is the relative difference of
  ellipticity at a given redshift to $z=0$. The redshift evolution of the
  ellipticity (especially at large radii, $r> R_{500}$) is driven by the larger
  amount of substructures at higher redshifts due to the increased mass
  accretion rate of group/cluster halos at these redshifts.  The
  pressure-weighted ellipticities track the density-weighted ellipticities well
  and show the same trends with redshift.}
\label{fig:eliptmassbins}
\end{figure*}

\subsection{Projected and intrinsic shapes}

In order to tie the underlying 3-dimensional structure of clusters to observable
2-dimensional projections, we compare the intrinsic 3D axis ratios to axis
ratios of random 2D projections, i.e. we project the DM density and gas
density/pressure distributions along a randomly chosen direction and then
compute the 2D moment-of-inertia tensor.  The results are shown in Figure
\ref{fig:2Dcompelipt}.  We find that the 2D axis ratios ($\ba$) for both, the
gas density and pressure are systematically closer to unity than the 3D ratios
$\ca$ by $\sim 5$--$10$\% (for the virial region and the central part). In the
case of the DM distribution, the projected (2D) axis ratios are on average a
$\sim 15$\% underestimate of the intrinsic (3D) axis ratios.  Using the linear
correlation coefficient statistic ($r_{\rmn{s}}$), we find that the random 2D
axis ratios are strongly correlated with the intrinsic 3D axis ratios with a
radially increasing correlation strength (cf. Fig \ref{fig:2Dcompelipt}). As
expected, $\ca$ serves as a limit to the observable projected (2D) axis ratio.
We find that the mean 2D axis ratio, as a function of cluster radius, closely
tracks $\ca$, modulo a roughly constant $\sim 5-10$\% bias (for the gas density
and pressure).

\subsection{Mass and redshift dependence of shape profiles}

Both the density- and pressure-weighted ellipticities show the same general
trends with radius and cluster mass. The ellipticity increases with increasing
cluster mass by $\sim$ 50\% over the mass ranges shown (cf. Fig
\ref{fig:eliptmassbins}).  On the right-hand side of
Fig. \ref{fig:eliptmassbins}, we show the redshift evolution of the cluster
shapes and find that the ellipticity is a stronger function of redshift than the
mass.  For increasing redshift, the break in the ellipticity profile moves to
smaller radii (when scaled to $R_{200}$). Both behaviors, the mass and redshift
dependence can be understood in the hierarchical picture for structure
formation, where clusters show increased mass accretion rates and hence
an increased level of substructure for larger clusters (at a given redshift) or,
equivalently, for a cluster of given mass at higher redshifts which probe on
average systematically younger systems.  Similar to the non-thermal pressure
support, the redshift evolution found in the ellipticities are lessened by a
different choice of scaling radius (cf. Appendix \ref{sec:rad}).  This result suggests that
using a single (constant) ellipticity profile for clusters is not
sufficient for percent level accuracy.

Pressure-weighted ellipticities are marginally more spherical than the
density-weighted ellipticities for $r<R_{500}$ (cf. Figs. \ref{fig:CAelipt} and
\ref{fig:eliptmassbins}). However, between $R_{500}$ and $2R_{500}$ the behavior
is reversed.  This is because the core region shows a smaller kinetic pressure
support implying that hydrostatic forces had time to act and to smooth out the
pressure distribution whereas at larger radii, pressure-weighted ellipticities
are affected more by infall caused by a noticeable pressure clumping at these
radii (cf. BBPS4).  We have shown that scales around $R_{500}$
are the most robust for studying the virial properties of clusters, since all
ellipticities show only $\sim 10$\% redshift evolution and mass dependence on
these scales. Taken together with the relatively modest degree of 
non-thermal pressure support on these scales found in \S~\ref{sec:Pkin} and that these scales are far enough out to
avoid the complications of the intricate ``short-distance'' physics of
the cluster core region, we can give further justification for what is already  the practice in the X-ray
cluster community, driven by the nature of the $X$-ray data, namely a focus on $R_{500}$ and environs for "global" cluster properties. 

\begin{figure}
\epsscale{1.20}
\plotone{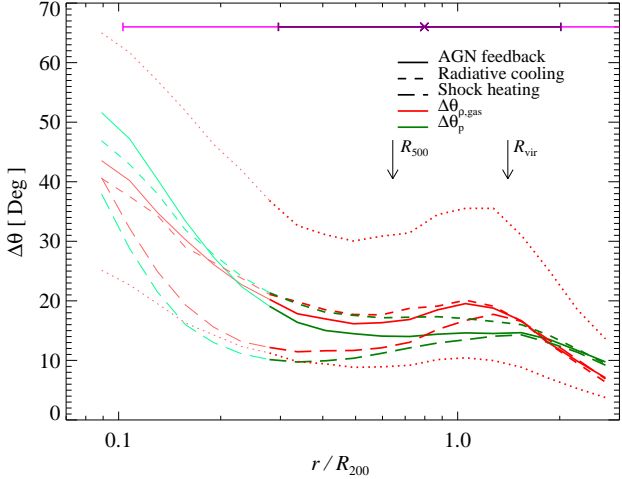}
\caption{The weighted median angles between the DM major axis and gas (red) and
  pressure (green) axes at each radius for all simulated physics
  models: AGN feedback (solid), radiative cooling (short-dashed), and shock
  heating-only (long-dashed). The 25$^{\rmn{th}}$ and 75$^{\rmn{th}}$ percentile
  values are shown for the gas density in the AGN feedback model (dotted).  On
  average the gas and pressure axes are misaligned by $20$ to $30$ degrees to
  the DM principle axis, independent of our simulated physics models. However,
  both simulations with radiative cooling show more misalignment in the inner
  regions than the non-radiative simulations. The light colors and lines
  represent the region where the average cluster shape is close to spherical
  ($c/a|_\rmn{DM} > 0.75$) such that the major axes are not well defined and
  their angles are approaching a random distribution. Note that we have weighted
  the average angles by $1 - \ca$ to down-weight the angles from the spherical
  ICM shapes and the cluster interiors. The horizontal purple and pink error
  bars have the same meaning as in Fig. \ref{fig:pratio_mass}. }
\label{fig:eliptallign}
\end{figure}

\subsection{Alignment variations and semi-analytical models}
\label{sec:twist}

Semi-analytic models for the baryon distribution in clusters include an
underlying assumption that baryons will arrange themselves along equipotential
surfaces (or in some cases the DM density-weighted surfaces).  Given the
importance of this assumption, we test its validity in our simulations.  In
Figure \ref{fig:CA_M_z}, we plot the ratio $\ca$ for both dark matter and gas as
function of cluster mass and redshift.  While $\ca$ for DM haloes decreases with
halo mass as expected \citep{2002ApJ...574..538J}, we find that $\ca$ is
constant for the gas distribution.  This is potentially a problem for
semi-analytic models of ICM gas \citep{2005ApJ...634..964O,2009ApJ...700..989B},
which solve for the resulting gas distribution in a DM-dominated gravitational
potential as obtained from dissipationless simulations while allowing for a
constant non-thermal pressure contribution (in the latter case).  However, the
gravitational potential from the DM is more spherical than the underlying matter
distribution \citep[e.g.][]{2011ApJ...734...93L}, and so the semi-analytic shape
estimates are not as discrepant as one might expect from Figure
\ref{fig:CA_M_z}.

A more important issue is the alignment of the gas or pressure with respect to
the DM.  We calculate the angular difference between the major axes of the DM
and those of the gas and pressure major axes at a given radius, using the
moment-of-inertia tensor eigenvectors $\vecbf{E}_{1,\rmn{DM}}(r) \cdot
\vecbf{E}_{1,\rmn{gas}}(r)$.  When calculating misalignment,
the major axes of nearly spherical objects are poorly defined quantities. To
avoid this problem we calculate, in each radial bin, a weighted average using $1
- \ca$ as the weight. Furthermore, we exclude the region inside $0.3R_{200}$
since the gas and pressure shapes are nearly spherical, with
$c/a|_\rmn{DM}>0.75$.  On average at a given radius, the cluster gas and
pressure are $20-30$ degrees misaligned from the major axis of the DM
(cf. Fig. \ref{fig:eliptallign}).  In the next section we show SZ measurements
of the total thermal energy in clusters, $Y$, strongly depend on the projection
axis through the cluster. Thus, misalignment between the semi-analytic baryon
distortion and the ``true'' distribution may cause biases when using
semi-analytic models to, \textit{e.g.}, tie weak-lensing and SZ observations
together.

\section{SZ Scaling Relation}
\label{sec:SZ}

In this section we explore the impact of AGN feedback, cluster shapes and
kinetic pressure support on the SZ-flux-to-mass relation, \YM, using our large
sample of clusters. We compute the SZ flux for all clusters for both, spherical
boundaries and cylindrical apertures ($Y_{\rmn{sph}}$ and $Y_{\rmn{cyl}}$). For
the cylindrical aperture calculations the total fluxes are computed along each
axis of the moment-of-inertia frame, measured at $R_{200}$, and additionally
along each axis of another randomly-oriented frame.  We choose the line of sight
boundaries for the cylindrical integrations to be three times the radius of the
aperture. This procedure enables quantifying the importance of substructure,
which we have already shown in Sections \ref{sec:Pkin} and \ref{sec:Elpt} to be
significant at radii beyond $R_{200}$. From the calculated $Y_{\Delta}$ values
we fit an average scaling relation,
\begin{equation}\label{eq:ym}
Y_{\Delta} = 10^B \, 
\left(\frac{M_{\Delta}}{3\times 10^{14}\, \hsev^{-1}\,M_{\sun}}\right)^A 
\,\hsev^{-1} \,\rmn{Mpc}^2,
\end{equation} 
\noindent where $A$ and $B$ are the fit parameters for the slope and
normalization, respectively. We weight each cluster by its $Y_{\Delta}$ when
fitting for $A$ and $B$ to keep the low-mass clusters from completely dominating
the fit.

\begin{figure}
\epsscale{1.20}
\plotone{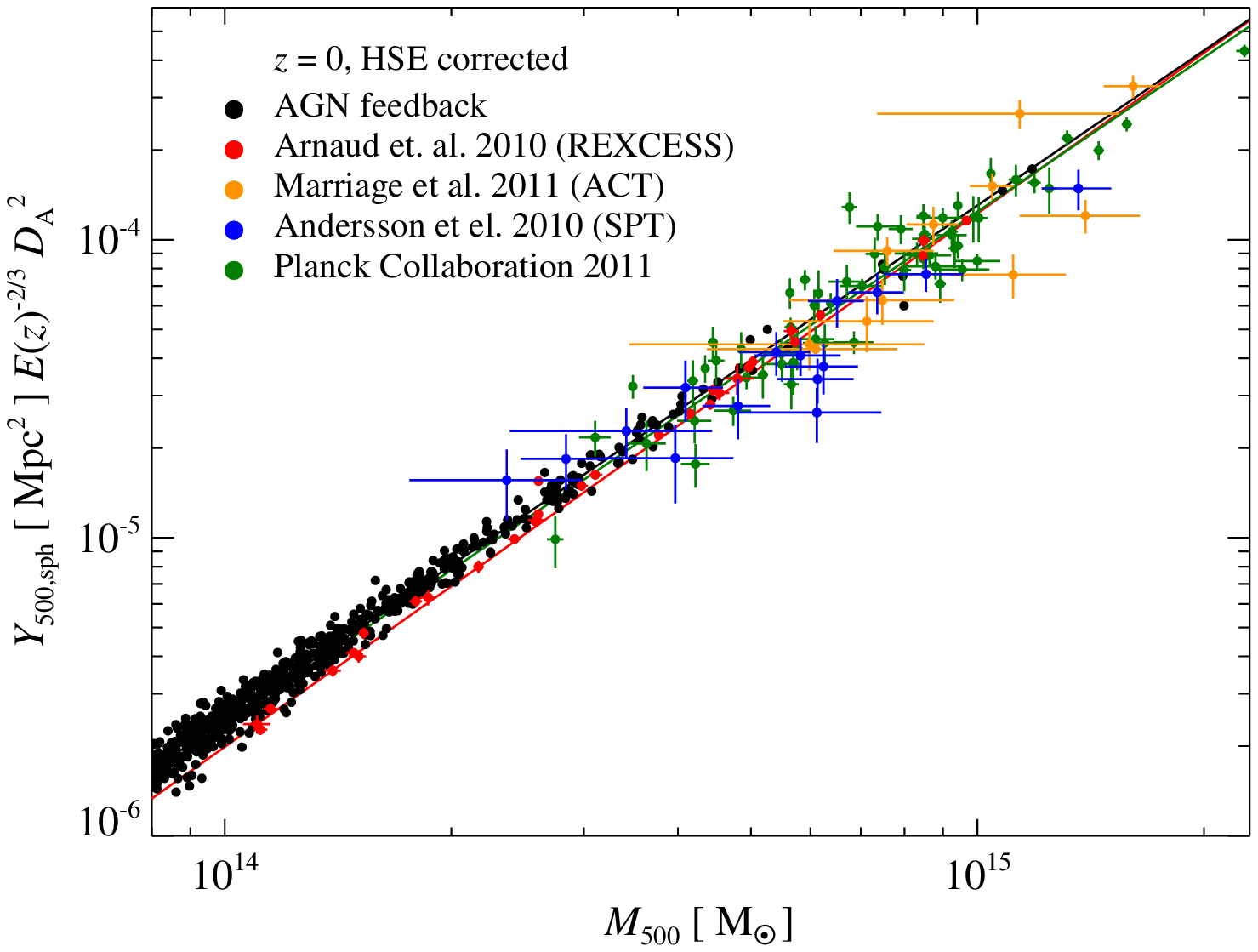}
\caption{The \YM scaling relation for the AGN feedback simulations compared
  to recent X-ray results from \citet{2010A&A...517A..92A} and SZ results from
  ACT \citep{2010arXiv1010.1065M}, SPT \citep{2010arXiv1006.3068A}, and {\em Planck}
  \citep{2011arXiv1101.2026P}. We have applied the 15\% correction to the X-ray
  $M_{\rmn{HSE}}$ from \citet{2006ApJ...650..128K}.  }
\label{fig:YMrelcomp}
\end{figure}

\subsection{Self-similar \YM scaling relation}

We review the expectations for $Y$ in the idealized case of a cluster in virial
equilibrium to help understand how possible deviations from the self-similar \YM
relation and the scatter about it may arise. Starting with Eq.~(\ref{eq:Ydelta}),
which has been rewritten as,
\begin{equation}
  \label{eq:Y}
  Y=\frac{\sigma_\rmn{T}}{m_\e c^2}\int_0^{R_{200}}\dd V P_\e=
\frac{(\gamma-1)\,\sigma_\rmn{T}}{m_\e c^2}\,x_\e\,X_\rmn{H}\,\mu\,E_\rmn{gas},
\end{equation}
where $x_\e$ is the electron fraction defined as the ratio of electron and
hydrogen number densities $x_\e = n_\e/n_\rmn{H} = (X_\rmn{H}+1)/(2\,X_\rmn{H})
= 1.158$, $\gamma=5/3$ is the adiabatic index, $\mu = 4/(3X_\rmn{H} + 1 + 4
X_\rmn{H} x_\e) = 0.588$ denotes the mean molecular weight for a fully ionized
medium of primordial abundance, and we assume equilibrium between the electron
and ion temperatures. Next, we define the characteristic temperature of the halo
\citep{2002MNRAS.336.1256K} as
\begin{equation}
  \label{eq:T200}
  kT_{200} = \frac{G M_{200} \, \mu\, m_\p}{3 R_{200}} = \frac{\mu\,m_\p}{3}
  \left[10\, G\, H_0\, M_{200}\,E(z)\right]^{2/3},
\end{equation}
\noindent so we can write the total thermal energy of the halo with Eq.~(\ref{eq:T200}) as
\begin{eqnarray}
  \label{eq:Eth}
  E_\rmn{gas} &=& \frac{3}{2}\,N_\rmn{gas}\,kT_{200} = 
  (1-f_*)\,f_\rmn{b}\,f_{\mathrm{c}}\,\frac{G M_{200}^2}{2\, R_{200}} \nonumber\\
  &=& (1-f_*)\,f_\rmn{b}\,f_{\mathrm{c}}\,\frac{G}{2}\,\left[800\,\rho_\rmn{cr}(z)\right]^{1/3}M_{200}^{5/3}.
\end{eqnarray}
Here $f_* \lesssim M_*/M_\rmn{b}$ is the stellar mass fraction within
the halo and $f_{\mathrm{c}}$ is the correction factor for the fraction
of missing baryons at a given overdensity. Then we insert
Eq.~(\ref{eq:Eth}) into Eq.~(\ref{eq:Y}) to get the integrated Compton-$y$
parameter within $R_{200}$,
\begin{eqnarray}
  \label{eq:Y2}
  Y &=& \frac{(\gamma-1)\sigma_\rmn{T}}{m_\e c^2}x_\e X_\rmn{H}\mu
  (1-f_*)f_\rmn{b}f_{\mathrm{c}}\,G
  \left[\frac{\pi}{3}100\rho_\rmn{cr}(z)\right]^{1/3}M_{200}^{5/3}\nonumber\\
  &=& 97.6\,\hsev^{-1}\,\rmn{kpc}^2E(z)^{2/3}
  \left(\frac{M_{200}}{10^{15}\,\hsev^{-1}\rmn{M}_\odot}\right)^{5/3}
  \frac{\Omega_\rmn{b}}{0.043}\frac{0.25}{\Omega_\rmn{m}}
  \label{eq:Ysim}
\end{eqnarray}
\noindent For Eq.~(\ref{eq:Ysim}), we set $f_*=0$, $f_{\mathrm{c}}=0.93$ (as
calculated from our shock heating simulations at $R_{200}$) and adopted the
cosmological parameters of our simulation. This simple analytical expression for
the \YM scaling relation allows one to explore the assumptions underlying its
derivation.  More specifically, we test the assumptions of spherical
gravitational potential, zero non-thermal pressure support, and constant
$f_\rmn{b}$ (and for simulation with star formation, constant $f_*$) at
$R_{\Delta}$, independent of cluster mass.

\begin{table*}
  \caption{\YM scaling relation fits for different simulated physics, sub-sampling in
    kinetic-to-thermal energy and ellipticity (of the density and pressure distribution), and along different projected
    axes yielding $Y_{\rmn{cyl}}$.}
\label{tab:YMs}
\begin{center}
   \leavevmode
\begin{tabular}{l|ccc|ccc|ccc}
  \hline
  \hline
$\Delta = 200$  & & $z = 0$ &  & &$z = 0.5$ &  & & $z = 1$ & \\
  \hline
      &  $B$ & $A$ & $\sigma_Y$&  $B$ & $A$ & $\sigma_Y$ &  $B$ & $A$ & $\sigma_Y$\\
  \hline
  Simulated physics & & &  & & &  & & & \\
  \hline
  Theory, Eq.~(\ref{eq:Ysim}) & -4.88 & 1.67 & -  & -4.81 & 1.67 & - & -4.74 & 1.67 &- \\
  Shock heating     & -4.87 $\pm$  0.01 &  1.64 $\pm$  0.03 & 0.115 $\pm$ 0.001 & -4.81 $\pm$  0.02 &  1.63 $\pm$  0.04 & 0.116 $\pm$ 0.001 & -4.76 $\pm$  0.05 &  1.61 $\pm$  0.07 & 0.111 $\pm$ 0.002 \\
  Radiative cooling & -4.94 $\pm$  0.01 &  1.67 $\pm$  0.03 & 0.119 $\pm$ 0.001 & -4.88 $\pm$  0.02 &  1.67 $\pm$  0.05 & 0.121 $\pm$ 0.002 & -4.83 $\pm$  0.05 &  1.66 $\pm$  0.09 & 0.121 $\pm$ 0.002\\
  AGN feedback      & -4.92 $\pm$  0.01 &  1.71 $\pm$  0.03 & 0.133 $\pm$ 0.001 & -4.87 $\pm$  0.02 &  1.72 $\pm$  0.05 & 0.136 $\pm$ 0.002 & -4.82 $\pm$  0.05 &  1.73 $\pm$  0.09 & 0.147 $\pm$ 0.003\\
  \hline
  $\KU^a$ & & &  & & &  & & & \\
  \hline
  Lower $3^{\mathrm{rd}}$  & -4.88 $\pm$  0.04 &  1.73 $\pm$  0.07 & 0.115 $\pm$ 0.002 & -4.83 $\pm$  0.06 &  1.73 $\pm$  0.11 & 0.122 $\pm$ 0.003 & -4.78 $\pm$  0.14 &  1.76 $\pm$  0.21 & 0.137 $\pm$ 0.004\\
  Middle $3^{\mathrm{rd}}$ & -4.92 $\pm$  0.02 &  1.73 $\pm$  0.05 & 0.116 $\pm$ 0.002 & -4.87 $\pm$  0.04 &  1.73 $\pm$  0.08 & 0.121 $\pm$ 0.003 & -4.82 $\pm$  0.07 &  1.73 $\pm$  0.13 & 0.138 $\pm$ 0.005\\
  Upper $3^{\mathrm{rd}}$  & -4.94 $\pm$  0.02 &  1.72 $\pm$  0.04 & 0.123 $\pm$ 0.002 & -4.89 $\pm$  0.03 &  1.73 $\pm$  0.07 & 0.122 $\pm$ 0.003 & -4.85 $\pm$  0.09 &  1.72 $\pm$  0.15 & 0.141 $\pm$ 0.005\\
  \hline
  $\ca$ (gas)$^a$ & & &  & & &  & & & \\
  \hline
  Lower $3^{\mathrm{rd}}$  & -4.94 $\pm$  0.02 &  1.72 $\pm$  0.05 & 0.131 $\pm$ 0.002 & -4.89 $\pm$  0.04 &  1.72 $\pm$  0.08 & 0.133 $\pm$ 0.003 & -4.84 $\pm$  0.09 &  1.73 $\pm$  0.15 & 0.147 $\pm$ 0.005\\
  Middle $3^{\mathrm{rd}}$ & -4.92 $\pm$  0.02 &  1.70 $\pm$  0.05 & 0.129 $\pm$ 0.002 & -4.86 $\pm$  0.05 &  1.73 $\pm$  0.09 & 0.133 $\pm$ 0.003 & -4.82 $\pm$  0.10 &  1.72 $\pm$  0.16 & 0.136 $\pm$ 0.004\\
  Upper $3^{\mathrm{rd}}$  & -4.90 $\pm$  0.02 &  1.72 $\pm$  0.05 & 0.119 $\pm$ 0.002 & -4.86 $\pm$  0.04 &  1.72 $\pm$  0.08 & 0.123 $\pm$ 0.003 & -4.80 $\pm$  0.09 &  1.73 $\pm$  0.15 & 0.138 $\pm$ 0.004\\
  \hline
  $\ca$ (pressure)$^a$ & & &  & & &  & & & \\
  \hline
  Lower $3^{\mathrm{rd}}$  & -4.94 $\pm$  0.02 &  1.71 $\pm$  0.05 & 0.132 $\pm$ 0.002 & -4.88 $\pm$  0.04 &  1.73 $\pm$  0.08 & 0.134 $\pm$ 0.003 & -4.84 $\pm$  0.08 &  1.73 $\pm$  0.14 & 0.152 $\pm$ 0.005\\
  Middle $3^{\mathrm{rd}}$ & -4.92 $\pm$  0.03 &  1.71 $\pm$  0.05 & 0.129 $\pm$ 0.002 & -4.87 $\pm$  0.04 &  1.72 $\pm$  0.08 & 0.128 $\pm$ 0.003 & -4.82 $\pm$  0.08 &  1.72 $\pm$  0.14 & 0.134 $\pm$ 0.004\\
  Upper $3^{\mathrm{rd}}$  & -4.90 $\pm$  0.03 &  1.72 $\pm$  0.05 & 0.122 $\pm$ 0.002 & -4.86 $\pm$  0.05 &  1.71 $\pm$  0.09 & 0.129 $\pm$ 0.003 & -4.78 $\pm$  0.12 &  1.76 $\pm$  0.19 & 0.137 $\pm$ 0.004\\
  \hline
  $\ba$ (gas)$^a$ & & &  & & &  & & & \\
  \hline
  Lower $3^{\mathrm{rd}}$  & -4.93 $\pm$  0.02 &  1.70 $\pm$  0.04 & 0.138 $\pm$ 0.007 & -4.88 $\pm$  0.04 &  1.72 $\pm$  0.07 & 0.140 $\pm$ 0.010 & -4.83 $\pm$  0.07 &  1.73 $\pm$  0.12 & 0.149 $\pm$ 0.016\\
  Middle $3^{\mathrm{rd}}$ & -4.92 $\pm$  0.02 &  1.71 $\pm$  0.05 & 0.130 $\pm$ 0.006 & -4.87 $\pm$  0.04 &  1.72 $\pm$  0.08 & 0.137 $\pm$ 0.009 & -4.83 $\pm$  0.11 &  1.72 $\pm$  0.18 & 0.149 $\pm$ 0.014\\
  Upper $3^{\mathrm{rd}}$  & -4.91 $\pm$  0.02 &  1.72 $\pm$  0.05 & 0.128 $\pm$ 0.002 & -4.86 $\pm$  0.04 &  1.72 $\pm$  0.07 & 0.132 $\pm$ 0.003 & -4.81 $\pm$  0.09 &  1.73 $\pm$  0.14 & 0.136 $\pm$ 0.004\\
  \hline
  $Y_{\rmn{cyl}}$ rotated$^a$ & & &  & & &  & & & \\
  \hline
  Minor axis  & -4.87 $\pm$  0.01 &  1.69 $\pm$  0.03 & 0.126 $\pm$ 0.001 & -4.83 $\pm$  0.02 &  1.69 $\pm$  0.05 & 0.136 $\pm$ 0.002 & -4.79 $\pm$  0.05 &  1.70 $\pm$  0.09 & 0.144 $\pm$ 0.003\\
  Middle axis & -4.87 $\pm$  0.01 &  1.69 $\pm$  0.03 & 0.128 $\pm$ 0.001 & -4.82 $\pm$  0.02 &  1.69 $\pm$  0.05 & 0.133 $\pm$ 0.002 & -4.78 $\pm$  0.05 &  1.69 $\pm$  0.09 & 0.147 $\pm$ 0.003\\
  Major axis  & -4.84 $\pm$  0.01 &  1.68 $\pm$  0.03 & 0.137 $\pm$ 0.002 & -4.79 $\pm$  0.02 &  1.67 $\pm$  0.05 & 0.152 $\pm$ 0.002 & -4.74 $\pm$  0.05 &  1.68 $\pm$  0.09 & 0.167 $\pm$ 0.003\\
  \hline
  $Y_{\rmn{cyl}}$ random$^a$ & & &  & & &  & & & \\
  \hline
  axis 1      & -4.86 $\pm$  0.01 &  1.69 $\pm$  0.03 & 0.129 $\pm$ 0.001 & -4.82 $\pm$  0.02 &  1.69 $\pm$  0.05 & 0.140 $\pm$ 0.002 & -4.77 $\pm$  0.05 &  1.70 $\pm$  0.09 & 0.146 $\pm$ 0.003\\
  axis 2      & -4.86 $\pm$  0.01 &  1.69 $\pm$  0.03 & 0.129 $\pm$ 0.001 & -4.82 $\pm$  0.02 &  1.69 $\pm$  0.05 & 0.138 $\pm$ 0.002 & -4.78 $\pm$  0.05 &  1.68 $\pm$  0.09 & 0.150 $\pm$ 0.003\\
  axis 3      & -4.86 $\pm$  0.01 &  1.69 $\pm$  0.03 & 0.130 $\pm$ 0.001 & -4.82 $\pm$  0.02 &  1.69 $\pm$  0.05 & 0.141 $\pm$ 0.002 & -4.77 $\pm$  0.05 &  1.70 $\pm$  0.09 & 0.152 $\pm$ 0.003\\
  \hline
\end{tabular}
\end{center}
\begin{quote} 
  $^a$ For fits to all sub-samples/projections, we use our AGN feedback model. Fit parameters are defined in Eq.~(\ref{eq:ym}).\\
  $^b$ For the $\ba$ sub-sampling of $Y_{\rmn{cyl}}$, we chose random axis 1.
\end{quote}
\end{table*}

\begin{figure*}
\begin{center}
  \hfill
  \resizebox{0.5\hsize}{!}{\includegraphics{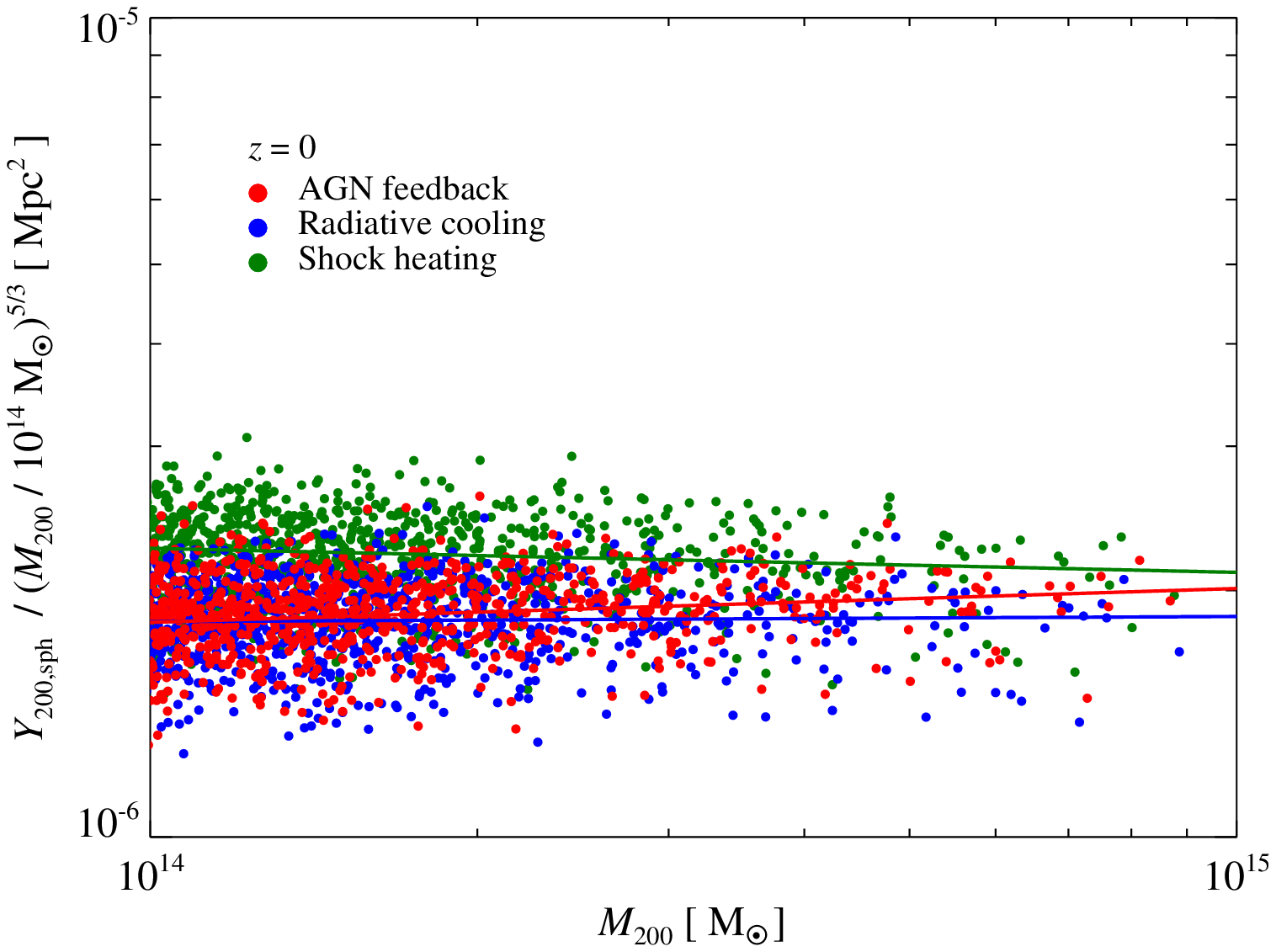}}%
  \resizebox{0.5\hsize}{!}{\includegraphics{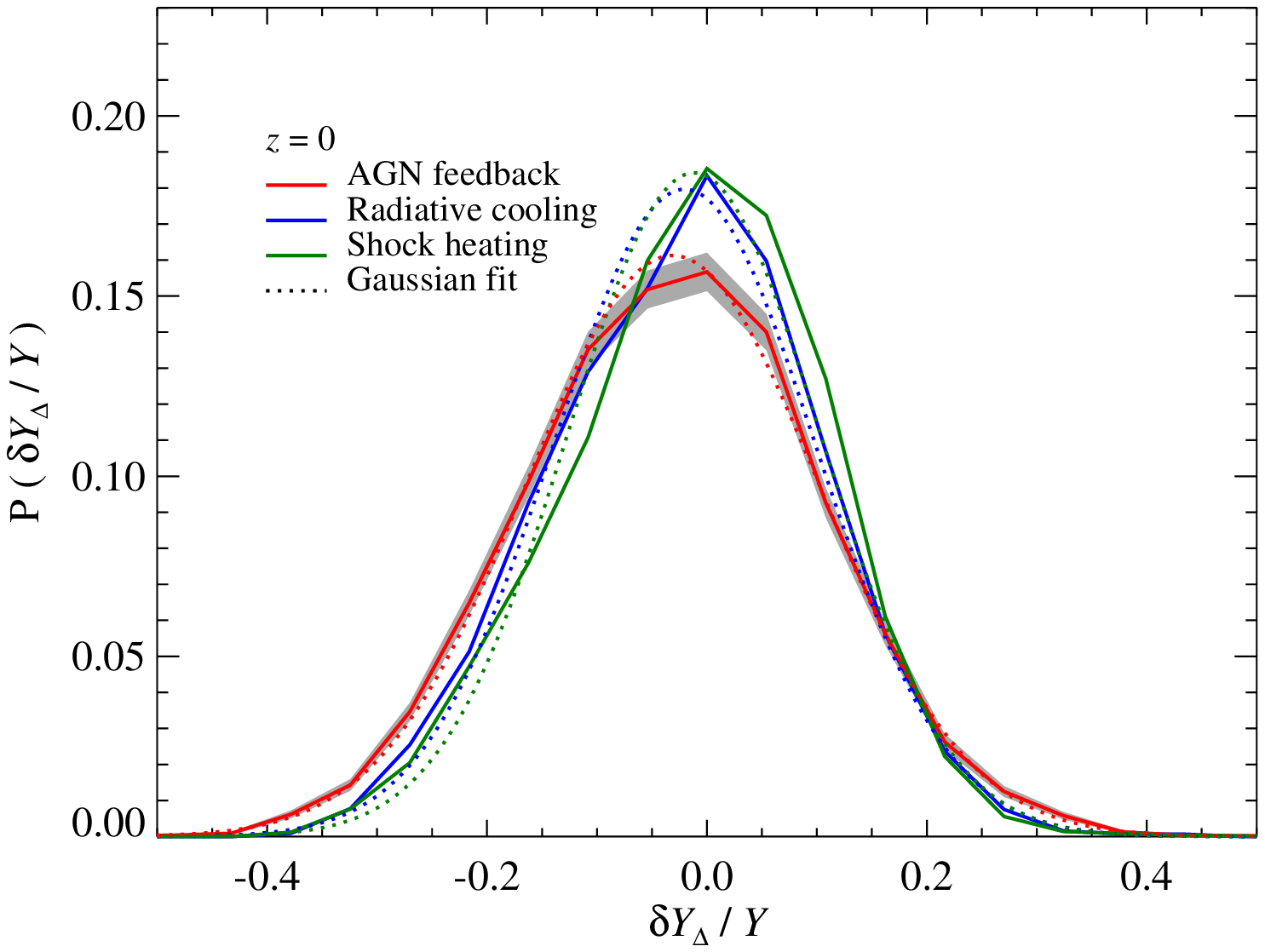}}\\
\end{center}
\caption{The normalization, slope and scatter of the \YM scaling relations all
  depend on the simulated physics. Left: The \YM scaling relations at $z=0$ for
  all simulated physics models: shock heating (green), radiative cooling (blue),
  and AGN feedback (red). The $y$-axis has been scaled by $M^{5/3}$ to highlight
  the deviations from self-similarity. Right: The probability distributions for
  the relative deviation, $\delta Y_\Delta / Y$, with respect to the best fits for all three physics
  models. We also show Gaussian fits (dotted lines) and include Poisson
  deviations for the AGN feedback simulations (grey band). We find that the AGN
  feedback simulations have the largest scatter and a steeper slope compared to
  the other simulations.}
\label{fig:YMrelphys}
\end{figure*}

\subsection{Comparison to data}

In Figure \ref{fig:YMrelcomp}, we compare $Y_{\rmn{sph}}$ for our simulated
clusters to the X-ray results from \citet{2010A&A...517A..92A}, and the SZ
results from ACT \citep{2010arXiv1010.1065M}, SPT \citep{2010arXiv1006.3068A},
and {\em Planck} \citep{2011arXiv1101.2026P}.  We adopt the 15\% correction to the
X-ray $M_{\rmn{HSE}}$ estimates from \citet{2006ApJ...650..128K} which is valid
for the respective observational sample selection criterion.  Our
$Y_{\rmn{sph}}$-M relation with AGN feedback is consistent with the current data
from X-ray and SZ observations.  However, at group scales, our simulations
slightly overpredict the SZ flux due to the too high gas fractions,
$f_\rmn{gas}$, in our simulations compared to X-ray observations
(cf. BBPS4). Potentially our simulations are missing some of the relevant physics
that governs $f_\rmn{gas}$ \citep[see, e.g.,][]{PCB} or underestimate the action
of AGN feedback on these mass scales.

The $Y_{\rmn{sph}}$ reported by SZ surveys for known clusters use an
X-ray-derived estimate of the aperture size.  This is useful because the cluster
radii are typically poorly measured in SZ, and so the X-ray aperture fixes the
SZ measurement along the otherwise degenerate aperture flux/aperture radius
relation.  However, this prior introduces correlations between the X-ray and SZ
observations, which makes comparisons between these observations difficult to
interpret.

\begin{figure*}
\begin{center}
  \hfill
  \resizebox{0.5\hsize}{!}{\includegraphics{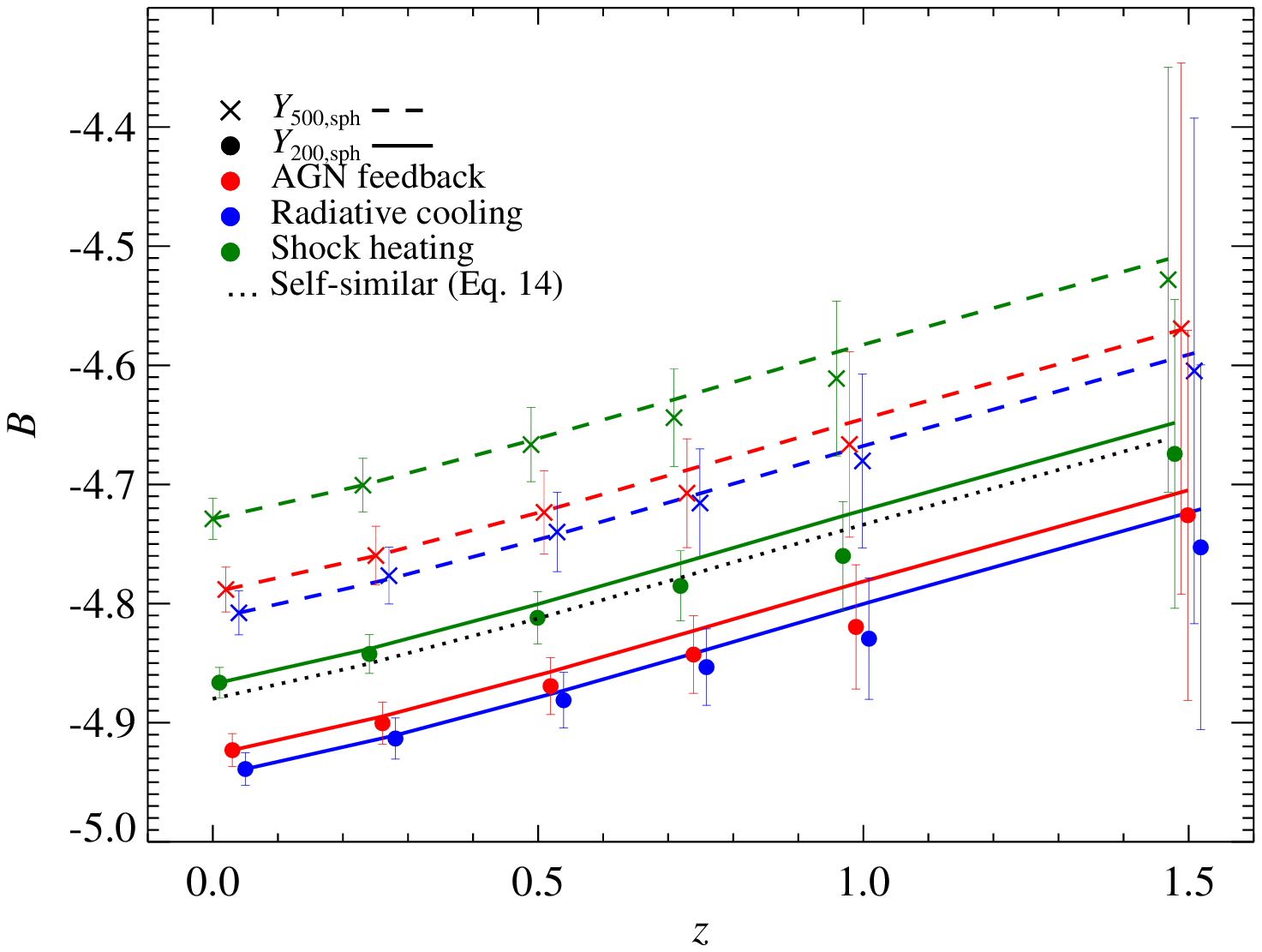}}%
  \resizebox{0.5\hsize}{!}{\includegraphics{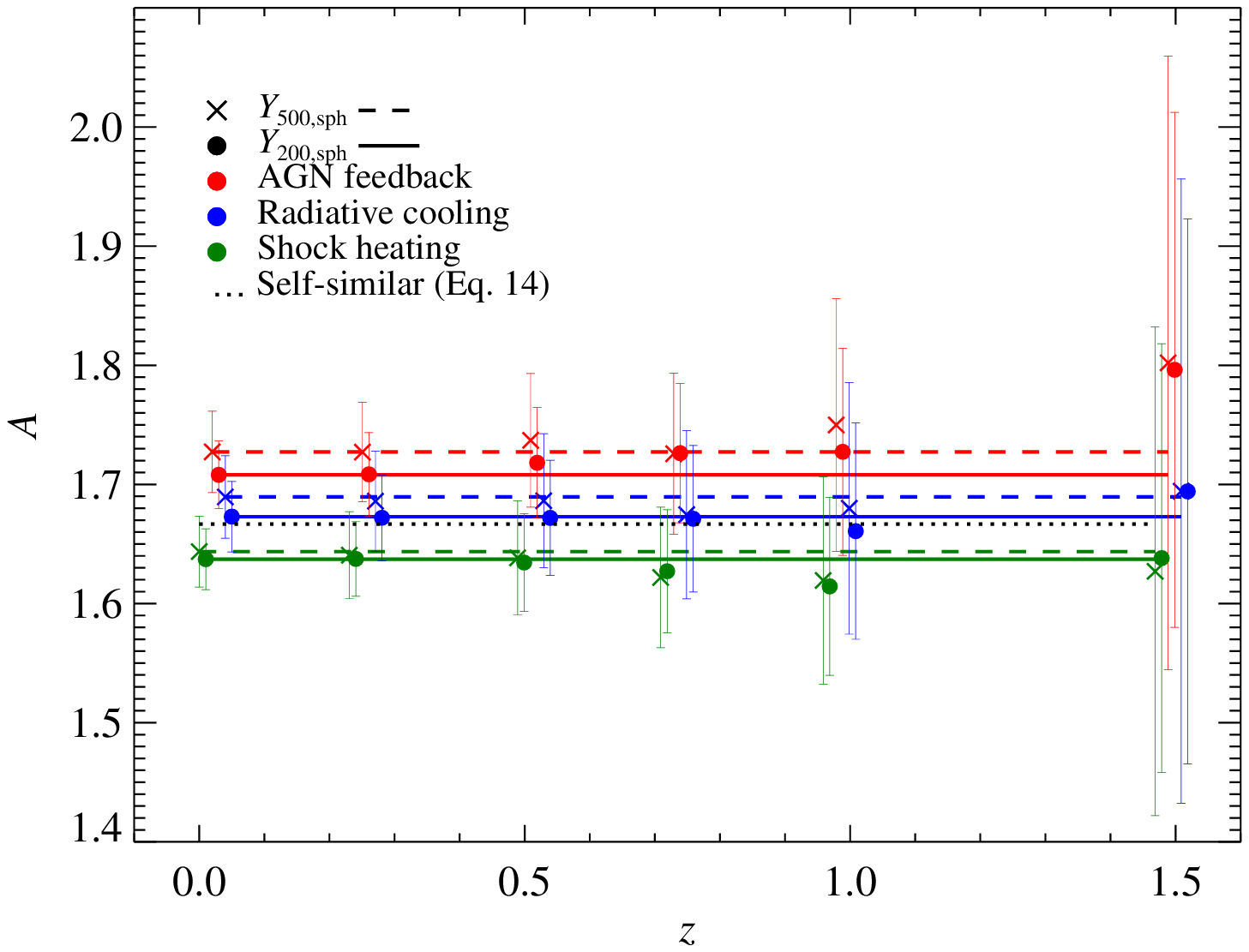}}\\
\end{center}
\caption{All simulated \YM scaling relations evolve self-similarly with redshift
  according to Eq.~(\ref{eq:Ysim}).  We show the \YM scaling relation fits for
  the normalization, $B$, (left panel) and slope, $A$, (right panel) as a
  function of redshift and for two different cluster masses $M_{200}$ and
  $M_{500}$, and compare those to the self similar prediction for $M_{200}$ (dotted
  black). The \YM relation of AGN feedback simulations has a different
  slope, but shows no anomalous redshift evolution relative to self-similar
  evolution.}
\label{fig:YM_zfunc}
\end{figure*}

\begin{figure*}
\begin{center}
  \hfill
  \resizebox{0.5\hsize}{!}{\includegraphics{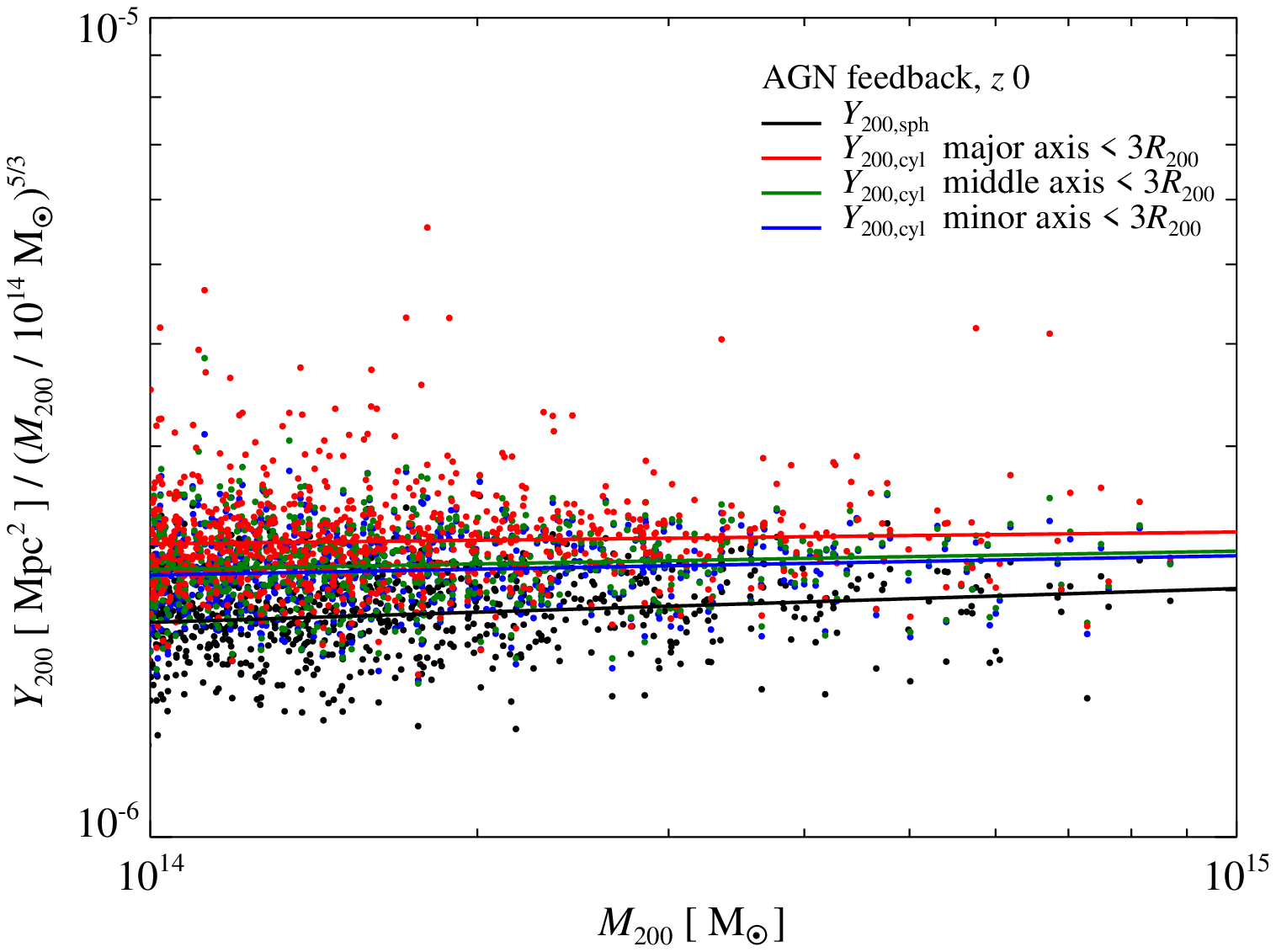}}%
  \resizebox{0.5\hsize}{!}{\includegraphics{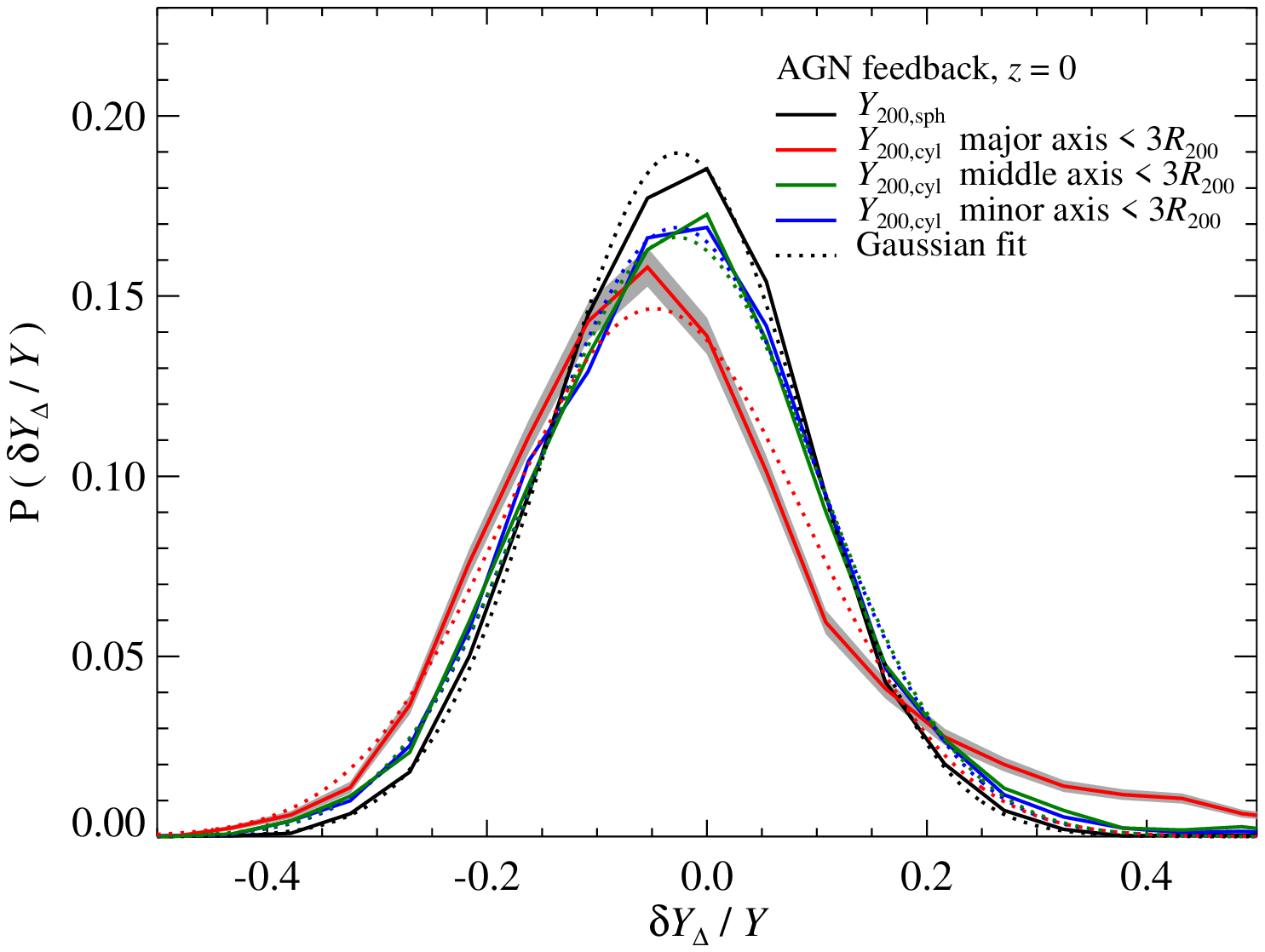}}\\
\end{center}
\caption{Rotating the clusters into their major, middle and minor axes and
  calculating projected (cylindrical) \YM relations shows the effect of infalling
  substructure. Left: The cylindrical \YM scaling relations from the AGN
  simulations for clusters that have been rotated into their major, middle, and
  minor axes defined by computing the (3D) moment-of-inertia tensor within
  $R_{200}$. Right: The probability distributions for the scatter, $\delta
  Y_\Delta$, relative to the best fits for all three distributions, each
  representing a distinctive rotation as well as the spherical distribution
  (black). We include the Gaussian fits (dotted lines) and the Poisson
  deviations for the major axis rotation (grey band). Rotating the clusters such
  that integration happens along the major axis increases the total $Y$ values,
  while further distorting and increasing the scatter (due to the large
  cluster-to-cluster variance in the infall regions). The
  $Y_{\rmn{cyl}}$ values are integrated along the given axis from $-3R_{200}$ to
  $3R_{200}$; hence for any given cluster $Y_{\rmn{cyl}}\ge Y_{\rmn{sph}}$.}
\label{fig:YM_CYL}
\end{figure*}

\begin{figure*}
\begin{center}
  \hfill
  \resizebox{0.5\hsize}{!}{\includegraphics{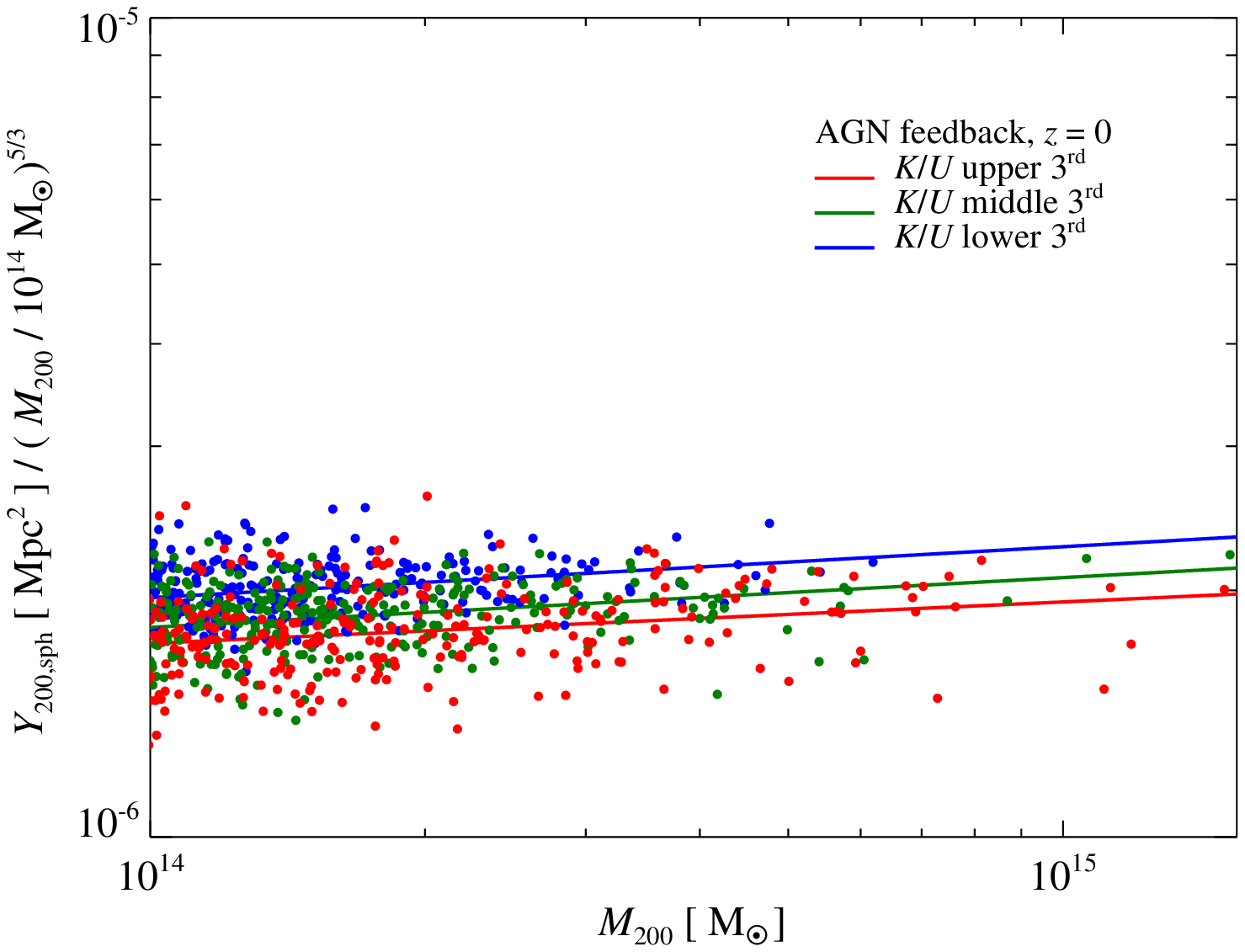}}%
  \resizebox{0.5\hsize}{!}{\includegraphics{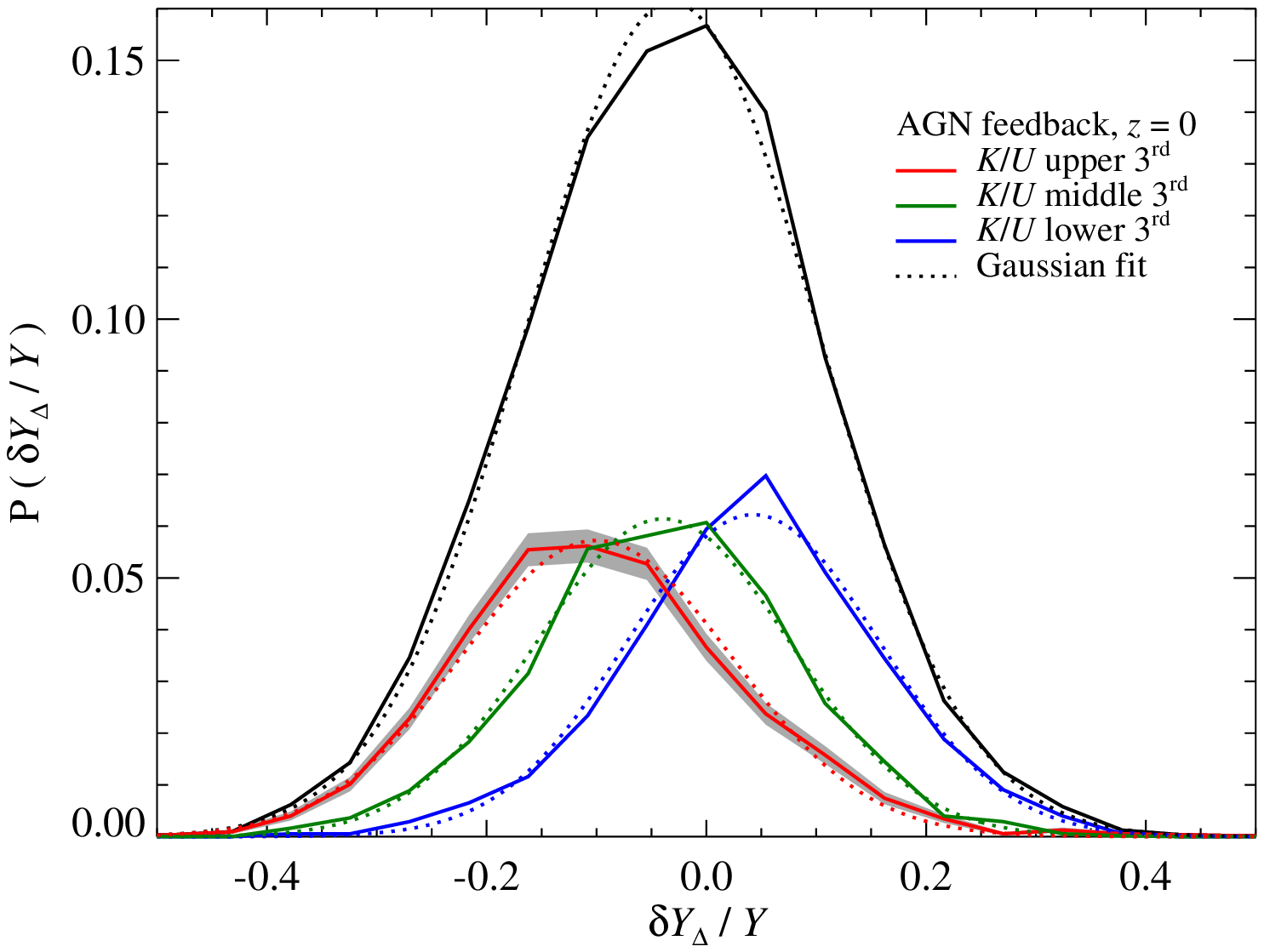}}\\
\end{center}
\caption{Sub-sampling the \YM relations by the kinetic-to-thermal energy ratio
  ($\KU$) for the AGN simulations. Left: The \YM scaling relation for the three
  $\KU$ sub-samples, upper $3^{\mathrm{rd}}$ (red), middle $3^{\mathrm{rd}}$
  (green) and lower $3^{\mathrm{rd}}$ (blue), with the corresponding slope
  fitted to those points. The $y$-axis has been scaled by $M^{5/3}$ to highlight
  the deviations from self-similarity. Right: The probability distributions for
  the relative deviation, $\delta Y_\Delta / Y$, with respect to the best fits for the three
  sub-samples and the total distribution (black), including the Gaussian fits
  (dotted lines) and the Poisson deviations for the upper $3^{\mathrm{rd}}$
  sub-sample (grey band). The sub-sample of $\KU$ with the largest kinetic
  pressure support (upper $3^{\mathrm{rd}}$) shows systematically lower total
  $Y$ values for a given mass as well as larger scatter, while the lower $\KU$
  sub-sample has the lowest scatter of $\sim 11$\%. This is expected because a
  larger kinetic pressure contribution implies a lower thermal pressure and
  hence decreases $Y$.}
\label{fig:YM_KU}
\end{figure*}

\subsection{Physics dependence of the \YM relation}

In Figure \ref{fig:YMrelphys} we show the dependence of the \YM relation on our
three simulated physics models, i.e., shock heating, radiative cooling and star
formation, and AGN feedback. The stark differences between the shock heating
simulation and the two radiative simulation models arise from the loss of
baryons in the ICM to star formation. The radiative cooling simulations show a
constant normalization offset of $\sim 20 \%$, which nearly matches the $f_*$
values for these simulations. In Table \ref{tab:YMs} we show that the
self-similar expectation of Eq.~(\ref{eq:Ysim}) almost completely captures the
cluster thermodynamics in our simulations when integrated over cluster-sized
apertures.  Including more physically motivated sub-grid models in the
simulations, we find that both, the shock heating and radiative cooling slopes
are consistent with this self-similar derivation for the \YM relation, while the
AGN feedback simulations have a steeper, mass-dependent slope. This break from
self-similarity in the AGN simulations arises from the suppression of star
formation in the higher mass clusters and a feedback-induced deficit of gas
inside the lower mass clusters.  Over the redshift ranges we explore ($z = 0$ to
$z = 1.5$) and for all simulated physics models, the \YM scaling relation
normalization changes as predicted by self-similar evolution and the slopes
remain essentially constant (cf. Fig.~\ref{fig:YM_zfunc}). So, the \YM relations
from AGN simulations are different at $z=0$, but evolve as predicted by
self-similar evolution. This result is independent of the two aperture sizes
chosen, which correspond to over-densities of 200 and 500 times the critical
density (Fig.~\ref{fig:YM_zfunc}). As we have repeatedly found in previous
sections, the clusters interior to the radii $R_{500}$ and even $R_{200}$ are
relatively well behaved, with only modest impact of cluster ellipticities and
kinetic pressure.

To quantify the scatter, we compute the relative deviation of each cluster from
the mean relation, $\delta Y_{\Delta} / Y = (Y_{\Delta} -
Y_{\Delta,\rmn{fit}})/Y_{\Delta,\rmn{fit}}$, and then fit this distribution with
a Gaussian probability distribution function (PDF),
\begin{equation}\label{eq:yvar}
G(\delta Y_{\Delta}/Y) = A_0 \rmn{exp} \left[\frac{-(\delta
Y_{\Delta}/Y)^2}{2\sigma_Y^2} \right].
\end{equation} 
\noindent Here the parameter $A_0$ is the normalization and $\sigma_Y$ is the
variance, which we will refer to as the scatter. Here we have chosen to model
the variation about the mean as a Gaussian, while previous work by
\citet{2010ApJ...715.1508S} showed that a log-normal distribution is also a
reasonable description of the scatter. In Appendix~\ref{sec:Gaussian}, we show
that within the (Poisson) uncertainties, the scatter is clearly Gaussian
distributed and only approximately log-normal. Forcing a log-normal distribution
introduces higher-order moments such as skewness and kurtosis as can be seen by
the tails in the distributions and their asymmetric shapes.

\begin{figure*}
\begin{center}
  \hfill
  \resizebox{0.5\hsize}{!}{\includegraphics{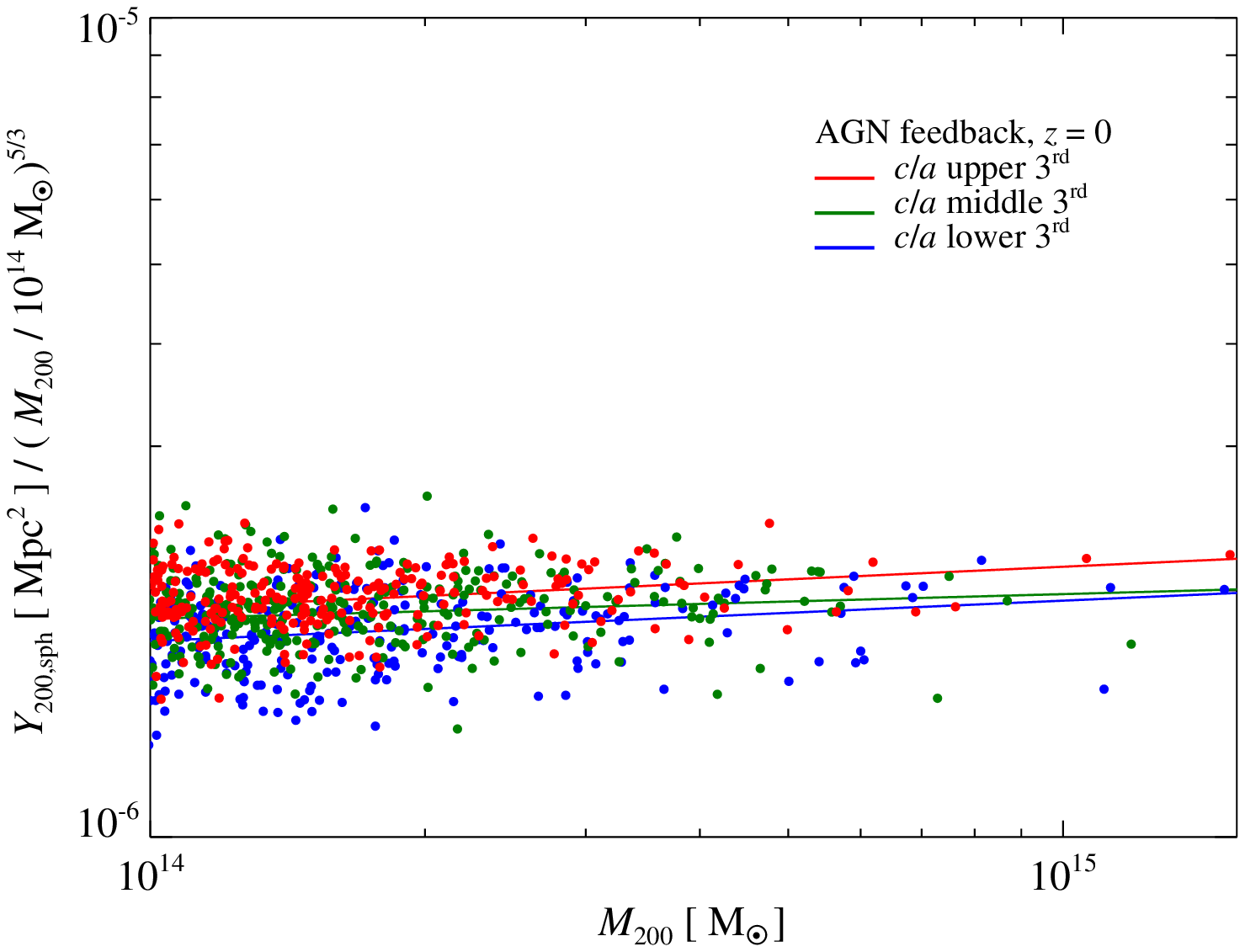}}%
  \resizebox{0.5\hsize}{!}{\includegraphics{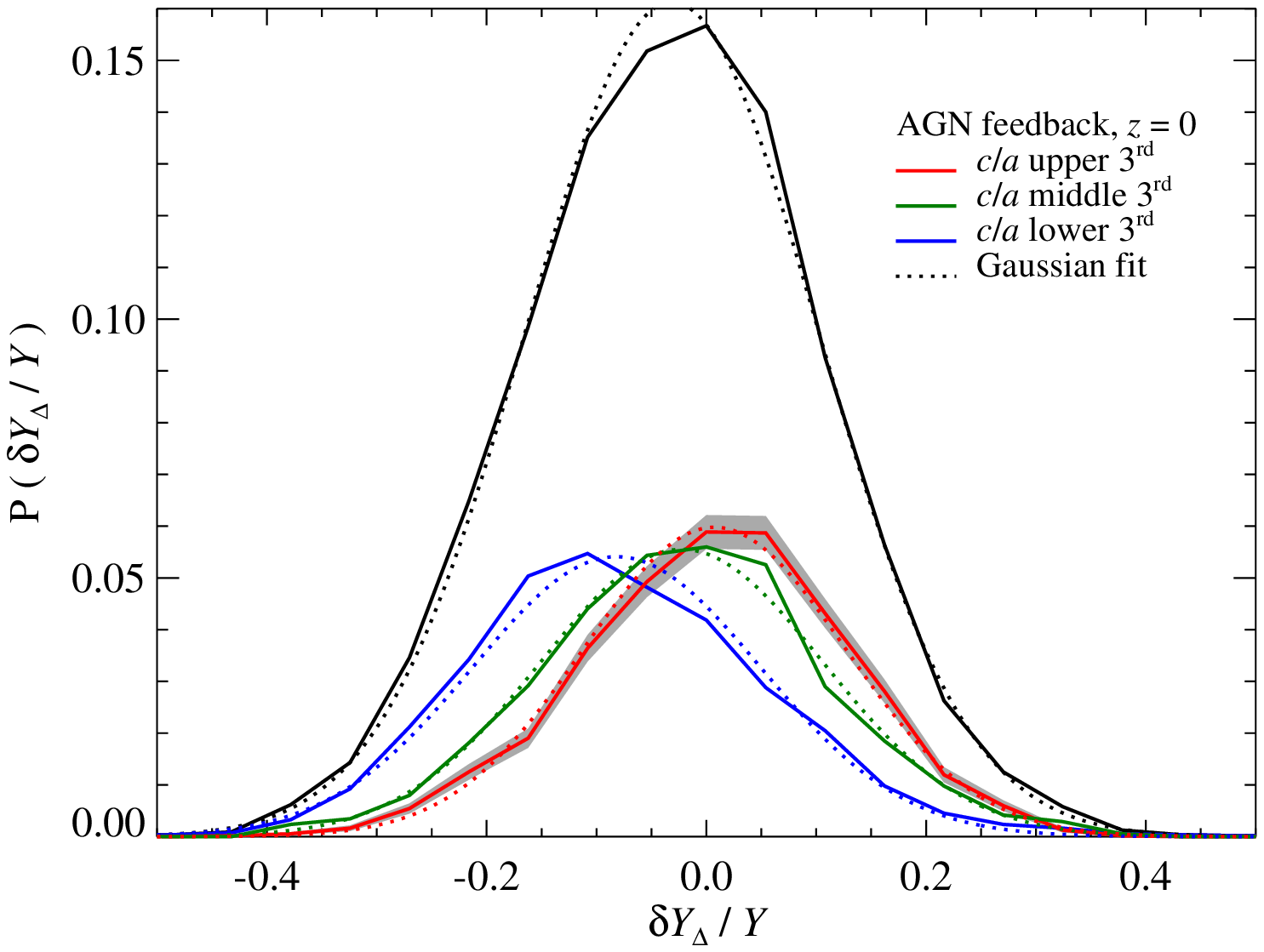}}\\
  \resizebox{0.5\hsize}{!}{\includegraphics{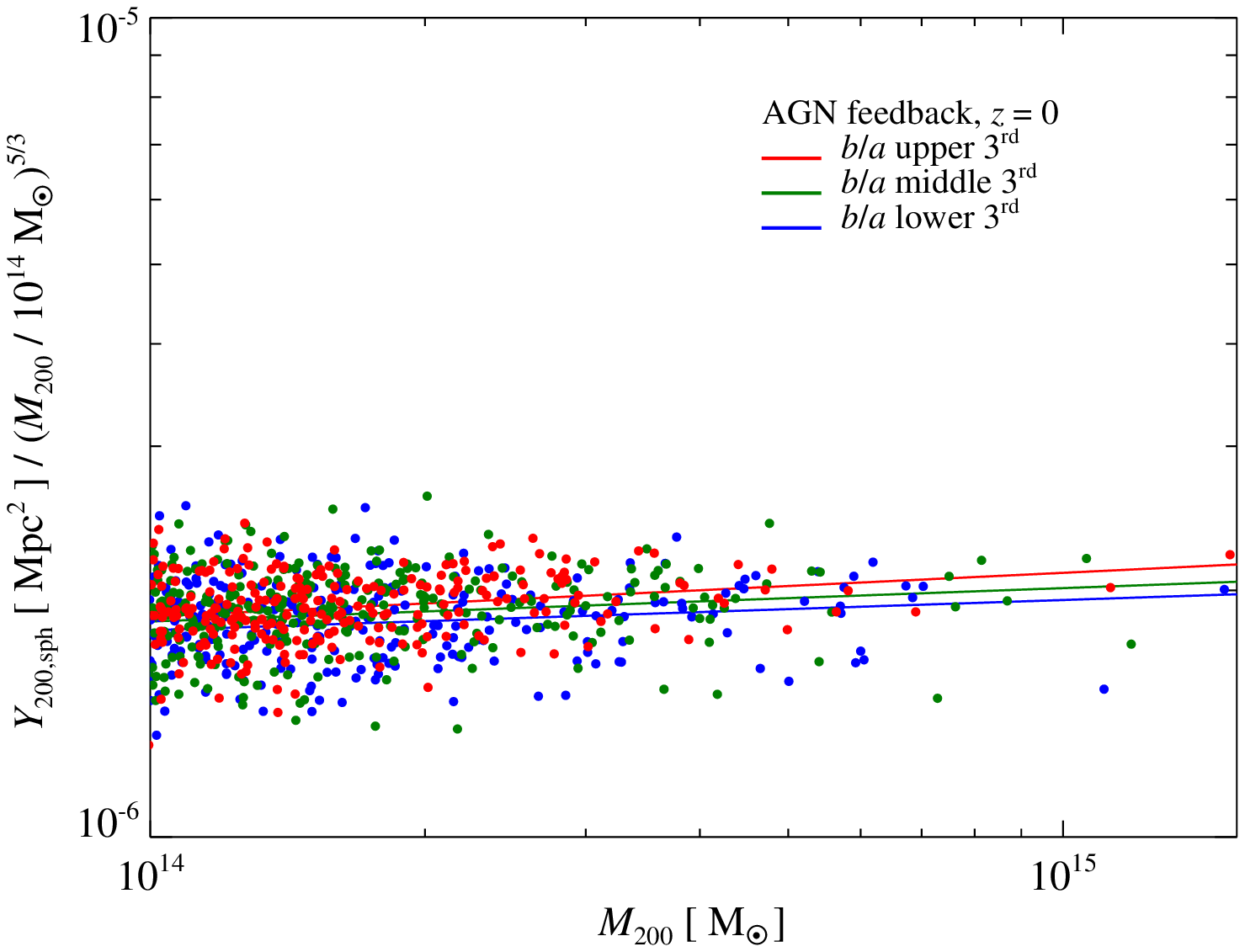}}%
  \resizebox{0.5\hsize}{!}{\includegraphics{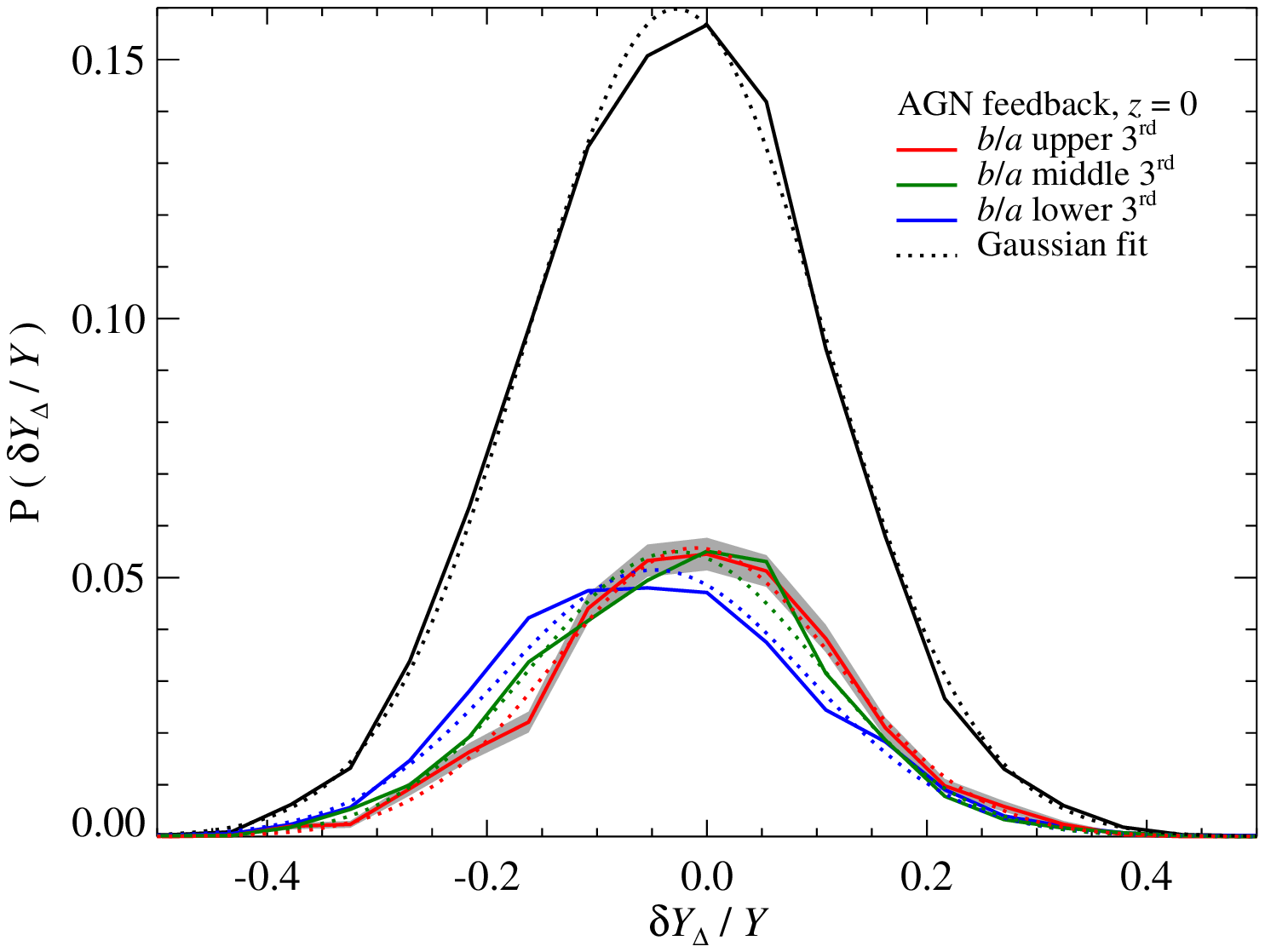}}\\
\end{center}
\caption{Sub-sampling the \YM relations by the gas $\ca$ axis ratio (upper
  panels) gas $\ba$ axis ratio (lower panels) for the AGN simulations. Left: The
  \YM scaling relation for the three $\ca$ sub-samples, upper $3^{\mathrm{rd}}$
  (red), middle $3^{\mathrm{rd}}$ (green) and lower $3^{\mathrm{rd}}$ (blue),
  with the corresponding slope fitted to those points. The y-axis has been
  scaled by $M^{5/3}$ to highlight the deviations from self-similarity. Right:
  The probability distributions for the relative deviation, $\delta Y_\Delta /
  Y$, with respect to the best fits for the three sub-samples and the total
  distribution (black), including the Gaussian fits (dotted lines) and the
  Poisson deviations for the upper $3^{\mathrm{rd}}$ sub-sample (grey band). The
  sub-sample of $\ca$ containing the lowest values (largest ellipticities) shows
  systematically lower total $Y$ values for a given mass and larger scatter,
  while the more spherical high $\ca$ sub-sample shows a lower scatter of $\sim
  12$\%. Additionally we find that the pressure $\ca$ axis ratio sub-sample has
  similar results (cf.Table \ref{tab:YMs}).  The sub-sample of $\ba$ containing
  the lowest values (largest projected ellipticities) shows larger scatter than
  the sub-sample with the highest values by $\sim 23$\%. There are
  significant changes in the \YM relation at $z=0$, however, smaller in
  comparison to the $\ca$ sub-sampling.}
\label{fig:YM_CA}
\end{figure*}

\begin{figure}
\epsscale{1.20}
\plotone{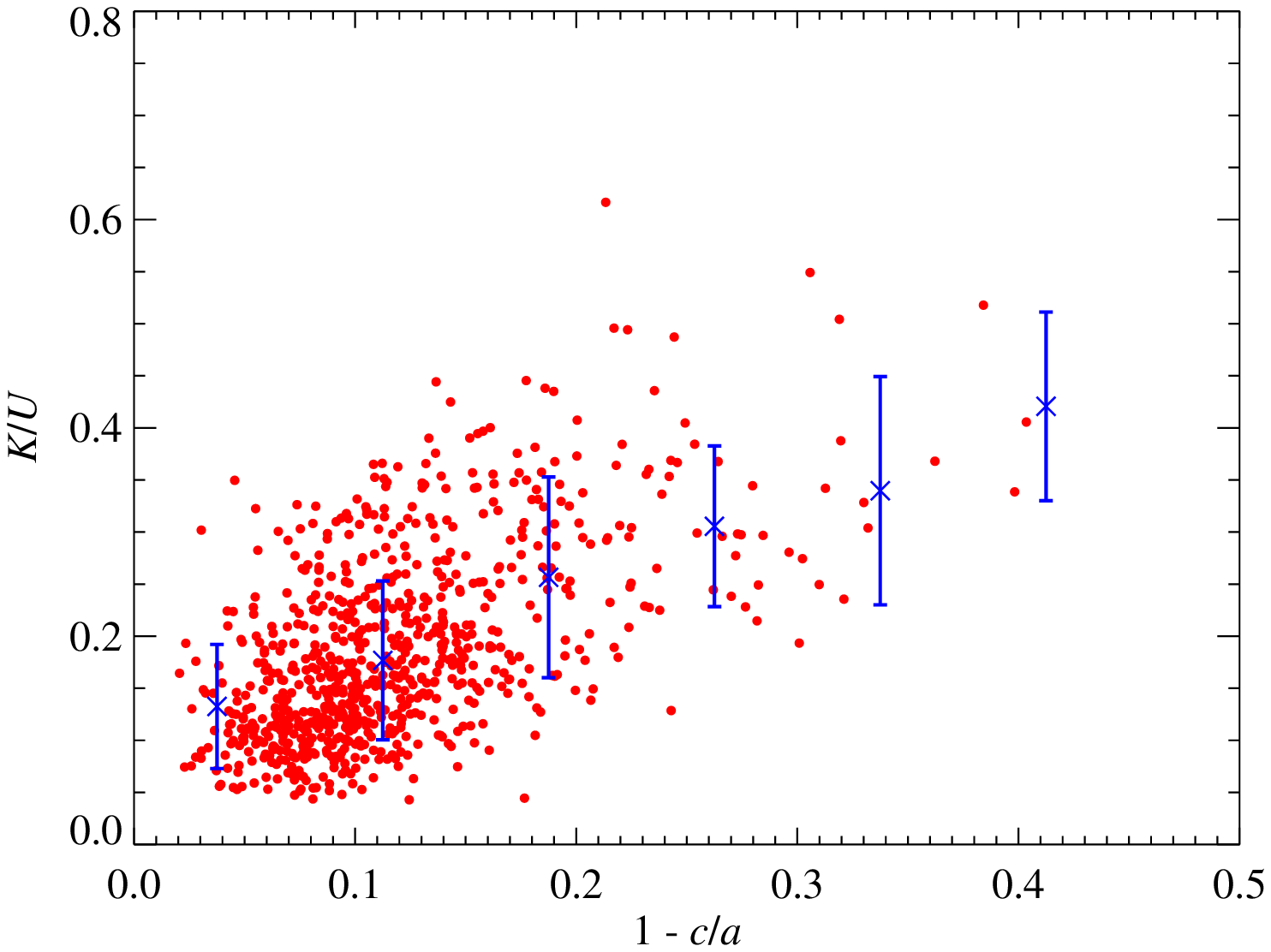}
\caption{The correlations between the $\KU$ ratio and $1\,-\ca$. Here
the red points represent each cluster in the simulations and the blue
crosses are average quantities. The linear correlation coefficient is 0.575}
\label{fig:CORREL}
\end{figure}

We find that the scatter, $\sigma_Y$, for the entire sample of clusters is
between 11 \% and 13 \% (cf. Fig. \ref{fig:YMrelphys} and Table \ref{tab:YMs}),
which is consistent with previous work
\citep{2006ApJ...650..538N,2008ApJ...686..206S,2010ApJ...715.1508S,2010ApJ...725.1124Y}. In
the simplest simulations with only shock heating the source for this scatter in
the \YM relation has been proposed to arise from the formation time, the
concentration, and the dynamical state of the cluster
\citep{2010ApJ...725.1124Y}. As our simulations include more sub-grid physics
models the scatter increase from $\sim 11$\% to $\sim 13$\% at $z=0$ and changes
further from $\sim 11$\% to $\sim 15$\% at $z=1$. Of the three different physics
models, the simulations with AGN feedback gives the largest scatter, which is
consistent with semi-analytic results \citep{2008ApJ...686..206S}. This model
for AGN feedback is self-regulated \citep{2010ApJ...725...91B} and injects $\sim
2/3$ of the energy before $z=1$ when the average cluster mass is significantly
smaller and the associated potentials are shallower so that a fixed energy
injection by AGNs may in principle have a stronger impact\footnote{Similar
  results were found by \citet{2011MNRAS.412.1965M} in simulations with a more
  detailed feedback prescription.}. Thus, the increased scatter in the \YM
relation from the AGN feedback simulations compared to the simulations without
feedback is a result of the energy injection, which heats and disturbs the
ICM. This statement is in accordance with previous results from
\citet{2010ApJ...725...91B}, where they showed the impact of AGN feedback on the
pressure profiles of clusters and found that simulations with feedback had a
shallower asymptotic pressure profile slopes than those without feedback. Thus,
the intermittent nature of energy injection into the group system early-on
results in a larger scatter in the \YM relation compared to simulations without
energetic feedback.

\subsection{Cylindrical Apertures}

For pointed SZ observations of clusters and SZ surveys, a natural,
model-independent observable is the projected flux, $Y_{\rmn{cyl}}$
\citep{2009ApJ...694.1034M,2011ApJ...728...39S}. We find
$Y_{\rmn{cyl}}>Y_{\rmn{sph}}$ in all cases, whether we chose the projection
along a principal or a random axis. This is due to the assumed extension along
the line-of-sight integration which we choose to be three times the aperture
radius; in observations, structure beyond this scale may additionally contribute
in some cases.  In fact, a projection integral out to $3R_{200}$ decreases the
\YM slope for the AGN feedback simulations such that it becomes consistent with
the self-similar slope (cf.{\ }Table \ref{tab:YMs}). We find no difference
between the random 2D projections and the integration along the middle or minor
axes with respect to the normalization and slope (cf.{\ }Fig.  \ref{fig:YM_CYL}
and Table \ref{tab:YMs}). The scatter for the random 2D projections is
marginally larger than the projections along middle and minor axes. Our results
show that the integration along the major axis yields dramatically different
results, both, for the normalization and scatter in comparison to projections
along the other axes. This has its origin in the more extended tails of the PDF
(cf.{\ }Fig \ref{fig:YM_CYL}). The normalization and scatter between the major
axis and the other axes increase by $\sim 7$\% and $\sim 9$\%, respectively. At
a higher redshifts, these differences are amplified and we find a $\sim 12$\%
increase in the normalization and an increase in scatter by $\sim 16$\%. This
indicates that substructure is preferentially aligned with major axis and that
substructure heavily influences the result from the moment-of-inertia tensor
beyond $R_{200}$.

\subsection{Toward a fundamental plane of \YM}

After quantifying the scatter of the entire sample, we aim at understanding its
origin. This may enable us to either construct a linear combination of
physically motivated observables that minimizes the scatter or to employ
sub-sampling of the full distribution according to some parameter so that the
resulting distribution exhibits a smaller intrinsic scatter and potentially
allows for tighter cosmological constraints \citep[e.g.,][]{2008ApJ...686..201A}.

In the previous sections we explored the average radial trends for kinetic pressure
support from bulk motions and gas density/pressure shapes of the ICM. Utilizing
this information, we rank order clusters according to their kinetic pressure
support and intrinsic shape information. We follow the same fitting procedure as
above for subsets of the lower $3^{\mathrm{rd}}$, middle $3^{\mathrm{rd}}$, and
upper $3^{\mathrm{rd}}$ of the correspondingly sorted distributions in order to
demonstrate the impact of kinetic pressure support and asphericity on the \YM
relation fits and scatter. For the rest of this section we concentrate our
analysis on the \YM relations of the AGN feedback simulations, since they show
the largest scatter (this will provide an upper limit on the scatter) and are
most likely our best representation of ``real'' clusters in comparison to the
other simulated physics models. We compute the ratio of kinetic-to-thermal
energy, $\KU$, within radial bins and use this ratio as a measure of dynamical
state for the galaxy clusters. We define the internal kinetic energy, $K$, and
thermal energy, $U$, of a cluster as
\begin{eqnarray}
\label{eq:KU}
K (<r) &\equiv& \sum_i \frac{3 m_{\rmn{gas},i}  P_{\mathrm{kin},i}}{2\rho_i}, \\
U (<r) &\equiv& \sum_i \frac{3 m_{\rmn{gas},i}  P_{\mathrm{th},i}} {2\rho_i},
\end{eqnarray} 
where $m$ and $\rho$ are the gas mass and the SPH density, respectively for all
particles $i$ less than radius $r$. The ratio $\KU$ is the volume integrated
analog of the ratio $P_{\mathrm{kin}}/P_{\mathrm{th}}$  shown in
\S~\ref{sec:Pkin} and hence is also an indicator of formation history and 
substructure. For the sub-sample with the highest ratio of
$\KU$, we find a smaller normalization (cf. Fig. \ref{fig:YM_KU} and Table
\ref{tab:YMs}) where the difference between this upper $3^{\mathrm{rd}}$ of the
distribution and lower $3^{\mathrm{rd}}$ is $\sim 15$\%. Here some of the
thermal pressure support has been compensated for by kinetic pressure support
resulting in lower integrated thermal electron pressure, thus, lowering
$Y$-values. More massive clusters are typically in the high
$\KU$ sample rather then the other two samples. We find that the sub-sample with
the smallest $\KU$ values shows the lowest scatter, $\sim 11$\% for the AGN
feedback simulations. Further sub-sampling of the smallest $\KU$ values (e.g.,
the lowest $6^{\mathrm{th}}$) does not decrease the scatter, which is limited to
$\sim 11$\%.

We also sort our cluster sample by the ratio of minor to major axis for both 3D
and projected 2D, $\ca$ and $\ba$\footnote{We show one projection for the $\ba$
  sub-sampling. The other projections yield similar results.} respectively, as
defined in Sec. \ref{sec:Elpt}. Following the same procedure as for the $\KU$
sub-sample and restricting ourselves to the AGN feedback simulations, we find
that splitting the clusters up by ellipticity, $\ca$, gives similar results in
comparison to $\KU$-splitting. The galaxy clusters with smaller ellipticities
have larger total $Y$ values and less scatter, while the more triaxial clusters
have lower total $Y$ and large scatter (cf. Fig. \ref{fig:YM_CA}). These trends
are reflected in the fit parameters of the sub-sample \YM relation shown in
Table \ref{tab:YMs}, where the differences between the upper $3^{\mathrm{rd}}$
and lower $3^{\mathrm{rd}}$ sub-samples normalization parameters is $\sim
12$\%. Additionally, we found that using the pressure shapes instead of the gas
shapes yield almost identical results (cf. Table \ref{tab:YMs}).  We find that
sub-sampling the clusters with the $\ba$ statistics has the greatest impact on
the scatter.  The sub-sample of clusters that appear to be elongated in the
plane of the sky have larger scatter than the more spherical clusters
(cf. Fig. \ref{fig:YM_CA} and Table \ref{tab:YMs}). Also, the $\ba$ sub-sampling
causes a similar bias in the \YM relation in comparison to the $\ca$
sub-sampling, but not as significant.

The results from sub-sampling clusters according $\KU$ and $\ca$ indicate that
there are correlations between these physical properties and the scatter in the
\YM relation. In Figure \ref{fig:CORREL} we show that larger $\KU$ ratios
correlate with larger $1-\,\ca$, i.e. larger triaxiality, with a linear
coefficient value, $r_{\rmn{s}} = 0.58$.  These correlations between kinetic
pressure support and ellipticity are the result of the growth of structure being
hierarchical.  This supports the argument that kinetic pressure support,
ellipticity and sub-structure are all tracers of the dynamical state and the
formation history of galaxy clusters, which is the ultimate cause of the
intrinsic scatter of the \YM relation. Similar results were found by
\citet{2011ApJ...729...45R} and \citet{2011arXiv1107.5740K} Previous work by
\citet{2010ApJ...725.1124Y} found mass trends in the measured scatter, which is
consistent with our findings after extrapolating their lower mass range to our
larger masses.  However, their conclusion is different from ours, since they
claim that the scatter is most sensitive to the DM concentration; a finding that
may partially be due to the insufficient resolution in their simulations.

\section{Discussion and Conclusions}
\label{sec:conclusion} 

In this paper we demonstrate that the spatial distribution of the ICM,
kinetic pressure support from bulk motions, and self-regulated thermal
energy feedback in clusters cores (that we refer to as AGN feedback)
all play very important roles for the thermal properties of galaxy
clusters. In particular, the observables for large SZ cluster
surveys, such as ACT, SPT and {\em Planck}, will be modified by these
processes. Below we highlight and expand on our main results.

{\em Non-thermal pressure support and cluster shapes:} The contribution to the
overall pressure support in clusters from bulk motions, $P_{\rmn{kin}}$,
increases substantially for larger radii and is a strong function of both,
cluster mass and redshift. Including AGN feedback marginally decreases
$P_{\rmn{kin}}/P_{\rmn{th}}$ in comparison to the other (more simplified)
simulation models, namely our shock heating-only model and that which
additionally includes radiative cooling, star formation, supernova feedback, and
CRs. However, the difference is not substantial enough to be statistically
inconsistent with the variance around the median of
$P_{\rmn{kin}}/P_{\rmn{th}}$. The mass dependence and redshift evolution of
$P_{\mathrm{kin}}/P_{\mathrm{th}}$ is governed by $P_{\mathrm{kin}}$ and a
direct result of the hierarchical growth of structure. Semi-analytic approaches
are just beginning to model $P_{\rmn{kin}}$. The full dependence on radius, mass
and redshift of this component is, by definition, self-consistently included in
hydrodynamic simulations. 

We find that the distribution of gas density and pressure are weak functions of
the simulated physics models within $R_{200}$ (excluding the cluster core) and
that AGN feedback mildly modifies the average gas shapes. The cluster mass
dependence of the ellipticity is more moderate in comparison to
$P_{\rmn{kin}}/P_{\rmn{th}}$. The ellipticity is small within $R_{500}$ with
little redshift evolution. In combination with the comparably small non-thermal
pressure support at these scales (which rises dramatically beyond this
characteristic radius), the small clumping factor measured in our simulations (cf. BBPS4),
and the small modification of our simulated cluster physics at these radii (in
particular of our implementation of AGN feedback), this result is reassuring for
X-ray observations of clusters which use $R_{500}$ to characterize clusters with
high-quality {\em Chandra} and {\em XMM Newton} observations. Hence, our
analysis theoretically supports this choice of radius (which was initially motivated by the simulations in \citet{1996ApJ...469..494E}) and justifies some of the main assumptions such as spherical symmetry and
an almost radius-independent hydrostatic mass bias of $\sim20-25$\% when using a
fair sample of clusters without morphological selection which may be applicable
for the future eROSITA sample.

We find substantial redshift evolution in different dynamical quantities, e.g.,
$P_{\rmn{kin}}/P_{\rmn{th}}$, the velocity anisotropy, and anisotropy parameters
such as ellipticities. This is in particular the case for the changes in the
power-law behaviors of the radial profile of these quantities such as the sudden
break in ellipticities which moves to smaller radii as the redshift increases
(when scaled to $R_{200}$). The break and the more pronounced ellipticities and
$P_{\rmn{kin}}/P_{\rmn{th}}$ outside a characteristic radius are a direct result
of increased level of substructure predicted by hierarchical structure formation
and the associated higher mass accretion rate at higher redshift. We explicitly
show (in the Appendix \ref{sec:rad}) that most of this redshift evolution is somewhat
artificial and can be absorbed in a re-definition of the virial radius: scaling
with the radius that contains a mean density of 200 times the average mass
density rather than the critical density of the universe considerably weakens
the observed trends with redshift. This also suggests a physical definition of
the virial radius in terms of dynamical quantities (that, however, remain
poorly defined observationally), e.g., the equipartition radius of thermal and
kinetic pressure, the region where the velocity anisotropy becomes strongly
radial, or the radius at which the ellipticity or substructure level increases
dramatically. These seemingly different criteria all select a rather similar
radius around $R_{200,\rmn{m}}$; almost independent of redshift.

On scales $>R_{500}$, stacking analyses of projected SZ cluster images can be
done with data from SZ experiments such as ACT and SPT, and, provided there is a
suitable sample size, one may be able to detect projected gas pressure shapes,
potentially even in bins of redshift.  The results on the randomly projected 2D
axis ratios represent the theoretical expectations.  Any statistics from the
intrinsic 3D distribution is highly correlated with the projected 2D
distribution; we find that the (more elliptical) intrinsic cluster shapes can on
average be inferred from their projected analogues by applying a $\sim 5-10$\%
correction on the ellipticity.  Another interesting outcome from our shape
analysis is that there is no direct and simple mapping of shapes and alignments
for DM spatial distribution to the gas and pressure distributions possible
mostly due to the difference in substructure distribution and dissipational
nature of the gas.  This result is troublesome for semi-analytic models which
use dissipationless simulations as a template to solve for the gas distributions
and pressure shapes. Such a method will produce additional triaxiality and
misalignment for such a semi-analytical model of the ICM. The overall magnitude
of the shape is reconciled by using the gravitational potential
\citep[e.g.,][]{2005ApJ...634..964O,2009ApJ...700..989B,2011ApJ...727...94T}
which has been shown to be less triaxial \citep{2011ApJ...734...93L} than the
DM. However, providing an algorithm to re-alignment these pseudo gas
distributions is a non-trivial task.

{\em \YM scaling relations:} Our simulations are in good agreement
with the current \YM scaling relations from both X-ray observations
and SZ surveys. However, to properly predict the \YM scaling relations
for an SZ experiment such as ACT, SPT or {\em Planck} without any
prior knowledge of cluster masses, careful mock observations are
needed. Those would have to include a simulation of the CMB sky with
associated experiment noise and adopt the relevant cluster selection
pipelines for the given experiment that employs the same cluster
profile used for matched filtering in order to include all the
systematics and potential biases that are intrinsic to the data
analysis, e.g., X-ray priors on the aperture size.

We find that the inclusion of AGN feedback causes a deviation from the
predictions of self-similar evolution for both the normalization and
slope of the \YM relation (as measured within $R_{200}$). However, we
recover the self-similar slope again in our projected \YM scaling
relations (where we integrate along a cylinder of half-height $3
R_{200}$), suggesting that AGN feedback pushes a fraction of its gas
beyond the virial radius and a larger aperture/projection radius is
able to recover the thermal energy from this larger reservoir of gas.

Including AGN feedback also increases scatter in the \YM relation compared to
simulations that include shock heating alone, from $\sim 11$ \% to $\sim 13$
\%. Interestingly, sorting the clusters into sub-samples of $\KU$ and $\ca$ will
reduce this scatter; e.g., $\KU$ sub-sampling reduces the scatter from $\sim
12$\% to $\sim 11$\%.  We find that sub-sampling introduces only a small
(predictable) bias in the normalizations on the order of a few percent.  This
suggests that observational proxies for the dynamical state and ellipticities
may be used to construct a fundamental plane of the \YM relation. The scatter
ultimately originates from the merging history with its redshift and mass
dependent accretion rates; those determine the non-thermal pressure support, the
level of substructure, and the ellipticity. While sub-sampling on one of these
secondary tracers may decrease the scatter, it is unlikely to decrease much more
if more tracers are used (as they probe the same underlying process, albeit with
a different weighting).  Conversely, our band-splitting analysis on the \YM
relations suggests that large outliers from the mean relations would be
interesting candidates for follow up with high resolution SZ observations, since
they are more likely to have larger kinetic pressure support and ellipticities.

A fundamental point to take away is that all results at larger radii
($>R_{200}$) for the kinetic pressure support and ICM shapes are
dominated by substructure. We also see the impact of substructure on
the cylindrical \YM scaling relation when integrating along the major
axis with which substructure is preferentially aligned.  Quantifying
substructure statistically is difficult because of the problem of
double-counting: the large volume contained within the radius that
contains 95\% of the total SZ flux, $4R_{200}$, necessarily leads to
overlapping volumes of neighboring clusters, especially at
high-redshift. Thus, this property remains challenging to model
phenomenologically or analytically.

As discussed previously in the literature
\citep[e.g.][]{2010ApJ...725...91B,2011ApJ...727L..49S} SZ galaxy
cluster may provide further insight into the interesting astrophysics
associated with the ICM of clusters. This however may
significantly complicate cosmological analyses in producing competitive
constraints. However, these are exciting prospects for studies of
feedback and other energy injection processes within clusters
especially at higher redshift since the selection function of SZ
cluster surveys probes  clusters which populate the massive
and high redshift end of the distribution.

\acknowledgments

We thank Mike Nolta, Norm~Murray,
Hy~Trac, Gus Evrard, Alexey Vikhlinin, Andrey Kravtsov, Laurie Shaw, Doug Rudd,
and Diasuke Nagai for useful discussions. Research in Canada is supported by
NSERC and CIFAR. Simulations were run on SCINET and CITA's Sunnyvale
high-performance computing clusters. SCINET is funded and supported by CFI,
NSERC, Ontario, ORF-RE and UofT deans.  C.P. gratefully acknowledges financial
support of the Klaus Tschira Foundation. We also thank KITP for their hospitality during the 2011 galaxy
cluster workshop. KITP is supported by National Science Foundation under Grant
No. NSF PHY05-51164. 

\bibliography{bibtex/nab}
\bibliographystyle{apj}

\clearpage

\begin{appendix}

\begin{figure}
\epsscale{1.20}
\plotone{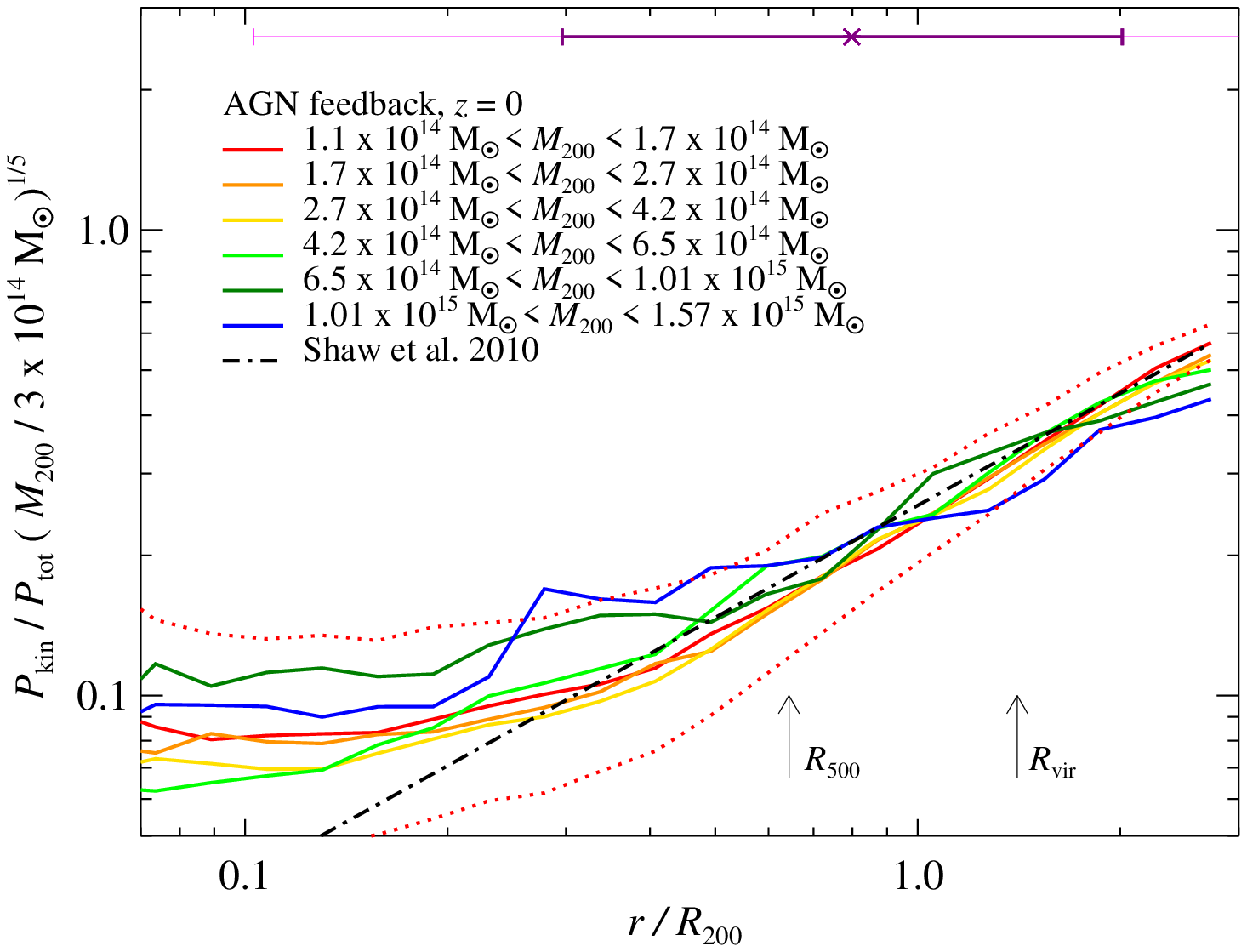}
\caption{The kinetic pressure-to-total pressure is weakly mass-dependent,
  $P_{\mathrm{kin}}/P_{\mathrm{tot}}\propto M_{200}^{1/5}$ as indicated by the
  scaling of the $y$-axis. Shown is the median of
  $P_{\mathrm{kin}}/P_{\mathrm{tot}}$ as a function of radius for the AGN
  feedback simulations for various mass bins with the 25$^{\rmn{th}}$ and
  75$^{\rmn{th}}$ percentile values illustrated by the dotted lines for the
  lowest mass bin at $z=0$. For comparison, we also show the model for 
  $P_{\mathrm{kin}}/P_{\mathrm{tot}}$ by \citet{2010ApJ...725.1452S}, which has
  been fit to match AMR simulations (dash-dotted). We illustrate the 1 and 2
  $\sigma$ contributions to $Y_{\Delta}$ centered on the median for the feedback
  simulation by horizontal purple and pink error bars. Therefore, ignoring this
  mass dependence results in a 60 \% difference in this ratio for an order of
  magnitude change in the cluster mass. The median of
  $P_{\mathrm{kin}}/P_{\mathrm{th}}$ scales as $M_{200}^{1/3}$, which results in
  a larger difference.}
\label{fig:pkinfit}
\end{figure}

\begin{figure}
\epsscale{1.20}
\plotone{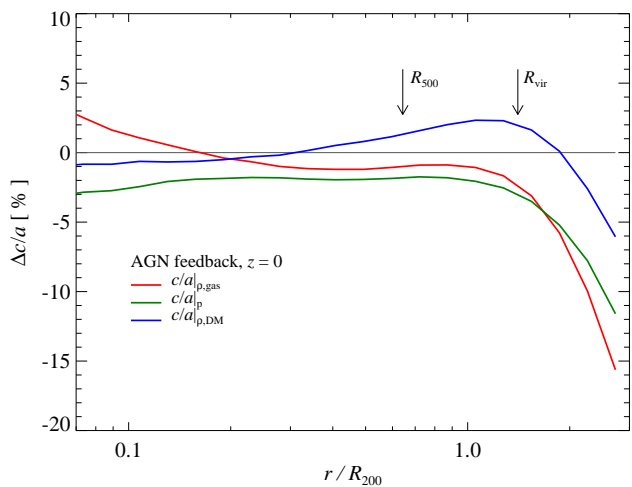}
\caption{Shown is the relative difference between axis ratios with and without
  the $r^{-2}$ weighting for the gas-density (red line), DM-density (blue line)
  and gas-pressure (green line) weightings.  Additionally including the $r^{-2}$
  weighting in the definition of the moment-of-inertia tensor down-weights the
  contribution at larger radii by $\lesssim 15$\%, thus reducing the effect of
  substructure. }
\label{fig:C_A_r2}
\end{figure}

\begin{figure*}
\begin{center}
  \hfill
  \resizebox{0.5\hsize}{!}{\includegraphics{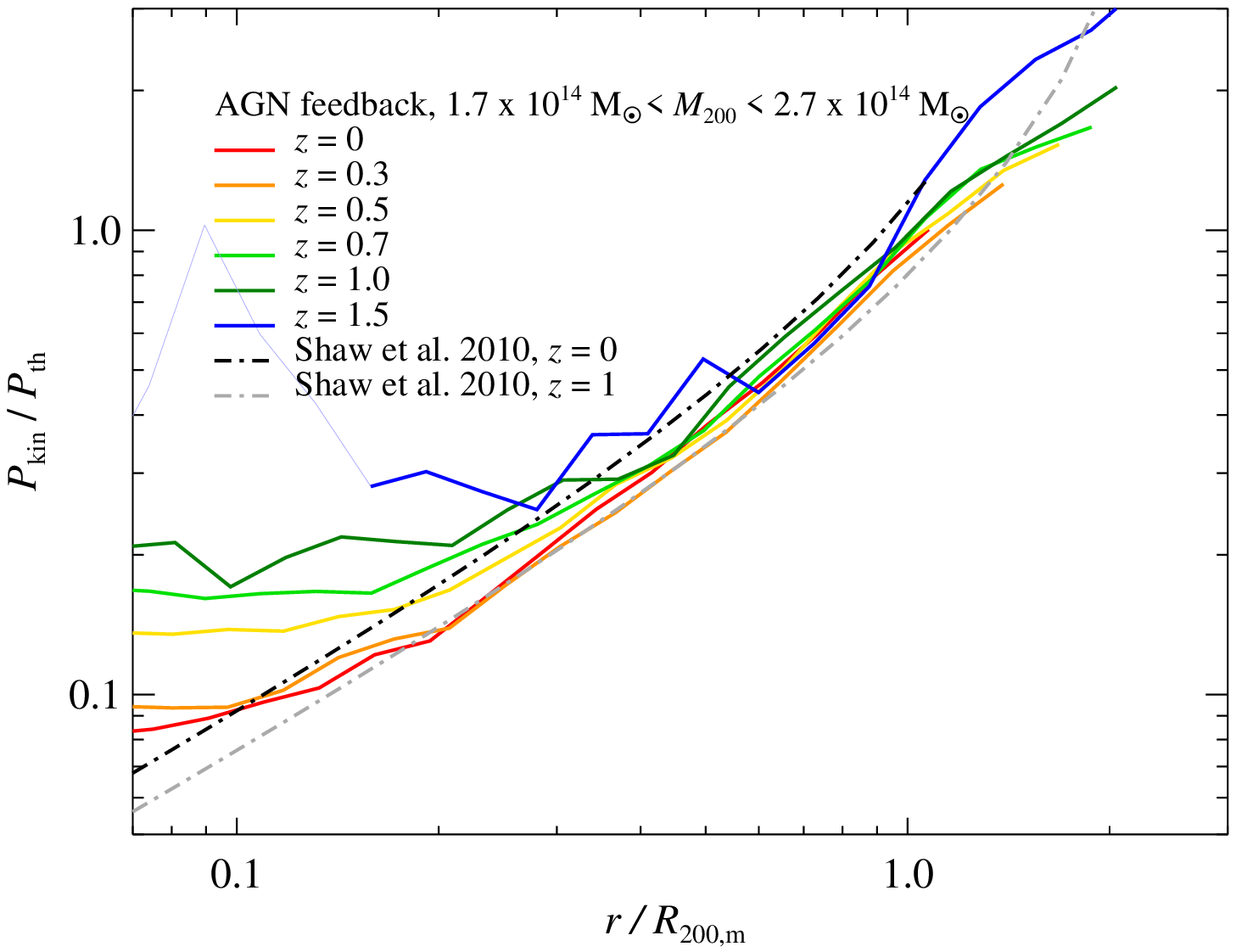}}%
  \resizebox{0.5\hsize}{!}{\includegraphics{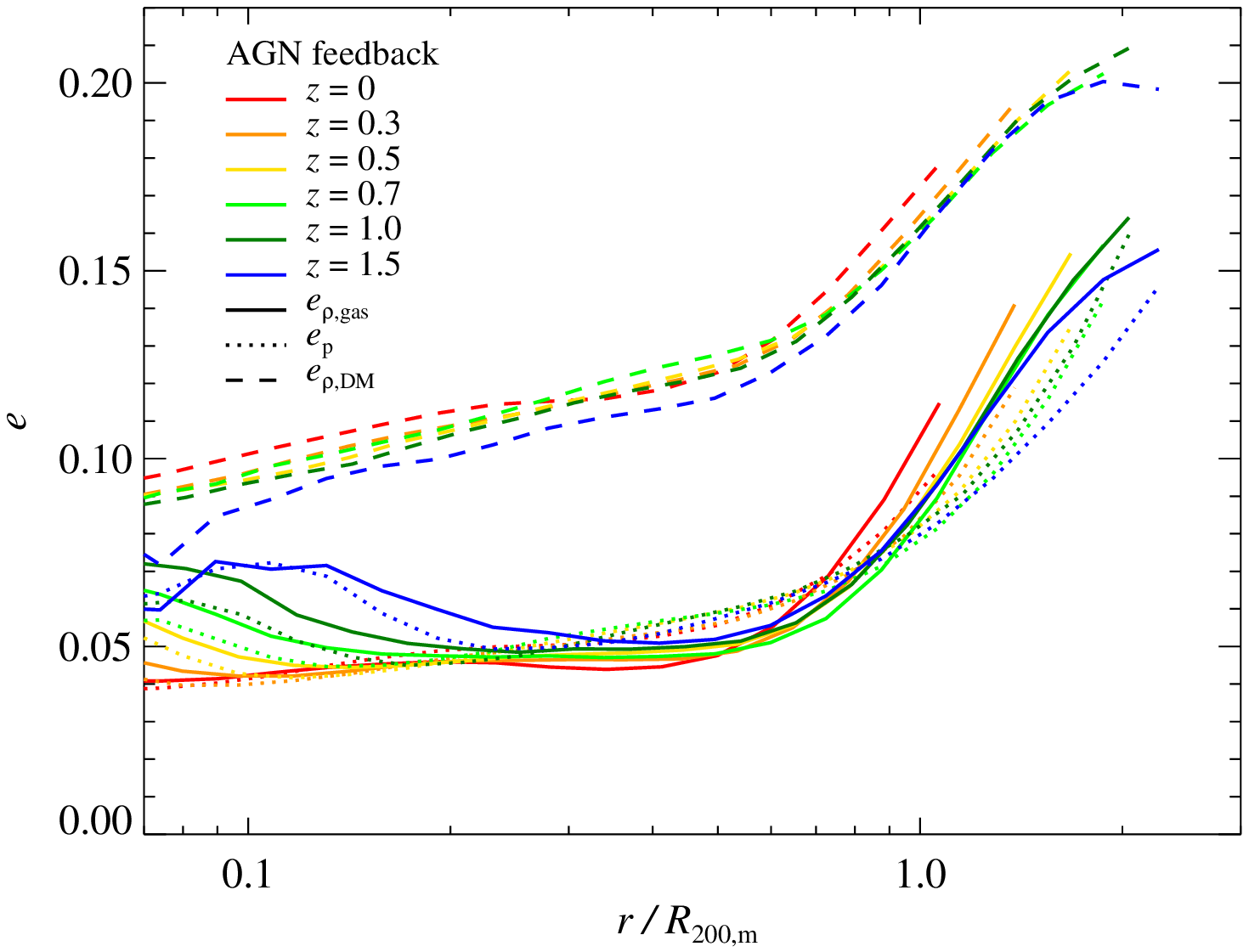}}\\
\end{center}
\caption{The choice for our working definition of virial radius has an impact on
  the redshift evolution of both, the kinetic pressure support (left) and
  ellipticity (right) of clusters. The figures shown here are the same as
  Fig. \ref{fig:pratio_mass} and Fig. \ref{fig:eliptmassbins} except that the
  dimensionless radius has been scaled by $R_{200,\mathrm{m}}$ instead of
  $R_{200}$. With this definition of virial radius, the redshift evolution of
  both, kinetic pressure support and ellipticity is weakened, especially in the
  outer regions.}
\label{fig:Pkin_e_zevo_renorm}
\end{figure*}

\section{Fitting function for $P_{\mathrm{kin}}/P_{\mathrm{tot}}$}
\label{sec:fit_Pkin_overPtot}

In Section \ref{sec:Pkin} we show that the ratio
$P_{\mathrm{kin}}/P_{\mathrm{th}}$ is a function of mass. However, the previous
empirical fitting function for $P_{\mathrm{kin}}/P_{\mathrm{tot}}$
\citep{2010ApJ...725.1452S} does not include a mass dependence,

\begin{equation}
\frac{P_{\mathrm{kin}}}{P_{\mathrm{tot}}} (r,z) = \alpha(z) \left(\frac{r}{R_{500}}\right)^{n_{\rmn{nt}}}\,
\left(\frac{M_{200}}{3\times10^{14} \rmn{M}_\sun}\right)^{n_M},
\label{eq:pkinfit}
\end{equation}

\noindent where $\alpha(z) \equiv \alpha_0 (1 + z)^{\beta}$ for low redshifts
($z \lesssim 1$) and the fit parameters are $\alpha_0=0.18\pm0.06$, $\beta=0.5$,
$n_{\rmn{nt}}=0.8\pm0.25$, and by construction, $n_M=0$. In
Fig. \ref{fig:pkinfit} we compare the fitting function for
Eq.~(\ref{eq:pkinfit}) and $P_{\mathrm{kin}}/P_{\mathrm{tot}}$, split by
different mass bin which have been scaled by $M_{200}^{1/5}$, i.e. $n_M = 1/5$,
that minimizes our $\bchi^2$. We chose a normalization of $3\times10^{14}
\rmn{M}_{\sun}$ to match the fitting function of
\citet{2010ApJ...725.1452S}. Thus, the median difference between
$P_{\mathrm{kin}}/P_{\mathrm{tot}}$ of a $10^{15} \rmn{M}_{\sun}$ and a
$10^{14} \rmn{M}_{\sun}$ cluster is $\sim 60$\%. The ratio in
Eq.~(\ref{eq:pkinfit}) is similar to what is shown Fig. \ref{fig:pratio_mass},
however, $P_{\mathrm{kin}}/P_{\mathrm{th}}$ depends more sensitively on mass. We
find that the mass dependence for this ratio amounts to $M_{200}^{1/3}$.

\section{Down-weighting the substructure in the moment-of-inertia tensor}
\label{sec:wr2}

For both the gas density- and pressure-weighting of the moment-of-inertia
tensor, the inclusion of an additional $x^{-2}$-weighting has a relatively minor
influence on cluster shapes (cf.  Fig. \ref{fig:C_A_r2}) and we do not see large
differences in the axis ratios at the larger radii. The $x^{-2}$-weighting does
lessen the influence of substructure which we have seen to be important at radii
beyond $R_{200}$, but it does not remove it or isolate its signal. This would be
a non-trivial task for any stacking analysis as it was recently suggested by
\cite{kravtsov2011recent}.

\section{Clusters in Velocity Space and a Dynamical Radius Definition}
\label{sec:rad}

\begin{figure}
\epsscale{1.20}
\plotone{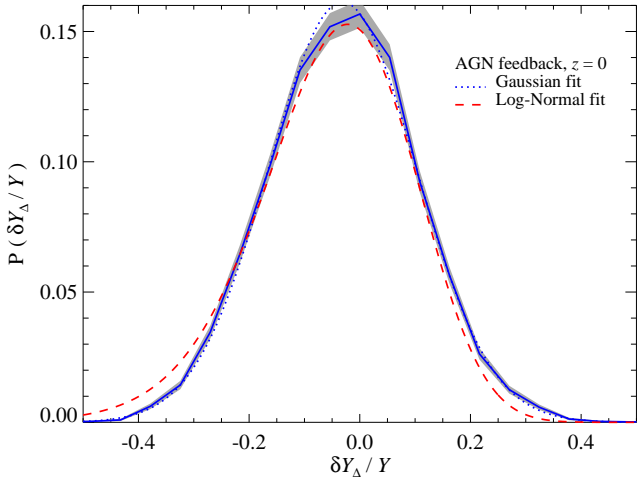}
\caption{Comparison of Gaussian and log-normal scatter relative to the
  best-fit \YM scaling relation at $z=0$. We show the distribution of
  the relative {\em linear} deviation from the mean relation $\delta
  Y_{\Delta} / Y$ (cf. Eq.~\ref{eq:yvar}) with the solid blue line and
  compare it to a Gaussian fit (blue dotted line) and log-normal fit
  ($\delta \log Y_{\Delta}$, red dashed line). The Poisson deviations
  are shown with the grey band.  Here we transformed the fit to the $\delta
  \log Y_{\Delta}$ distribution into $\delta Y_{\Delta} / Y$ so they
  could be shown together. The $\delta Y_{\Delta} / Y$ distribution is
  fit by a Gaussian better than the $\delta \log Y_{\Delta}$
  distribution, with $\bchi^2 (\delta Y_{\Delta} / Y) \sim 1$ and
  $\bchi^2 (\delta \log Y_{\Delta}) \sim 7$.  Forcing a log-normal
  distribution introduces higher-order moments such as skewness and
  kurtosis as can be seen by the asymmetric shapes of the tails in the
  log-normal fit.}
\label{fig:Gauss-lognorm}
\end{figure}

The radial trends over redshift seen in Figures \ref{fig:pratio_mass} and
\ref{fig:eliptmassbins} call for re-examination of the choice for the working
definition of radius, which is directly related to the definition of the cluster
mass \citep[see][for a more thorough discussion of cluster mass definitions in
dissipationless simulations]{2002ApJS..143..241W}. It has been the common choice
by both observers and theorists to define the mass within an radii where the
average overdensity is greater than a large multiple of a given background
density, such as $\rho_\rmn{cr}(z)$ and $\bar{\rho}_\rmn{m}(z)$. For low
redshift observations, the more popular definition has been the
$\rho_\rmn{cr}(z)$ as the iso-density surface, since no prior knowledge of
$\Omega_{\rmn{m}}$ is required. The question remains what definition is
physically more intuitive when comparing across various redshifts. At late times
$(z < 1)$, clearly the inclusion of the dark energy greatly influences the
redshift evolution of the critical density compared to the mean matter
density. For a hypothetical isolated non-accreting cluster using the
$R_{\Delta}$ definition will result in the cluster radius shrinking as time
approaches present day. Using the $R_{\Delta,\rmn{m}}$ scaling we find that the
radial regions at which kinetic pressure is in equipartition with thermal
pressure and the sharp break found in the ICM ellipticity align at $\sim
200R_{\Delta,\rmn{m}}$ (cf. Fig. \ref{fig:Pkin_e_zevo_renorm}).  In BBPS3 we
show that the velocity anisotropy has the same radial trends as in Figures
\ref{fig:pratio_mass} and \ref{fig:eliptmassbins} and that
$200R_{\Delta,\rmn{m}}$ traces a distinct dynamical region of clusters, the
splash-back radius, i.e., is caused by the turn-around of earlier collapsed
shells which minimizes the radial velocity component such that the tangential
components dominate the velocity.

\section{Gaussian or Log-normal scatter?}
\label{sec:Gaussian} 

Previous approaches quantified the scatter around the best-fit \YM scaling
relation with a log-normal distribution, i.e. they characterized the
distribution of $\delta \log Y_{\Delta} = \log Y_{\Delta} - \log
Y_{\Delta,\rmn{fit}}$ with a Gaussian. Deviations from this log-normal
distribution were computed with the Edgeworth expansion, introducing substantial
higher order moments, such as skewness and kurtosis
\citep[e.g.,][]{2010ApJ...725.1124Y}. Using a non-linear least squares approach we fit a Gaussian to both the   $\delta \log Y_{\Delta}$ and $\delta
Y_{\Delta} / Y$ distributions.
In Fig.~\ref{fig:Gauss-lognorm} we show
that $\delta
Y_{\Delta} / Y$ distribution is a better fit by a Gaussian within the (Poisson) uncertainties than the $\delta \log Y_{\Delta}$ distribution, with $\bchi^2 \sim 1$ compared to $\bchi^2 \sim 7$, respectively. Hence we
suggest to use relative {\em linear} deviation instead of log-normal scatter for future
characterizations of the scatter in \YM relation.

\end{appendix}

\end{document}